\documentclass[11pt,a4paper]{article}
\pdfoutput=1
\usepackage{jheppub}
\usepackage{graphicx}
\usepackage{dcolumn}
\usepackage{bm}
\usepackage{amsmath,amssymb,amscd}
\usepackage{longtable}
 \usepackage{slashed}

 \usepackage{caption}
\usepackage{subcaption}

\usepackage[all]{xy}

\newcommand{\be}{\begin{eqnarray}}
\newcommand{\ee}{\end{eqnarray}}

\newcommand{\bn}{\begin{enumerate}}
\newcommand{\en}{\end{enumerate}}

\newcommand{\beq}{\begin{equation}}
\newcommand{\eeq}{\end{equation}}
\newcommand{\bea}{\begin{equation}\begin{aligned}}
\newcommand{\eea}{\end{aligned}\end{equation}}

\def\CN{{\cal N}}

\newcommand {\CalI} {\mathcal I}

\def\dd{{\rm d}}

\def\det{{\rm det}}

\DeclareMathOperator{\Tr}{Tr}
\DeclareMathOperator{\im}{\mathbb{I}m}

\title{2D Seiberg-like dualities for orthogonal gauge groups}

\author[a,b,c]{Hyungchul Kim,}
\author[d]{Sungjoon Kim,}
\author[d]{Jaemo Park}

\affiliation[a]{
Center of Mathematical Sciences and Applications, Harvard University, \\
Cambridge, 02138, USA
}
\affiliation[b]{
Jefferson Physical Laboratory, Harvard University, \\
Cambridge, MA 02138, USA
}
\affiliation[c]{
Device Solution Business, Samsung Electronics Co., LTD, \\
Pyeongtaek, 17786, Korea
}
\affiliation[d]{
Department of Physics, POSTECH, \\
Pohang 790-784, Korea
}

\emailAdd{hyungchul\_kim@g.harvard.edu, \, h7.kim@samsung.com}
\emailAdd{sjkim0305@postech.ac.kr}
\emailAdd{jaemo@postech.ac.kr}

\abstract{
We consider the analogue of Seiberg duality for two-dimensional $\CN=(2,2)$ gauge theory with orthogonal gauge groups and with fundamental chiral multiplets proposed by Hori. Following Hori, when we consider $O(k)$ gauge group as the (semi)-direct product of $SO(k) \ltimes Z_2$, we have to consider two kinds of the theories $O_{\pm}(k)$ depending on the orbifold action of $Z_2$. We give the evidences for the proposed dualities
by working out the elliptic genus of dual pair. The matching of the elliptic genus is worked out perfectly for the proposed dualities.}

\begin{document}

\maketitle

\section{Introduction}
Recently there has been much progress on the understanding the Seiberg-like dualities in lower dimensions than 4.
With the help of the recently developed localization results, substantial evidences were cumulated in 2-dimensions and 3-dimensions.
There's a close relation between the dualities in 4-dimensions, 3-dimensions and 2-dimensions\cite{Aharony:2013dha}.
In 2-dimensions, Seiberg-like dualities  for $\CN= (2,2)$ $U(k)$ gauge theory with fundamental chiral multiplets  with/without anti-fundamental chiral multiplets
 were studied in \cite{Hori:2006dk, Benini:2012ui, Benini:2013nda, Benini:2013xpa, Benini:2014mia, Gadde:2013ftv, Gomis:2014eya, Gadde:2015wta}, and the elliptic genus was computed to give the evidences for such dualities. The peculiar feature is that such duality holds for asymptotically free theories as well, while in higher dimensions
 the duality holds for superconformal field theories (SCFT).
Also the analogue of the so-called Kutasov-Schwimmer-Seiberg dualities are studied by \cite{Cho:2017bhd} for $U(N)$ gauge group following the conjecture about such dualities for 2-dimensional theories in \cite{Gomis:2014eya} in the context of AGT correspondence. The authors of \cite{Gomis:2014eya}  gives the evidences for such dualities by working out the $S^2$ partition function.
 Furthermore these 2-dimensional  dualities are shown to be derived from their 3-dimensional or 4-dimensional analogues.

 When we consider the dualities involving orthogonal groups, many of the subtleties arise. Given the Lie algebra of $so(k)$, there are
 many possibilities of gauge groups such as $spin(k), SO(k), O(k)$. In 4-dimensions, we need more refinements due to the existence of
 different line operators. Thus there are rich structures of dualities involving $so(k)$ Lie-algebra and via dimensional reduction, equally
 rich structures of dualities in 3-dimensions \cite{Aharony:2013dha}. We also expect the similar rich structures in 2-dimensions concerning the dualities of orthogonal
 groups. Here we study the $N=(2,2)$ Seiberg-like dualities of orthogonal gauge groups with fundamental chiral multiplets. In fact such duality was proposed and studied by Hori some while ago \cite{Hori:2011pd} and recently studied in \cite{Closset:2017vvl}. Especially when we regard $O(k)$ gauge group as the (semi)-direct product of $SO(k) \ltimes Z_2$, there are two choices of $Z_2$ orbifold actions. Thus we have two kinds of $O(k)$ theories, denoted by
 $O_{\pm}(k)$ theories depending on the $Z_2$ actions.
 The pattern of the proposed dualities are
 \begin{align}
O_+(k)~&\longleftrightarrow ~ SO(N-k+1)\nonumber\\
SO(k)~&\longleftrightarrow ~ O_+(N-k+1)\\
O_-(k)~&\longleftrightarrow ~ O_-(N-k+1).\nonumber
\end{align}
where $N$ is the number of chiral multiplets.
We work out the elliptic genus of the proposed dualities for a few cases and show that the elliptic genus perfectly matches, thereby
providing important evidences for such dualities.
One subtlety arising for 2-dimensional theory is the existence of the non-compact Coulomb branches.
Whenever such non-compact branches exist, the computation of the elliptic genus is subtle and the blind computation of the elliptic
genus does not give the sensible answer. When we deal with dualities of $U(k)$ gauge group, we turn on FI-term or $\theta$ term to lift such
non-compact Coulomb branches. When we deal with $SO(k)/O(k)$ groups, we can turn on the discrete  $\theta$ angle.
Depending on $k, N$ some of the theories exhibit the non-compact Coulomb branches. They are called the irregular theories. If the non-compact
 Coulomb branches are lifted such theory is called regular. All tests of dualities have been checked for regular theories.
 For irregular theories, elliptic genus does not give the same answer for the candidate dual theories. The Witten index as a limit of
 the elliptic genus typically gives rational number for irregular theories. In this paper, we provide evidences for the dualities of the regular theories with orthogonal gauge groups.

 Also additional difficulty arises when evaluating the elliptic genus of the theory with $SO(k)/O(k)$ groups.
 For $U(k)$ theory with fundamental flavors and adjoint chiral multiplets, the general formulae of the elliptic genus are
 known since the classification of the nontrivial JK-residues are possible\cite{Cho:2017bhd}. However for  the theory
 with $SO(k)/O(k)$ gauge group with $N$ fundamental multiplets, such general results are not known. Thus we have to work out
 each case separately  so that check of the dualities are done for small rank of the gauge groups and small number of matter
 multiplets. Furthermore for  the $U(k)$ theory, equality of the elliptic genus of the dual pair can be shown analytically \cite{Cho:2017bhd}. Here we manage to prove the equality of the elliptic genus of the dual
 pair for small $k, N$ and numerically check the equality of the elliptic genus of the dual pair for more complicated cases.

 The contents of the paper are as follows.
 In the section 1, we introduce the $Z_2$ orbifold of the massive chiral multiplet as a warm-up of the study of $O(k)$ gauge group.
 In the section 2, we work out the elliptic genus for the proposed dual pairs of orthogonal gauge group and find the perfect match.
 In the appendix A, we work out the elliptic genus for the pure $SO(3)$ gauge theory. Depending on the mod 2 $\theta$ angle we can have
 both regular and irregular theory. We indicate the subtleties of the elliptic genus arising due to the existence of noncompact
 branches. Some useful formulae for the elliptic genus are also collected at the appendix B and C. The chiral ring structure and operator matching
 between the dual pairs are discussed at the appendix D. It's more convenient to use the superconformal index, which imposes NS boundary conditions on fermions, to enumerate the gauge
 invariant operators, though superconformal index and the elliptic genus are related by the spectral flow.
 At the appendix E, we provide the analytic proof of the equality of elliptic genus of the dual pairs for small $k, N$.

 {\bf Note} As this work is completed, we receive the paper by Aharony et al \cite{Aharony:2017adm}, which overlaps partially with our paper.
 See the appendix B of their paper. Also as the 2nd revision of our paper is completed, which includes the analytic proof of the elliptic
 genus of dual pairs with gauge group $SO(1), O_{\pm}(1), SO(2), O_{\pm}(2)$, the paper by Avraham et al \cite{Avraham} appears where the analytic proof of similar theories is worked out. On the third revision, we extend the analytic proof to simple cases of theories with $SO(3), O_{\pm}(3), SO(4), O_{\pm}(4)$ gauge group.

\section{Massive field and $\mathbf Z_2$ orbifolds}
Let us consider a free chiral superfield $\Phi$ with either a complex mass or a twisted mass. Because the theory has the mass gap, the Witten index of the theory is well-defined and count the number of supersymmetric vacua. We consider its $\mathbf Z_2$ orbifolds. There are two choices in the $\mathbf Z_2$ orbifold. The standard $\mathbf Z_2$ orbifold defined in \cite{Hori:2011pd} which acts as $\Phi \rightarrow -\Phi$ has the number of vacua
\begin{align}
\Tr (-1)^F = \left\{\begin{array}{ll}
1&\mbox{~for complex mass}\\
2&\mbox{~for twised mass}.
\end{array}\right.
\end{align}
 The theory of a chiral superfield with complex mass has one supersymmetric vacuum in twisted sector while the theory with a twisted mass has two vacua, one in twisted sector and one in untwisted sector. The other non-standard $\mathbf Z_2$ symmetry is denoted by $\mathbf Z_2(-1)^{F_s}$ where $ (-1)^{F_s} $ is the operator that acts as the sign flip of all states in the untwisted RR sector. The nonstandard $\mathbf Z_2(-1)^{F_s}$ orbifold theory of a chiral multiplet has the Witten index,
\begin{align}
\Tr (-1)^F = \left\{\begin{array}{ll}
2&\mbox{~for complex mass}\\
1&\mbox{~for twised mass}.
\end{array}\right.
\end{align}

We carry out a consistency check with the elliptic genus. The elliptic genus of a chiral supermultiplet with a complex mass $W=m \Phi^2$ is given by
\begin{align}
Z^{\Phi,m}(\tau,z,\xi) = \frac{\theta_1(\tau|-\frac{z}{2})}{\theta_1(\tau|\frac{z}{2})}=-1
\end{align}
where $ q=e^{2\pi i \tau} $, $ y=e^{2\pi i z} $, $ \tau $ is the complex structure of torus and $ z $ is a holonomy for the left-moving $ U(1) $ R-symmetry.
The value $ 1 $ represents a trivial vacuum where the sign depends on the definition of the fermion number of the vacuum. The elliptic genus of the $\mathbf Z_2$ orbifold theory of the massive chiral multiplet is given by
\begin{align}
Z^{\Phi/\mathbf Z_2,m}(\tau,z,\xi) =  \frac{1}{2}\sum_{a,b=0}^{1} Z^{\Phi/\mathbf Z_2,m}_{(ab)}(\tau,z,\xi) \label{EG:Z2_ComplexMass}
\end{align}
where
\begin{align}
Z^{\Phi/\mathbf Z_2,m}_{(ab)}(\tau,z,\xi) = y^{-b/2}\frac{\theta_1(\tau|-\frac{z}{2} +\frac{a+b\tau}{2})}{\theta_1(\tau|\frac{z}{2} +\frac{a+b\tau}{2})}=(-1)^{(a+1)(b+1)}~.\label{Z2ComplexMassive}
\end{align}
The factor $y^{-b/2}$ is needed for the elliptic genus to show the sensible modular behavior.
Thus we have
\begin{align}
Z^{\Phi/\mathbf Z_2,m}(\tau,z,\xi)=1
\end{align}
Contributions from untwisted sector is
\begin{align}
Z^{\Phi/\mathbf Z_2,m,\text{untwisted}}(\tau,z,\xi) =
& \frac{1}{2}\sum_{a=0}^{1}Z^{\Phi/\mathbf Z_2,m}_{(a0)}
=0
\end{align}
and that of twisted sector is
\begin{align}
Z^{\Phi/\mathbf Z_2,m,\text{twisted}}(\tau,z,\xi) =
& \frac{1}{2}\sum_{a=0}^{1}Z^{\Phi/\mathbf Z_2,m}_{(a1)}
=1
\end{align}
Thus only twisted sector ground state survives. This $ \mathbf Z_2 $ orbifold in \eqref{EG:Z2_ComplexMass} is the standard one in which the ground states of untwisted sector are projected out as defined in \cite{Hori:2011pd}.

The elliptic genus of the non-standard $ \mathbf Z_2(-1)^{F_s} $ is given by
\begin{align}
Z^{\Phi/\mathbf Z_2(-1)^{F_s},m}(\tau,z,\xi) =  \frac{1}{2}\left( -Z^{\Phi/\mathbf Z_2,m}_{(00)} +Z^{\Phi/\mathbf Z_2,m}_{(10)} +Z^{\Phi/\mathbf Z_2,m}_{(01)}+Z^{\Phi/\mathbf Z_2,m}_{(11)} \right)=2\label{EG:Z2n_ComplexMass}
\end{align}
Now let's consider a chiral multiplet with a twisted mass. If a theory has generic twisted masses and has no non-compact Coulomb branch, namely a gapped theory, its Witten index can be obtained by a limit of the elliptic genus of the theory with generic flavor holonomies on the torus, which regulate bosonic zero modes. Witten index of one free chiral multiplet having a twisted mass can be obtained by
\begin{align}
&Z^{\Phi}(\tau,z,\xi) = \frac{\theta_1(\tau|-z +\xi)}{\theta_1(\tau|\xi)}
\\
&\Tr (-1)^F=\lim_{z\rightarrow 0} Z_{\Phi}(\tau,z,\xi) =1
\end{align}
where $\xi$ is a holonomy for $U(1)$ flavor symmetry. Let us compute the Witten index of the standard $\mathbf Z_2$ orbifold.
\begin{align}
&Z^{\Phi/\mathbf Z_2}(\tau,z,\xi) = \frac{1}{2}\sum_{a,b=0}^{1} Z^{\Phi/\mathbf Z_2}_{(ab)}(\tau,z,\xi)\label{EG:Z2_FreeChiral}
\end{align}
where
\begin{align}
Z^{\Phi/\mathbf Z_2}_{(ab)}(\tau,z,\xi)=y^{-b/2}\frac{\theta_1(\tau|-z +\xi +\frac{a+b\tau}{2})}{\theta_1(\tau|\xi+\frac{a+b\tau}{2})}
\end{align}
We have
\begin{align}
\Tr (-1)^F=\lim_{z\rightarrow 0} Z^{\Phi/\mathbf Z_2}(\tau,z,\xi) =2,
\end{align}
thus both the untwisted RR vacuum and the twisted one survive. The $ \mathbf Z_2 (-1)^{F_s} $ orbifold has the elliptic genus
\begin{align}
Z^{\Phi/\mathbf Z_2(-1)^{F_s}}(\tau,z,\xi) =  \frac{1}{2}\left( -Z^{\Phi/\mathbf Z_2}_{(00)} +Z^{\Phi/\mathbf Z_2}_{(10)} +Z^{\Phi/\mathbf Z_2}_{(01)}+Z^{\Phi/\mathbf Z_2}_{(11)} \right)\label{EG:Z2n_FreeChiral}
\end{align}
so we have
\begin{align}
\Tr (-1)^F=\lim_{z\rightarrow 0} Z^{\Phi/\mathbf Z_2(-1)^{F_s}}(\tau,z,\xi) =1
\end{align}
Notice the sign flip of the $(00)$ sector. $(-1)^F$ of the untwisted RR ground state is the opposite to that of the twisted RR ground state.

\section{Hori duality}
Hori proposed the dualities between $\mathcal N=(2,2)$ theories with gauge groups,
\begin{align}
O_+(k)~&\longleftrightarrow ~ SO(N-k+1)\nonumber\\
SO(k)~&\longleftrightarrow ~ O_+(N-k+1)\\
O_-(k)~&\longleftrightarrow ~ O_-(N-k+1).\nonumber
\end{align}
for $N \geq k$. We refer the theory with the gauge group on the left hand as the \textit{theory A} and the corresponding dual theory on the right hand side as the \textit{theory B}. Theory A has $N$ massless chiral multiplets $Q_\alpha$, $\alpha=1,\ldots,N$ in the fundamental representation
with no superpotential. Theory B has $N$ massless chiral multiplets $q^\alpha$ in the fundamental representation and $\frac{N(N+1)}{2}$ gauge singlet chiral multiplets $M_{\alpha\beta}$, $M_{\alpha\beta}=M_{\beta\alpha}$, with the superpotential,
\begin{align}
W=\sum_{\alpha,\beta=1}^{N}M_{\alpha\beta}q^{\alpha}q^{\beta}~.
\end{align}
The mesons $Q_\alpha Q_\beta$ in the original theory correspond to $M_{\alpha\beta}$.
The global charges of the fields are given by the Table 1.
\begin{table}[ht]
	\centering
	\begin{tabular}{|c|ccc|}
		\hline
		& $U(1)_L$ & $U(1)$ & $SU(N)$  \\
		\hline
		$Q$ & $ 0 $ & 1 &$ \mathbf N $ \\
		$M$ & $ 0 $ & 2 &$ \mathbf{\frac{N(N+1)}{2}} $ \\
		$q$ & $ \frac{1}{2} $ & -1 &$ \overline{\mathbf N} $ \\
		\hline
	\end{tabular}
	\caption{\label{} The quantum numbers of the chiral superfields for the left-moving R-symmetry $U(1)_L$ and a flavor symmetry $U(N)$}
\end{table}

The $O(k)$ gauge theory can be treated as the $\mathbf Z_2$ orbifold of the $SO(k)$ gauge theory. There are two versions of $\mathbf Z_2$ orbifold for the theory. For a different choice of $\mathbf Z_2$ orbifold the untwisted RR ground states can  survive or be projected out while the twisted RR ground states always survive. Following \cite{Hori:2011pd} we call "standard" for the $\mathbf Z_2$ orbifold under which the untwisted RR ground states survive. The other "non-standard" $\mathbf Z_2$ orbifold is denoted by $\mathbf Z_2 (-1)^{F_s}$ where $ (-1)^{F_s}$ is the operator that acts as the sign flip of all states in the untwisted RR sector while the states in the twisted RR sector are invariant. This is similar to
the terminologies for the case of one chiral multiplet with twisted mass.

In order to define $O_+(k)$ and $O_-(k)$ gauge theory, the twisted mass deformation is useful because it gives discrete quantum vacua. The number of ground states, especially $\mathbf Z_2$ symmetric vacua, depends on $k$ and $N$ so do definitions of $O_+(k)$ and $O_-(k)$. When $k$ is odd, the theory with generic twisted masses has $\mathbf Z_2$ symmetric vacuum thus the spectrum in this sector is sensitive to the choice of orbifold. When $N$ is even, $O_+(k)$  gauge theory is the one with standard $\mathbf Z_2$ orbifold while $O_-(k)$ theory has the non-standard one.
For odd $N$, the role of $O_+(k)$ and $O_-(k)$ are interchanged. When $k$ is even and $N$ is odd, the theory has $\mathbf Z_2$ symmetric vacua and $O_+(k)$ (resp. $O_-(k)$) gauge theory is the one with standard $\mathbf Z_2$ (resp. non-standard) orbifold. When $k$ and $N$ are even, the theory does not have $\mathbf Z_2$ symmetric vacua so we extend the definition of $O_+(k)$ and $O_-(k)$ theory with odd $N$ to even $N$. In other words, $O_+(k)$ (resp. $O_-(k)$) gauge theory with even $N$ is $O_+(k)$ (resp. $O_-(k)$) gauge theory with $N+1$ chiral fields where $N$ massless and one massive chiral fields with a complex mass. The standard orbifold has twice as many $\mathbf Z_2$ symmetric vacua as the non-standard one.  The total number of vacua for $O_+(k)$ and $O_-(k)$ is summarized in Table 2.

Hori also claimed that the $ SO(k) $ gauge theory with $ N=k-1 $ massless fundamental chiral multiplets flows to a free theory of $ \frac{k(k-1)}{2} $ mesons, $ M_{\alpha\beta} $. Then the non-standard $ \mathbf Z_2 (-1)^{F_s} $ orbifold, $ O_-(k) $ gauge theory with $ N=k-1 $ has no vacuum in the untwisted sector and its twisted sector is dual to the free theory of mesons.	On the other hand, the standard $ \mathbf Z_2 $ orbifold, $ O_+(k) $ gauge theory with $ N=k-1 $, flows to two copies of the free theory of mesons, one in each of the untwisted and twisted sectors.

When $N < k-1$ , the
supersymmetry is broken.
The duality is valid when the theories do not have non-compact Coulomb branch, which are referred to as regular theories \cite{Hori:2011pd}. Regularity is determined by vacuum solutions of the effective twisted superpotential for the adjoint scalar field in the vector multiplets of the form, for even $ k $ (resp. odd $ k $)
\begin{align}
i\left(\begin{array}{ccccc}
&-\sigma_1&&&\\
\sigma_1&&&&\\
&&\ddots&&\\
&&&&-\sigma_{\frac{k}{2}}\\
&&&\sigma_{\frac{k}{2}}&
\end{array}\right)\qquad\mbox{\it resp.}\quad
i\left(\begin{array}{ccccc|c}
&-\sigma_1&&&&\\
\sigma_1&&&&&\\
&&\ddots&&&\\
&&&&-\sigma_{\frac{k-1}{2}}&\\
&&&\sigma_{\frac{k-1}{2}}&&\\
\hline
&&&&&0
\end{array}\right).
\label{AdjScalar}
\end{align}
When the chiral fields have no twisted masses the effective twisted superpotential only shifts the theta angle: $\theta_\text{eff} = N \pi$ for $k$ even or $\theta_\text{eff} = (N+1) \pi $ for $k$ odd for each $U(1)$ factor. If effective theta angle is zero modulo $2\pi$, the theory has non-compact Coulomb branch. Therefore, the theory is regular if and only if one of following is true,
\begin{align}
&1.~ \text{$N-k$ is odd and $\theta=0$}
\\
&2.~ \text{$N-k$ is even and $\theta=\pi$}
\end{align}
where $\theta$ is the tree level mod 2 theta angle. The mod 2 theta angle can be introduced by adding a fundamental chiral field with a complex mass. As the massive chiral field is integrated out it yields the theta angle $\pi$, which can be seen from the effective superpotential.

\begin{table}[h]
	\begin{center}\hspace{-1cm}
		\begin{tabular}{|c|c|c|c|c|}
			\hline
			$k$ & $N$ & $O_+(k)$ & $O_-(k)$ & $SO(k)$\\
			\hline
			even & even & $\displaystyle \binom{\frac{N}{2}}{\frac{k}{2}}$ & $\displaystyle \binom{\frac{N}{2}}{\frac{k}{2}}$ &  $\displaystyle 2\binom{\frac{N}{2}}{\frac{k}{2}}$\\			
			even & odd & $\displaystyle \binom{\frac{N-1}{2}}{\frac{k}{2}} + 2\binom{\frac{N-1}{2}}{\frac{k}{2}-1}$ & $\displaystyle \binom{\frac{N-1}{2}}{\frac{k}{2}} + \binom{\frac{N-1}{2}}{\frac{k}{2}-1}$ & $\displaystyle 2\binom{\frac{N-1}{2}}{\frac{k}{2}} + \binom{\frac{N-1}{2}}{\frac{k}{2}-1}$\\			
			odd & even  & $\displaystyle 2\binom{\frac{N}{2}}{\frac{k-1}{2}}$ & $\displaystyle \binom{\frac{N}{2}}{\frac{k-1}{2}}$ & $\displaystyle \binom{\frac{N}{2}}{\frac{k-1}{2}}$\\			
			odd & odd & $\displaystyle \binom{\frac{N-1}{2}}{\frac{k-1}{2}}$ & $\displaystyle 2\binom{\frac{N-1}{2}}{\frac{k-1}{2}}$ & $\displaystyle \binom{\frac{N-1}{2}}{\frac{k-1}{2}}$\\			\hline
		\end{tabular}
	\end{center}\caption{The total number of ground states of the $O_\pm(k)$ theories.}
\end{table}

\subsection{$O_{\pm}(1)$ gauge theories}
Let us consider a $O_{\pm}(1)$ gauge theory or a free theory ($SO(1)$) with $N$ massless fundamental chiral multiplets, $Q_\alpha$. As discussed before, definitions of $O_{\pm}(1)$ group can be summarized as follows.
\begin{table}[ht]
	\centering
	\begin{tabular}{|c|c|c|}
		\hline
		$N$ & $O_{+}(1)$ & $O_{-}(1)$  \\
		\hline
		even & $\mathbf Z_2 $ & $\mathbf Z_2(-1)^{F_s}$ \\
		odd & $\mathbf Z_2(-1)^{F_s}$ & $\mathbf Z_2$ \\
		\hline
	\end{tabular}
	\caption{Definitions of $O_{\pm}(1)$ group \label{}}
\end{table}

The standard $\mathbf Z_2$ acts as simultaneous sign flip of all chiral fields
\begin{align}
\mathbf Z_2: ~Q_\alpha\longmapsto -Q_\alpha
\end{align}
and it keeps the ground states in the untwisted RR sector. On the other hand, the non-standard $\mathbf Z_2(-1)^{F_s}$ orbifold projects out the untwisted RR ground states.

The elliptic genera of the theories can be written in terms of contributions from four sectors
\begin{align}
Z^{\text{A},\,SO(1),N}(\tau,z,\xi)&= Z^{O(1),N}_{(00)}
\\
Z^{\text{A},\,O_+(1),N}(\tau,z,\xi)&= \frac{1}{2}\left((-1)^{N} Z^{O(1),N}_{(00)} +Z^{O(1),N}_{(10)} +Z^{O(1),N}_{(01)} +Z^{O(1),N}_{(11)}\right) \label{EG_A_O(1)+N}
\\
Z^{\text{A},\,O_-(1),N}(\tau,z,\xi)&= \frac{1}{2}\left((-1)^{N+1} Z^{O(1),N}_{(00)} +Z^{O(1),N}_{(10)} +Z^{O(1),N}_{(01)} +Z^{O(1),N}_{(11)}\right) \label{EG_A_O(1)-N}
\end{align}
where
\begin{align}
&Z^{O(1),N}_{(ab)}(\tau,z,\xi)= y^{-\frac{N}{2}b}\prod_{\alpha=1}^{N}\frac{\theta_1(\tau|-z +\xi_\alpha +\frac{a+b\tau }{2})}{\theta_1(\tau|\xi_\alpha +\frac{a+b\tau }{2})}
\end{align}
The sign factors in front of $Z^{O(1),N}_{(00)}$ in \eqref{EG_A_O(1)+N} and \eqref{EG_A_O(1)-N} are related to the type of orbifolds. If a sign factor is $+1$ (resp. $-1)$ then the theory corresponds to $\mathbf Z_2$ (resp. $\mathbf Z_2(-1)^{F_s}$) orbifold.\footnote{ The additional minus sign can be introduced by  introducing one  massive chiral field with complex mass, which maps $ \mathbf Z_2 $ to $ \mathbf Z_2(-1)^{F_s} $ and vice versa. For example, $\mathbf Z_2$ orbifold of $ N $ massless and one massive chiral fields is the same as $\mathbf Z_2(-1)^{F_s}$ orbifold of $ N $ massless chiral fields.}

Let's compute the Witten index of the theories.
\begin{align}
\lim_{z\rightarrow 0} Z^{\text{A},\,SO(1),N}(\tau,z,\xi)&= 1
\\
\lim_{z\rightarrow 0} Z^{\text{A},\,O_+(1),N}(\tau,z,\xi)&= \frac{(-1)^N+1}{2}+1=\left\{\begin{array}{ll}
2&\quad\mbox{for $ N $ even}\\
1&\quad\mbox{for $ N $ odd}
\end{array}\right.
\\
\lim_{z\rightarrow 0} Z^{\text{A},\,O_-(1),N}(\tau,z,\xi)&= \frac{(-1)^{N+1}+1}{2}+1=\left\{\begin{array}{ll}
1&\quad\mbox{for $ N $ even}\\
2&\quad\mbox{for $ N $ odd}
\end{array}\right.
\end{align}
The result agrees with the number of ground states of the theories. The standard $\mathbf Z_2$ orbifold ($ O_+(1) $ for even $ N $ or $ O_-(1) $ for odd $ N $) has one vacuum in each of the untwisted and twisted sectors while the non-standard $ \mathbf Z_2(-1)^{F_s} $ orbifold has only one vacuum in the twisted sector.

Let us consider the theory B which is $O_{\pm}(1)$ theory or a $SO(1)$ theory. The elliptic genera of the dual theories can be written as
\begin{align}
Z^{\text{B},\,SO(1),N}(\tau,z,\xi)&=  \widetilde Z^{O(1),N}_{(00)}
\\
Z^{\text{B},\,O_+(1),N}(\tau,z,\xi)&= \frac{1}{2}\left((-1)^{N}  \widetilde Z^{O(1),N}_{(00)} +\widetilde Z^{O(1),N}_{(10)} +\widetilde Z^{O(1),N}_{(01)} +\widetilde Z^{O(1),N}_{(11)}\right)
\\
Z^{\text{B},\,O_-(1),N}(\tau,z,\xi)&= \frac{1}{2}\left((-1)^{N+1} \widetilde Z^{O(1),N}_{(00)} +\widetilde Z^{O(1),N}_{(10)} +\widetilde Z^{O(1),N}_{(01)} +\widetilde Z^{O(1),N}_{(11)}\right)
\end{align}
where
\begin{align}
&\widetilde Z^{O(1),N}_{(ab)}(\tau,z,\xi)=Z^{M,N}(\tau,z,\xi)\times Z^{O(1),N}_{(ab)}(\tau,z,-\xi+z/2)
\\
&Z^{M,N}(\tau,z,\xi)=\left(\prod_{\alpha=1}^{N}\prod_{\beta=\alpha}^{N}\frac{\theta_1(\tau|-z +\xi_\alpha +\xi_\beta)}{\theta_1(\tau|\xi_\alpha +\xi_\beta)}\right)
\end{align}
$ Z^{M,N} $ is the contribution of the mesonic singlet fields and $ Z^{O(1),N}_{(ab)}(\tau,z,-\xi+z/2) $ comes from fundamental fields in the dual theory.

Let us check the dualities for $N=1$, which require that
\begin{align}
Z^{\text{A},\,SO(1),1}&= Z^{\text{B},\,O_+(1),1}
\\
Z^{\text{A},\,O_+(1),1}&= Z^{\text{B},\,SO(1),1}
\\
Z^{\text{A},\,O_-(1),1}&= Z^{\text{B},\,O_-(1),1}
\end{align}
and in terms of contributions of sectors,
\begin{align}
&Z^{O(1),1}_{(00)}=\frac{1}{2}\left(-\widetilde Z^{O(1),1}_{(00)} +\widetilde Z^{O(1),1}_{(10)} +\widetilde Z^{O(1),1}_{(01)} +\widetilde Z^{O(1),1}_{(11)} \right)\label{EG:SO(1)N=1/O+(1)}
\\
&\widetilde Z^{O(1),1}_{(00)} =\frac{1}{2}\left(-Z^{O(1),1}_{(00)} +Z^{O(1),1}_{(10)} +Z^{O(1),1}_{(01)} +Z^{O(1),1}_{(11)} \right)  \label{EG:O+(1)N=1/SO(1)}
\\
&Z^{O(1),1}_{(00)} +Z^{O(1),1}_{(10)} +Z^{O(1),1}_{(01)} +Z^{O(1),1}_{(11)} \label{EG:O-(1)N=1/O-(1)}
=\widetilde Z^{O(1),1}_{(00)} +\widetilde Z^{O(1),1}_{(10)} +\widetilde Z^{O(1),1}_{(01)} +\widetilde Z^{O(1),1}_{(11)}
\end{align}
Note that the equalities \eqref{EG:SO(1)N=1/O+(1)} and \eqref{EG:O+(1)N=1/SO(1)} imply \eqref{EG:O-(1)N=1/O-(1)}.\footnote{This is related to the fact that the duality between $O_-(k)$ theory can be derived from the duality between $SO(k)$ and $O_-(N-k+1)$ theory by a suitable
$Z_2$ orbifolding. \cite{Hori:2011pd}} The duality between two $O_-(1)$ theories have additional mapping of states that the twisted (resp. untwisted) sector of the theory A corresponds to the untwisted (resp. twisted) sector of the theory B. At the appendix E, we prove
analytically
\begin{align}
&Z^{O(1),1}_{(00)} +Z^{O(1),1}_{(10)} =\widetilde Z^{O(1),1}_{(01)} +\widetilde Z^{O(1),1}_{(11)}
\\
&Z^{O(1),1}_{(01)} +Z^{O(1),1}_{(11)} =\widetilde Z^{O(1),1}_{(00)} +\widetilde Z^{O(1),1}_{(10)}.
\end{align}
Furthermore, we prove an additional identity,
\begin{align}
Z^{O(1),1}_{(01)} -Z^{O(1),1}_{(11)}
=-\widetilde Z^{O(1),1}_{(01)} +\widetilde Z^{O(1),1}_{(11)}.
\end{align}
Because we have four independent equations, the contributions four sectors can be written in terms of those of dual theories,
\begin{align}\label{RelationO(1)O(1)1}
\left(\begin{array}{c}
Z^{O(1),1}_{(00)}\\
Z^{O(1),1}_{(10)}\\
Z^{O(1),1}_{(01)}\\
Z^{O(1),1}_{(11)}\\
\end{array}\right)
=
\frac{1}{2}\left(\begin{array}{cccc}
-1&1&1&1\\
1&-1&1&1\\
1&1&-1&1\\
1&1&1&-1\\
\end{array}\right)
\left(\begin{array}{c}
\widetilde Z^{O(1),1}_{(00)}\\
\widetilde Z^{O(1),1}_{(10)}\\
\widetilde Z^{O(1),1}_{(01)}\\
\widetilde Z^{O(1),1}_{(11)}\\
\end{array}\right)
\end{align}

\subsection{$O_{\pm}(2)$ gauge theories}
Let us consider $O_{\pm}(2)$ theories with $N$ fundamental chiral multiplets. The $O(2)$ gauge theory can be treated as the $\mathbf Z_2$ orbifold of the $SO(2)$ gauge theory where $\mathbf Z_2$ is generated by a group element,
\begin{align}
h=\left(\begin{array}{cc}
1&0\\
0&-1
\end{array}\right)
\end{align}
There are two versions of the orbifold \cite{Hori:2011pd}. For $N$ odd, the semi-classical vacuum equation of the theory with generic twisted masses for all flavors has $\frac{N-1}{2}$ pair of solutions which are mapped within themselves ($ \sigma_1\neq 0 $ in \eqref{AdjScalar}) and one solution which is $\mathbf Z_2$ symmetric, $ \sigma_1=0 $. We have two choices for the $\mathbf Z_2$ symmetric solution as in $O(1)$ cases. If $N$ is odd $O_+(2)$ theory is defined to have two vacua from one twisted and one untwisted sector while $O_-(2)$ theory is defined to have only one vacua in the twisted sector. Thus the total number of ground states is
\begin{align}
\Tr (-1)^F = \left\{\begin{array}{ll}
\frac{N-1}{2}+2 = \frac{N+3}{2}&\quad\mbox{in $O_+(2)$ theory}\\
\frac{N-1}{2}+1 = \frac{N+1}{2}&\quad\mbox{in $O_-(2)$ theory}
\end{array}\right. \label{O2Vacua1}
\end{align}
for odd $N$. The $O(2)$ theory with even $N$ flavors and generic twisted masses have only $\frac{N}{2}$ vacua which break the $\mathbf Z_2$ symmetry ($\sigma\neq 0$) so we have
\begin{align}
\Tr (-1)^F =
\frac{N}{2}&\quad \mbox{in $O_\pm(2)$ theory} \label{O2Vacua2}
\end{align}
for even $N$. Since the vacua break $\mathbf Z_2$ we cannot use the twisted mass deformation to define $O_\pm(2)$ theory. In \cite{Hori:2011pd}, $O_\pm(2)$ theory with an even number of massless flavors is defined from the $O_\pm(2)$ theory with odd number of flavors; even number of massless flavors and one flavor with a complex mass.

Let us compute the elliptic genus of the $O_\pm(2)$ theories with this definition. All flat connections of $O(2)$ gauge theory on $T^2$ are explained in \cite{Benini:2013nda}, \cite{Kim:2014dza}. The moduli space of the flat connections consists of seven components. One corresponds to the flat connection of $SO(2)$ group, $u \in \mathbb C /(\mathbb Z +\tau \mathbb Z)$ modulo $u\equiv -u$. Its contribution to the elliptic genus is given by
\begin{align}
Z^{O(2),N}_{(00)}=&- \sum_{\beta=1}^N \frac{i \eta(q)^3}{\theta_1(\tau|-z)}
\oint_{-\xi_\beta} \hspace{-.5em} \dd u\,
\prod_{\alpha=1}^N \frac{ \theta_1 \big(\tau \big| -z + u + \xi_\alpha \big)}{ \theta_1 \big(\tau \big| u + \xi_\alpha \big)} \,
\frac{ \theta_1 \big(\tau \big| -z - u + \xi_\alpha \big) }{ \theta_1 \big( \tau \big| - u + \xi_\alpha \big)}
\\
=& \sum_{\beta=1}^{N}\left(\prod_{\alpha=1, \alpha\neq \beta}^N \frac{ \theta_1 \big(\tau \big| -z +\xi_\alpha -\xi_\beta \big) }{ \theta_1 \big( \tau \big|\xi_\alpha -\xi_\beta \big)} \right) \left(\prod_{\gamma=1}^N \frac{ \theta_1 \big(\tau \big| -z +\xi_\gamma +\xi_\beta \big) }{ \theta_1 \big( \tau \big|\xi_\gamma +\xi_\beta \big)} \right)
\end{align}
The other six components are represented by discrete holonomies labeled by $s=(k,l,\pm)$ for $(k,l)=(1,0), (0,1), (1,1)$ and each contributions to the elliptic genus is given by
\begin{align}
&Z^{O(2),N}_s=\frac12 y^{-\frac{(N-1)}{2}l} \frac{\theta_1(\tau|a^1_s+a^2_s)}{ \theta_1(\tau|-z+a^1_s+a^2_s)} \prod_{\alpha=1}^N \frac{\theta_1 \big( \tau \big| (-z+a^1_s + \xi_\alpha \big) }{ \theta_1 \big(\tau \big| a^1_s + \xi_\alpha \big)}
\, \frac{ \theta_1 \big( \tau \big| -z+a^2_s + \xi_\alpha \big) }{ \theta_1 \big(\tau \big|a^2_s + \xi_\alpha \big)} \label{O2_Disc}
\end{align}
where $ y^{-\frac{(N-1)}{2}l} $ can be derived from modular properties of the elliptic genus and
\begin{align}
&(a^1_{(1,0,+)}, a^2_{(1,0,+)}) = \left(0,\frac{1}{2}\right),\quad
(a^1_{(1,0,-)}, a^2_{(1,0,-)}) = \left(-\frac{\tau}{2},\frac{1+\tau}{2}\right)
\\
&(a^1_{(0,1,+)}, a^2_{(0,1,+)}) = \left(0,\frac{\tau}{2}\right),\quad
(a^1_{(0,1,-)}, a^2_{(0,1,-)}) = \left(-\frac{1}{2},\frac{1+\tau}{2}\right)
\\
&(a^1_{(1,1,+)}, a^2_{(1,1,+)}) = \left(0,\frac{1+\tau}{2}\right),\quad
(a^1_{(1,1,-)}, a^2_{(1,1,-)}) = \left(\frac{1}{2},\frac{\tau}{2}\right)
\end{align}

Elliptic genus of $O(2)$ gauge theory can be written as
\begin{align}
Z^{O(2),N}=\frac{1}{2}\left(Z_{(00)}^{O(2),N}+\sum_{s}(-1)^{n_s} Z^{O(2),N}_{s}\right) \label{EG_O(2)_sign}
\end{align}
where we took a positive sign for $Z^{O(2),N}_{(00)}$ and the sign factors $(-1)^{n_s}$, $n_s\in \mathbb Z$ come from a choice of the two orbifolds and mod 2 $\theta$ angle \cite{Hori:2011pd}. We determine the sign factors as follows. The contributions of components $s=(1,0,+)$, $(0,1,+)$, $(1,1,+)$ should have the same phase and those of components  $s=(1,0,-)$, $(0,1,-)$, $(1,1,-)$ should also have the same phase. This can be seen from the modular property of the elliptic genus,
\begin{align}
&Z^{O(2),N}_{(00)}\left(-\frac{1}{\tau}, \frac{z}{\tau}, \frac{\xi_\alpha}{\tau}\right) = \mathcal A Z^{O(2),N}_{(00)}\left(\tau, z, \xi_\alpha\right)
,\quad Z^{O(2),N}_{(10\pm)}\left(-\frac{1}{\tau}, \frac{z}{\tau}, \frac{\xi_\alpha}{\tau}\right) = \mathcal A Z^{O(2),N}_{(01\pm)}\left(\tau, z, \xi_\alpha\right)\nonumber
\\
&Z^{O(2),N}_{(01\pm)}\left(-\frac{1}{\tau}, \frac{z}{\tau}, \frac{\xi_\alpha}{\tau}\right) = \mathcal A Z^{O(2),N}_{(10\pm)}\left(\tau, z, \xi_\alpha\right)
,\quad Z^{O(2),N}_{(11\pm)}\left(-\frac{1}{\tau}, \frac{z}{\tau}, \frac{\xi_\alpha}{\tau}\right) = \mathcal A Z^{O(2),N}_{(11\pm)}\left(\tau, z, \xi_\alpha\right)
\end{align}
where $\mathcal A=\exp{\frac{i\pi}{\tau}((2N-1)z^2 -4z\sum_\alpha \xi_\alpha)}$ and
\begin{align}
&Z^{O(2),N}_{(00)}\left(\tau+1, z, \xi_\alpha\right) = Z^{O(2),N}_{(00)}\left(\tau, z, \xi_\alpha\right)
,\quad Z^{O(2),N}_{(10\pm)}\left(\tau+1, z, \xi_\alpha\right) = Z^{O(2),N}_{(10\pm)}\left(\tau, z, \xi_\alpha\right)\nonumber
\\
&Z^{O(2),N}_{(01\pm)}\left(\tau+1, z, \xi_\alpha\right) = Z^{O(2),N}_{(11\pm)}\left(\tau, z, \xi_\alpha\right)
,\quad Z^{O(2),N}_{(11\pm)}\left(\tau+1, z, \xi_\alpha\right) = Z^{O(2),N}_{(01\pm)}\left(\tau, z, \xi_\alpha\right)
\end{align}
Note that $ Z^{O(2),N}_{(10\pm)} $ and $ Z^{O(2),N}_{(01\pm)} $ are exchanged under the transformation $\tau\rightarrow -\frac{1}{\tau}$ and $ Z^{O(2),N}_{(01\pm)} $ and $ Z^{O(2),N}_{(11\pm)} $ are exchanged under $ \tau \rightarrow \tau +1 $. Thus, we have only two independent sign factors in \eqref{EG_O(2)_sign},
\begin{align}
Z^{O(2),N}=\frac{1}{2}\left(Z_{(00)}^{O(2),N}+ \sum_{k,l} \left((-1)^{n_{+}}Z^{O(2),N}_{(k,l,+)} +(-1)^{n_{-}}Z^{O(2),N}_{(k,l,-)} \right)\right)
\end{align}

 In order to fix the signs we consider $O_+(2)$ theory with $N$ odd. It is a standard $\mathbf Z_2$ orbifold theory whose states in the untwisted sector survive. Furthermore, it is a regular theory by setting the mod 2 theta angle zero. Therefore all sign factors in the elliptic genus $Z^{O_+(2),N}$ with $N$ odd are positive,
\begin{align}
Z^{O_+(2),N}=\frac{1}{2}\left(Z_{(00)}^{O(2),N}+ \sum_{k,l} \left(Z^{O(2),N}_{(k,l,+)} +Z^{O(2),N}_{(k,l,-)}\right) \right)~.
\end{align}
For $N$ even, $O_+(2)$ theory is defined as a IR theory of $O_+(2)$ theory with $N+1$ fundamental chiral fields where $N$ massless and one massive chiral having a complex mass. The elliptic genus of $O_+(2)$ theory with $N$ even is given by
\begin{align}
Z^{O_+(2),N}&=\frac{1}{2}\left(Z_{(00)}^{O(2),N}Z_{(00), \textrm{massive}}^{O(2), 1}+ \sum_{k,l} \left(Z^{O(2),N}_{(k,l,+)}Z_{(k,l,+), \textrm{massive}}^{O(2), 1} +Z^{O(2),N}_{(k,l,-)}Z_{(k,l,-), \textrm{massive}}^{O(2), 1}\right) \right)\\
&=\frac{1}{2}\left(Z_{(00)}^{O(2),N}+ \sum_{k,l} \left(-Z^{O(2),N}_{(k,l,+)} +Z^{O(2),N}_{(k,l,-)}\right) \right)
\end{align}
where $Z_{(00), \textrm{massive}}^{O(2), 1}$, $Z_{(k,l,\pm), \textrm{massive}}^{O(2), 1}$ are the contribution of massive chiral whose left-moving $R$-charge is $\frac{1}{2}$.
As the massive chiral field is integrated out it changes the type of $\mathbf Z_2$ orbifold as well as it shifts the mod 2 theta angle by $\pi$.
Therefore, we can write the elliptic genus of $O_+(2)$ theory for any $N$ as
\begin{align}
&Z^{O_+(2),N}=\frac{1}{2}\left(Z_{(00)}^{O(2),N}+ \epsilon\sum_{k,l} \left(Z^{O(2),N}_{(k,l,+)} +e^{i\theta}Z^{O(2),N}_{(k,l,-)}\right) \right)
\end{align}
where $\epsilon=e^{i\theta}=(-1)^{N+1}$. We identify the factor $\epsilon$ as a choice of the two types of orbifolds from the fact that $\epsilon$ determines a relative phase between sectors $(00)$ and $(10)$, which are the contribution of untwisted sector. Next, we identify the other sign factor $e^{i\theta}$ as the mod 2 theta angle, which changes as the massive chiral is integrated out.

After we identified $\epsilon$ and $e^{i\theta}$ it is straight forward to get elliptic genus of $O_-(2)$ theory. For $N$ odd, $O_-(2)$ theory is a non-standard orbifold so $\epsilon=-1$ and it is a regular theory for $\theta=0$. The elliptic genus of $O_-(2)$ theory for any $N$ is given by
\begin{align}
&Z^{O_-(2),N}=\frac{1}{2}\left(Z_{(00)}^{O(2),N}+ \epsilon\sum_{k,l} \left(Z^{O(2),N}_{(k,l,+)} +e^{i\theta}Z^{O(2),N}_{(k,l,-)}\right) \right)
\end{align}
where $\epsilon=(-1)^{N}$, $e^{i\theta}=(-1)^{N+1}$. Therefore, the elliptic genus of $SO(2)$ and $O_\pm(2)$ theories can be written as
\begin{align}
Z^{\text{A},\,SO(2),N}(\tau, z, \xi_\alpha)=&
Z^{O(2),N}_{(00)} \label{EG_SO(2)}
\\
Z^{\text{A},\,O_+(2),N}(\tau, z, \xi_\alpha)=&
\frac{1}{2}\left(Z^{O(2),N}_{(00)} + (-1)^{N+1} \left( Z^{O(2),N}_{(10)} + Z^{O(2),N}_{(01)} + Z^{O(2),N}_{(11)}\right)\right) \label{EG_O+(2)}
\\
Z^{\text{A},\,O_-(2),N}(\tau, z, \xi_\alpha)=&\frac{1}{2}\left(
Z^{O(2),N}_{(00)} +(-1)^{N} \left(Z^{O(2),N}_{(10)} +Z^{O(2),N}_{(01)} +Z^{O(2),N}_{(11)}\right)\right) \label{EG_O-(2)}
\end{align}
where
\begin{align}
Z^{O(2),N}_{(kl)} =& Z^{O(2),N}_{(k,l,+)} + (-1)^{N+1} Z^{O(2),N}_{(k,l,-)} \label{EG_O(2)_N_kl}
\end{align}

Let's compute the Witten index.
It is obtained by a limit $y\rightarrow 1$,
\begin{align}
\lim_{y\rightarrow 1}Z^{\text{A},\,SO(2),N}=& N \label{WI_SO(2)}
\\
\lim_{y\rightarrow 1}Z^{\text{A},\,O_+(2),N}=& \left(\frac{N}{2} +\frac{(-1)^{N+1}+1}{4}\right) +\frac{(-1)^{N+1}+1}{2}
=\left\{\begin{array}{ll}
\frac{N}{2}&\quad\mbox{for $ N $ even}\\
\frac{N+1}{2}+1&\quad\mbox{for $ N $ odd}
\end{array}\right.\label{WI_O+(2)}
\\
\lim_{y\rightarrow 1}Z^{\text{A},\,O_-(2),N}=&
\left(\frac{N}{2} -\frac{(-1)^{N+1}+1}{4}\right) -\frac{(-1)^{N+1}+1}{2}
=\left\{\begin{array}{ll}
\frac{N}{2}&\quad\mbox{for $ N $ even}\\
\frac{N-1}{2}-1&\quad\mbox{for $ N $ odd}
\end{array}\right. \label{WI_O-(2)}
\end{align}
where contributions in the parentheses come from the untwisted sector and the others come from the twisted sector.
When $N$ is even it reproduces \eqref{O2Vacua1} and \eqref{O2Vacua2}. However, when $ N $ is odd \eqref{WI_O-(2)} is different from the number of ground states (the sum of ground states in the untwisted and twisted sectors) of $ O_-(2) $ gauge theory due to  the relative sign difference for $(-1)^F$ between the untwisted and twisted sector ground states.

Let us consider the theory B with $ O_\pm(2) $ or $SO(2)$ gauge group. The elliptic genus is given by
\begin{align}
Z^{\text{B},\,SO(2),N}(\tau, z, \xi_\alpha)=&
\widetilde Z^{O(2),N}_{(00)} \label{EG_SO(2)_dual}
\\
Z^{\text{B},\,O_+(2),N}(\tau, z, \xi_\alpha)=&
\frac{1}{2}\left(\widetilde Z^{O(2),N}_{(00)} +(-1)^{N+1}\left(\widetilde  Z^{O(2),N}_{(10)} + \widetilde Z^{O(2),N}_{(01)} + \widetilde Z^{O(2),N}_{(11)}\right)\right) \label{EG_O+(2)_dual}
\\
Z^{\text{B},\,O_-(2),N}(\tau, z, \xi_\alpha)=&
\frac{1}{2}\left(\widetilde Z^{O(2),N}_{(00)} +(-1)^{N}\left(\widetilde Z^{O(2),N}_{(10)} + \widetilde Z^{O(2),N}_{(01)} + \widetilde Z^{O(2),N}_{(11)} \right)\right)\label{EG_O-(2)_dual}
\end{align}
where
\begin{align}
\widetilde Z^{O(2),N}_{(kl)}(\tau,z,\xi)&=\left(\prod_{\alpha=1}^{N}\prod_{\beta=\alpha}^{N}\frac{\theta_1(\tau|-z +\xi_\alpha +\xi_\beta)}{\theta_1(\tau|\xi_\alpha +\xi_\beta)}\right)\times Z^{O(2),N}_{(kl)}(\tau,z,-\xi+z/2)
\end{align}

Let us check the dualities for $N=1, 2, 3$. When $N=1$ the dual theories are free theories and we have
\begin{align}
&Z^{\text{A},\,SO(2),1}=Z^{O(2),1}_{(00)}  =Z^{M,1} = \frac{\theta_1(\tau|-z +2\xi)}{\theta_1(\tau|2\xi)} \label{EG_SO(2)N=1}
\\
&Z^{\text{A},\,O_+(2),1}=\frac{1}{2}\left(Z^{O(2),1}_{(00)} + Z^{O(2),1}_{(10)} + Z^{O(2),1}_{(01)} + Z^{O(2),1}_{(11)} \right) =2Z^{M,1}\label{EG_O+(2)N=1}
\\
&Z^{\text{A},\,O_-(2),1}=\frac{1}{2}\left(Z^{O(2),1}_{(00)} - Z^{O(2),1}_{(10)} - Z^{O(2),1}_{(01)} - Z^{O(2),1}_{(11)} \right) =-Z^{M,1} \label{EG_O-(2)N=1}
\end{align}
where $ Z^{O(2),1}_{(kl)} = Z^{O(2),1}_{(k,l,+)} +Z^{O(2),1}_{(k,l,-)}$ and $ Z^{M,1} $ is the elliptic genus of one free chiral meson $M$.
At appendix E, we prove analytically
\begin{align}
Z^{M,1}=Z^{O(2),1}_{(00)} = Z^{O(2),1}_{(10)} = Z^{O(2),1}_{(01)} = Z^{O(2),1}_{(11)}.
\end{align}

When $N=2$ the elliptic genera for the dualities are
\begin{align}
Z^{\text{A},\,SO(2),2}&= Z^{\text{B},\,O_+(1),2}   \label{eqn=2}
\\
Z^{\text{A},\,O_+(2),2}&= Z^{\text{B},\,SO(1),2}
\\
Z^{\text{A},\,O_-(2),2}&= Z^{\text{B},\,O_-(1),2}
\end{align}
and in terms of contributions of sectors,
\begin{align}
&Z^{O(2),2}_{(00)} = \frac{1}{2}\left(\widetilde Z^{O(1),2}_{(00)} +\widetilde Z^{O(1),2}_{(10)} +\widetilde Z^{O(1),2}_{(01)} +\widetilde Z^{O(1),2}_{(11)}\right) \label{EG_SO(2)N=2}
\\
& \widetilde Z^{O(1),2}_{(00)} =\frac{1}{2}\left( Z^{O(2),2}_{(00)} - Z^{O(2),2}_{(10)} - Z^{O(2),2}_{(01)} - Z^{O(2),2}_{(11)} \right)  \label{EG_O+(2)N=2}
\\
&Z^{O(2),2}_{(00)} + Z^{O(2),2}_{(10)} + Z^{O(2),2}_{(01)} + Z^{O(2),2}_{(11)} =-\widetilde Z^{O(1),2}_{(00)} +\widetilde Z^{O(1),2}_{(10)} +\widetilde Z^{O(1),2}_{(01)} +\widetilde Z^{O(1),2}_{(11)} \label{EG_O-(2)N=2}
\end{align}
where \eqref{EG_SO(2)N=2} and \eqref{EG_O+(2)N=2} implies \eqref{EG_O-(2)N=2}. But the twisted sector and the untwisted sector are exchanged under the duality so we also have
\begin{align}
Z^{O(2),2}_{(00)} + Z^{O(2),2}_{(10)} &=\widetilde Z^{O(1),2}_{(01)} +\widetilde Z^{O(1),2}_{(11)}
\\
Z^{O(2),2}_{(01)} + Z^{O(2),2}_{(11)} &= -\widetilde Z^{O(1),2}_{(00)} +\widetilde Z^{O(1),2}_{(10)}
\end{align}
With additional equality
\begin{align}
-Z^{O(2),2}_{(01)} +Z^{O(2),2}_{(11)}
=\widetilde Z^{O(1),2}_{(01)} -\widetilde Z^{O(1),2}_{(11)}~,
\end{align}
the contribution of each sector can be written as
\begin{align}
\left(\begin{array}{c}
Z^{O(2),2}_{(00)}\\
Z^{O(2),2}_{(10)}\\
Z^{O(2),2}_{(01)}\\
Z^{O(2),2}_{(11)}\\
\end{array}\right)
=
\frac{1}{2}\left(\begin{array}{cccc}
1&1&1&1\\
-1&-1&1&1\\
-1&1&-1&1\\
-1&1&1&-1\\
\end{array}\right)
\left(\begin{array}{c}
\widetilde Z^{O(1),2}_{(00)}\\
\widetilde Z^{O(1),2}_{(10)}\\
\widetilde Z^{O(1),2}_{(01)}\\
\widetilde Z^{O(1),2}_{(11)}\\
\end{array}\right)~.   \label{eqn=22}
\end{align}
At appendix E, we prove analytically the eq. (\ref{eqn=22}).

When $N=3$ the elliptic genera for the dualities are
\begin{align}
Z^{\text{A},\,SO(2),3}&= Z^{\text{B},\,O_+(2),3}
\\
Z^{\text{A},\,O_+(2),3}&= Z^{\text{B},\,SO(2),3}
\\
Z^{\text{A},\,O_-(2),3}&= - Z^{\text{B},\,O_-(2),3}
\end{align}
where  the theory B with $ O_-(2) $ group has additional sign flip for all states.
In terms of contributions of sectors, It is written as
\begin{align}
&Z^{O(2),3}_{(00)} =\frac{1}{2}\left( \widetilde Z^{O(2),3}_{(00)} +\widetilde Z^{O(2),3}_{(10)} +\widetilde Z^{O(2),3}_{(01)} +\widetilde Z^{O(2),3}_{(11)}\right) \label{EG_SO(2)N=3}
\\
& \widetilde Z^{O(2),3}_{(00)} = \frac{1}{2}\left(Z^{O(2),3}_{(00)} + Z^{O(2),3}_{(10)} + Z^{O(2),3}_{(01)} + Z^{O(2),3}_{(11)}  \right) \label{EG_O+(2)N=3}
\\
&Z^{O(2),3}_{(00)} -Z^{O(2),3}_{(10)} -Z^{O(2),3}_{(01)} -Z^{O(2),3}_{(11)} =-\Big(\widetilde Z^{O(2),3}_{(00)} -\widetilde Z^{O(2),3}_{(10)} -\widetilde Z^{O(2),3}_{(01)} -\widetilde Z^{O(2),3}_{(11)}\Big) \label{EG_O-(2)N=3}
\end{align}
We checked following relations numerically up to $q^3$,
\begin{align}
\left(\begin{array}{c}
Z^{O(2),3}_{(00)}\\
Z^{O(2),3}_{(10)}\\
Z^{O(2),3}_{(01)}\\
Z^{O(2),3}_{(11)}\\
\end{array}\right)
=
\frac{1}{2}\left(\begin{array}{cccc}
1&1&1&1\\
1&1&-1&-1\\
1&-1&1&-1\\
1&-1&-1&1\\
\end{array}\right)
\left(\begin{array}{c}
\widetilde Z^{O(2),3}_{(00)}\\
\widetilde Z^{O(2),3}_{(10)}\\
\widetilde Z^{O(2),3}_{(01)}\\
\widetilde Z^{O(2),3}_{(11)}\\
\end{array}\right)
\end{align}

We comment on the irregular theories. For irregular choice of mod 2 theta angle the Witten index as the limit of the elliptic genus is generally not an integer. For example, when $ N $ is even (resp. odd) the $ O_+(2) $ theory is irregular for $ \theta=0 $ (resp. $ \theta=\pi $) and the Witten index from its elliptic genus can be obtained by replacing $ (-1)^{N+1} $ with $ (-1)^{N} $ in \eqref{EG_O(2)_N_kl}. We have
\begin{align}
\lim_{y\rightarrow 1}Z^{\text{A},\,O_+(2),N}_{\text{irregular}}=& \frac{1}{2}\left(N +(-1)^{N+1}\frac{3}{2}\left(1+(-1)^N\right) \right)
=\left\{\begin{array}{ll}
\frac{N-3}{2}&\quad\mbox{for $ N $ even}\\
\frac{N}{2}&\quad\mbox{for $ N $ odd}
\end{array}\right.
\end{align}
The half-integer value can be understood as the contribution of bosonic zero mode parameterizing the non-compact Coulomb branch. We have checked that for irregular choice of the theta angle, the elliptic genera do not match for the potential dual pairs. .

\subsection{$O_{\pm}(3)$ gauge theories}
$O(3)$ group is a direct product, $SO(3)\times \mathbf Z_2$ where $\mathbf Z_2$ is generated by a group element $ -\mathbf 1_{3\times 3}$.
As defined before there are two versions of $\mathbf Z_2$ orbifold depending on a choice of the untwisted RR sector. Let us summarize the definition of $O_\pm(3)$ gauge group and the number of vacua in Table.\eqref{O(3)Vacua}.
\begin{table}[h]
	\begin{center}\hspace{-1cm}
		\begin{tabular}{|c|c|c|c|}
			\hline
			$N$ & $O_+(3)$ & $O_-(3)$ & $SO(3)$\\
			\hline
			even  & $N$ & $\frac{N}{2}$ & $\frac{N}{2}$\\			
			odd & $\frac{N-1}{2}$ & $N-1$ & $\frac{N-1}{2}$\\			
			\hline
		\end{tabular}
	\end{center}\caption{The total number of ground states of the $O_\pm(3)$ theories. The standard $\mathbf Z_2$ orbifold has twice as many vacua as the non-standard one. For even $N$, $O_+(3)$ is the standard $\mathbf Z_2$ orbifold while $O_-(3)$ is the non-standard $\mathbf Z_2(-1)^{F_s}$ orbifold. For odd N, the role of $O_+(3)$ and $O_-(3)$ are reversed.\label{O(3)Vacua}}
\end{table}

Let us compute the elliptic genus of the $O_\pm(3)$ theories using flat connections explained in \cite{Kim:2014dza}. The moduli space of the flat connections consists of eight components. Four of them are represented by continuous holonomies
\begin{align}
(a^1_{(k,l,+)}, a^2_{(k,l,+)}, a^3_{(k,l,+)}) = \left(u,-u,\frac{k+l\tau}{2}\right)~.
\end{align}
Contribution of each sector to the elliptic genus is
\begin{align}
&Z^{O(3),N}_{(k,l,+)}(\tau, z, \xi) \\
&=-\frac{1}{2} \frac{i \eta(q)^3}{\theta_1(\tau|-z)} y^{-\frac{N-2}{2}l} \sum_{u_* \,\in\, \mathfrak M^+_\text{sing}} \oint_{u=u_*} \dd u\, \frac{\theta_1(\tau|u+\frac{k+l\tau}{2})}{\theta_1(\tau| -z +u+\frac{k+l\tau}{2})} \, \frac{\theta_1(\tau|-u+\frac{k+l\tau}{2})}{\theta_1(\tau| -z -u+\frac{k+l\tau}{2})} \nonumber
\\
&\qquad\qquad\qquad\qquad\times \prod_{\alpha=1}^N \frac{\theta_1 \big( \tau \big| -z+u+\xi_\alpha \big) }{ \theta_1 \big(\tau \big| u + \xi_\alpha \big)} \frac{\theta_1 \big( \tau \big| -z-u+\xi_\alpha \big) }{ \theta_1 \big(\tau \big| -u + \xi_\alpha \big)} \frac{\theta_1 \big( \tau \big| -z+\xi_\alpha +\frac{k+l\tau}{2} \big) }{ \theta_1 \big(\tau \big| \xi_\alpha +\frac{k+l\tau}{2}\big)}\nonumber
\end{align}
where $ k,l=0,1 $ and $ \mathfrak M^+_\text{sing} =\Big\{z-\frac{k+l\tau}{2}, -\xi_\alpha\Big\} $. Evaluating the Jeffrey-Kirwan (JK) residue we obtain
\begin{align}
&Z^{O(3),N}_{(k,l,+)}(\tau, z, \xi)
\\&=-\frac{1}{2} y^{-\frac{N-2}{2}l} \frac{\theta_1(\tau|\!-\!z \!\!+\!\!k\!\!+\!\!l\tau)}{\theta_1(\tau| \!-\!2z \!\!+\!\!k\!\!+\!\!l\tau)} \prod_{\alpha=1}^{N} \frac{\theta_1(\tau|\xi_\alpha\!\!-\!\!\frac{k+l\tau}{2})}{\theta_1(\tau| z \!\!+\!\!\xi_\alpha\!\!-\!\!\frac{k+l\tau}{2})} \frac{\theta_1(\tau|\!-\!2z \!\!+\!\!\xi_\alpha\!\!+\!\!\frac{k+l\tau}{2})}{\theta_1(\tau| \!-\!z \!\!+\!\!\xi_\alpha\!\!+\!\!\frac{k+l\tau}{2})} \frac{\theta_1(\tau|\!-\!z \!\!+\!\!\xi_\alpha\!\!+\!\!\frac{k+l\tau}{2})}{\theta_1(\tau| \xi_\alpha\!\!+\!\!\frac{k+l\tau}{2})}\nonumber
\\&\quad+\frac{1}{2} y^{-\frac{N-2}{2}l} \sum_{\alpha=1}^{N} \frac{\theta_1(\tau|\xi_\alpha+\frac{k+l\tau}{2})}{\theta_1(\tau| -z +\xi_\alpha+\frac{k+l\tau}{2})} \frac{\theta_1(\tau|-\xi_\alpha+\frac{k+l\tau}{2})}{\theta_1(\tau| -z -\xi_\alpha+\frac{k+l\tau}{2})} \nonumber
\\
&\qquad \left(\prod_{\beta=1, \beta\neq \alpha}^{N} \frac{\theta_1(\tau|-z -\xi_\alpha +\xi_\beta)}{\theta_1(\tau| -\xi_\alpha+\xi_\beta)}\right) \left(\prod_{\gamma=1}^{N} \frac{\theta_1(\tau|-z +\xi_\alpha +\xi_\gamma)}{\theta_1(\tau| \xi_\alpha +\xi_\gamma)} \frac{\theta_1(\tau|-z +\xi_\gamma +\frac{k+l\tau}{2})}{\theta_1(\tau| \xi_\gamma +\frac{k+l\tau}{2})}\right)\nonumber
\end{align}
The other four are represented by discrete holonomies $ a^i_{(k,l,-)} $ and their contribution to the elliptic genus is given by
\begin{align}
&Z^{O(3),N}_{(k,l,-)}(\tau, z, \xi) \\
&=\frac{1}{4} y^{-\frac{N-2}{2}l} \left(\prod_{i=1}^{3}\prod_{j=i+1}^{3}\frac{\theta_1(\tau|a^i_{(k,l,-)}+a^j_{(k,l,-)})}{\theta_1(\tau|-z +a^i_{(k,l,-)}+a^j_{(k,l,-)})}\right)
\left(\prod_{\alpha=1}^{N}\prod_{i=1}^{3}\frac{\theta_1(\tau|-z+a^i_{(k,l,-)}+\xi_\alpha)}{\theta_1(\tau|a^i_{(k,l,-)}+\xi_\alpha)}\right)\nonumber
\end{align}
where
\begin{align}
&(a^1_{(0,0,-)}, a^2_{(0,0,-)}, a^3_{(0,0,-)}) = \left(\frac{1}{2},-\frac{1+\tau}{2},\frac{\tau}{2}\right)
\\
&
(a^1_{(1,0,-)}, a^2_{(1,0,-)}, a^3_{(1,0,-)}) = \left(-\frac{\tau}{2},\frac{1+\tau}{2},0\right)
\\
&(a^1_{(0,1,-)}, a^2_{(0,1,-)}, a^3_{(0,1,-)}) = \left(-\frac{1}{2},\frac{1+\tau}{2},0\right)
\\
&(a^1_{(1,1,-)}, a^2_{(1,1,-)}, a^3_{(1,1,-)}) = \left(\frac{1}{2},\frac{\tau}{2},0\right)
\end{align}

As in the $ O_\pm(2) $ cases we fix the mod 2 theta angle contribution by introducing a massive fundamental chiral field. As a result we define
\begin{align}
Z^{O(3),N}_{(kl)}=Z^{O(3),N}_{(k,l,+)}+(-1)^{N}Z^{O(3),N}_{(k,l,-)}
\end{align}
where $ (k,l)=(0,0),\,(1,0),\,(0,1),\,(1,1) $ and mod 2 theta angle factor $(-1)^{N} $ is fixed to be a regular theory.
The elliptic genus of the theories is given by
\begin{align}
Z^{\text{A},\,SO(3),N}(\tau, z, \xi_\alpha)=&
Z^{O(3),N}_{(00)}  \label{EG_SO(3)}
\\
Z^{\text{A},\,O_+(3),N}(\tau, z, \xi_\alpha)=&
\frac{1}{2}\left((-1)^N Z^{O(3),N}_{(00)} + Z^{O(3),N}_{(10)} + Z^{O(3),N}_{(01)} + Z^{O(3),N}_{(11)}\right) \label{EG_O+(3)}
\\
Z^{\text{A},\,O_-(3),N}(\tau, z, \xi_\alpha)=&
\frac{1}{2}\left((-1)^{N+1} Z^{O(3),N}_{(00)} + Z^{O(3),N}_{(10)} + Z^{O(3),N}_{(01)} + Z^{O(3),N}_{(11)} \right)\label{EG_O-(3)}
\end{align}
where the sign factors in front of $Z^{O(3),N}_{(00)}$ correspond to a choice of the two type of orbifolds as in $O(1)$ theories.

Let's check the Witten index limit $y\rightarrow 1$.
\begin{align}
\lim_{y\rightarrow 1}Z^{\text{A},\,SO(3),N}(\tau, z, \xi_\alpha)=& \frac{N}{2} -\frac{1}{4} + \frac{ (-1)^N}{4} =\left\{\begin{array}{ll}
\frac{N}{2}&\quad\mbox{$ N $ even}\\
\frac{N-1}{2}&\quad\mbox{$ N $ odd}
\end{array}\right.\label{WI_SO(3)}
\\
\lim_{y\rightarrow 1}Z^{\text{A},\,O_+(3),N}(\tau, z, \xi_\alpha)=&
\left(\frac{N}{2} -\frac{1}{4} +\frac{ (-1)^N}{4}\right)\left(\frac{(-1)^{N}+1}{2} +1\right)
=\left\{\begin{array}{ll}
N&\quad\mbox{for $ N $ even}\\
\frac{N-1}{2}&\quad\mbox{for $ N $ odd}
\end{array}\right.\label{WI_O+(3)}
\\
\lim_{y\rightarrow 1}Z^{\text{A},\,O_-(3),N}(\tau, z, \xi_\alpha)=&
\left(\frac{N}{2} -\frac{1}{4} + \frac{(-1)^N}{4}\right)\left(\frac{(-1)^{N+1}+1}{2} +1\right)  =\left\{\begin{array}{ll}
\frac{N}{2}&\quad\mbox{for $ N $ even}\\
N-1&\quad\mbox{for $ N $ odd}
\end{array}\right.\label{WI_O-(3)}
\end{align}
where in $ O_+(3) $ (resp. $ O_-(3) $) theory the contribution $ \frac{(-1)^{N}+1}{2} $ (resp. $ \frac{(-1)^{N+1}+1}{2} $) and $ +1 $ in the second parentheses come from the untwisted sector and the twisted sector respectively. It reproduces the number of vacua in Table \ref{O(3)Vacua}.

Let us compute the elliptic genus of the theory B which has the gauge group, $O_\pm(3)$ or $SO(3)$.
\begin{align}
Z^{\text{B},\,SO(3),N}(\tau, z, \xi_\alpha)=&
\widetilde Z^{O(3),N}_{(00)} \label{EG_SO(3)_dual}
\\
Z^{\text{B},\,O_+(3),N}(\tau, z, \xi_\alpha)=&
\frac{1}{2}\left((-1)^{N} \widetilde Z^{O(3),N}_{(00)} + \widetilde Z^{O(3),N}_{(10)} + \widetilde Z^{O(3),N}_{(01)} +\widetilde  Z^{O(3),N}_{(11)} \right)\label{EG_O+(3)_dual}
\\
Z^{\text{B},\,O_-(3),N}(\tau, z, \xi_\alpha)=&
\frac{1}{2}\left((-1)^{N+1} \widetilde Z^{O(3),N}_{(00)} +\widetilde Z^{O(3),N}_{(10)} +\widetilde Z^{O(3),N}_{(01)} + \widetilde Z^{O(3),N}_{(11)} \right)\label{EG_O-(3)_dual}
\end{align}
where
\begin{align}
\widetilde Z^{O(3),N}_{(kl)}(\tau,z,\xi)&=Z^{M,N}(\tau,z,\xi) Z^{O(3),N}_{(kl)}(\tau,z,-\xi+z/2)
\end{align}
and contribution of singlets $Z^{M,N}(\tau,z,\xi)= \left(\prod_{\alpha=1}^{N}\prod_{\beta=\alpha}^{N}\frac{\theta_1(\tau|-z +\xi_\alpha +\xi_\beta)}{\theta_1(\tau|\xi_\alpha +\xi_\beta)}\right) $.

We have checked the dualities for $N=2, 3, 4, 5$ cases. For $ N=2 $, the theory A is dual to a free theory and we proved analytically at appendix E that
\begin{align}
Z^{M,2}=Z^{O(3),2}_{(00)} = Z^{O(3),2}_{(10)} = Z^{O(3),2}_{(01)} = Z^{O(3),2}_{(11)}
\end{align}
up to $q^3$.
Thus we have
\begin{align}
&Z^{\text{A},\,SO(3),2}=Z^{M,2}
\label{EG_SO(3)N=2}
\\
&Z^{\text{A},\,O_+(3),2}=2Z^{M,2} \label{EG_O+(3)N=2}
\\
&Z^{\text{A},\,O_-(3),2}=Z^{M,2} \label{EG_O-(3)N=2}
\end{align}

For $N=3, 5$, the dualities imply the following identities of the elliptic genera
\begin{align}
&Z^{O(3),N}_{(00)} = \frac{1}{2}\left(-\widetilde Z^{O(N-2),N}_{(00)} +\widetilde Z^{O(N-2),N}_{(10)} +\widetilde Z^{O(N-2),N}_{(01)} +\widetilde Z^{O(N-2),N}_{(11)}\right) \label{EG_SO(3)N/O+(N-2)Odd}
\\
&\widetilde Z^{O(N-2),N}_{(00)} = \frac{1}{2}\left(-Z^{O(3),N}_{(00)} + Z^{O(3),N}_{(10)} + Z^{O(3),N}_{(01)} + Z^{O(3),N}_{(11)} \right)  \label{EG_O+(3)N/SO(N-2)Odd}
\\
&Z^{O(3),N}_{(00)} + Z^{O(3),N}_{(10)} + Z^{O(3),N}_{(01)} +Z^{O(3),N}_{(11)} =\widetilde Z^{O(N-2),N}_{(00)} +\widetilde Z^{O(N-2),N}_{(10)} +\widetilde Z^{O(N-2),N}_{(01)} +\widetilde Z^{O(N-2),N}_{(11)} \label{EG_O-(3)N/O-(N-2)Odd}
\end{align}
where \eqref{EG_SO(3)N/O+(N-2)Odd} and \eqref{EG_O+(3)N/SO(N-2)Odd} imply \eqref{EG_O-(3)N/O-(N-2)Odd}.  Furthermore, we checked the following relations
\begin{align}
\left(\begin{array}{c}
Z^{O(3),N}_{(00)}\\
Z^{O(3),N}_{(10)}\\
Z^{O(3),N}_{(01)}\\
Z^{O(3),N}_{(11)}\\
\end{array}\right)
=
\frac{1}{2}\left(\begin{array}{cccc}
-1&1&1&1\\
1&-1&1&1\\
1&1&-1&1\\
1&1&1&-1\\
\end{array}\right)
\left(\begin{array}{c}
\widetilde Z^{O(N-2),N}_{(00)}\\
\widetilde Z^{O(N-2),N}_{(10)}\\
\widetilde Z^{O(N-2),N}_{(01)}\\
\widetilde Z^{O(N-2),N}_{(11)}\\
\end{array}\right)
\end{align}
up to $q^3$ for $N=3,\, 5$ with some specific values of flavor holonomies such as $\xi_\alpha = \frac{\alpha}{N+1}$ for computational simplicity.

For $N=4$, the dualities imply the following identities of the elliptic genera
\begin{align}
&Z^{O(3),N}_{(00)} =\frac{1}{2}\left( \widetilde Z^{O(N-2),N}_{(00)} -\widetilde Z^{O(N-2),N}_{(10)} -\widetilde Z^{O(N-2),N}_{(01)} -\widetilde Z^{O(N-2),N}_{(11)} \right) \label{EG_SO(3)N/O+(N-2)Even}
\\
&\widetilde Z^{O(N-2),N}_{(00)} = \frac{1}{2}\left(Z^{O(3),N}_{(00)} + Z^{O(3),N}_{(10)} + Z^{O(3),N}_{(01)} + Z^{O(3),N}_{(11)} \right)  \label{EG_O+(3)N/SO(N-2)Even}
\\
&-Z^{O(3),N}_{(00)} + Z^{O(3),N}_{(10)} + Z^{O(3),N}_{(01)} +Z^{O(3),N}_{(11)} =\widetilde Z^{O(N-2),N}_{(00)} +\widetilde Z^{O(N-2),N}_{(10)} +\widetilde Z^{O(N-2),N}_{(01)} +\widetilde Z^{O(N-2),N}_{(11)} \label{EG_O-(3)N/O-(N-2)Even}
\end{align}
where \eqref{EG_SO(3)N/O+(N-2)Even} and \eqref{EG_O+(3)N/SO(N-2)Even} implies \eqref{EG_O-(3)N/O-(N-2)Even}. We further check the following relations
\begin{align}
\left(\begin{array}{c}
Z^{O(3),N}_{(00)}\\
Z^{O(3),N}_{(10)}\\
Z^{O(3),N}_{(01)}\\
Z^{O(3),N}_{(11)}\\
\end{array}\right)
=
\frac{1}{2}\left(\begin{array}{cccc}
1&-1&-1&-1\\
1&-1&1&1\\
1&1&-1&1\\
1&1&1&-1\\
\end{array}\right)
\left(\begin{array}{c}
\widetilde Z^{O(N-2),N}_{(00)}\\
\widetilde Z^{O(N-2),N}_{(10)}\\
\widetilde Z^{O(N-2),N}_{(01)}\\
\widetilde Z^{O(N-2),N}_{(11)}\\
\end{array}\right)
\end{align}
up to $q^3$ for $N=4$ with some specific values of flavor holonomies.

\subsection{$ O_\pm (4) $ gauge theories}
In this section we compute the elliptic genus for the $ O_+(4) $, $ O_-(4) $, $ SO(4) $ gauge theories with $ N $ fundamental chiral fields and test the dualities. For odd $ N $, $ O_+(4) $ (resp. $ O_-(4) $) gauge theory is the standard $ \mathbf Z_2 $ (resp. the non-standard $ \mathbf Z_2 (-1)^{F_s}$) orbifold theory of $ SO(4) $ gauge theory. $ \mathbf Z_2 $ can be generated by a group element $ \text{diag}(1,1,1,-1) $. $ O_+(4) $ and $ O_-(4) $ gauge theory have the same $ \binom{(N-1)/2}{2} $ vacua which break $ \mathbf Z_2 $ ($ \sigma_1\neq0, \sigma_2\neq0 $ in \eqref{AdjScalar}) and $ \frac{N-1}{2} $ vacua which preserve $ \mathbf Z_2 $ ($ \sigma_1\neq0, \sigma_2=0 $). $ O_+(4) $ gauge theory has $ \frac{N-1}{2} $ vacua in each of the untwisted and twisted sector while $ O_-(4) $ gauge theory has $ \frac{N-1}{2} $ vacua in only in the twisted sector. When $ N $ is even the theories have only $ \binom{N/2}{2} $ $ \mathbf Z_2 $ breaking vacua.
\begin{table}[h]
	\begin{center}\hspace{-1cm}
		\begin{tabular}{|c|c|c|c|}
			\hline
			$N$ & $O_+(4)$ & $O_-(4)$ & $SO(4)$\\
			\hline
			odd  & $\binom{\frac{N-1}{2}}{2} + N-1$ & $\binom{\frac{N-1}{2}}{2} + \frac{N-1}{2}$ & $2\binom{\frac{N-1}{2}}{2} + \frac{N-1}{2}$\\			
			even & $\binom{\frac{N}{2}}{2}$ & $\binom{\frac{N}{2}}{2}$ & 2$\binom{\frac{N}{2}}{2}$\\			
			\hline
		\end{tabular}
	\end{center}\caption{The total number of ground states of the $O_\pm(4)$ theories.\label{O(4)Vacua}}
\end{table}

Let us compute the elliptic genus of the $O_\pm(4)$ theories with flat connections explained in \cite{Kim:2014dza}. The moduli space of the flat connections consists of eight components, $ a_{k,l,\pm}= (a^1_{k,l,\pm},a^2_{k,l,\pm},a^3_{k,l,\pm},a^4_{k,l,\pm})$
\begin{align}
&a_{(0,0,+)} = \left(u_1,-u_2,u_2,-u_2\right),\quad &&
a_{(0,0,-)} = \left(0,-\frac{1}{2},\frac{1+\tau}{2},-\frac{\tau}{2}\right) \label{O4_Holonomy}
\\
&a_{(1,0,+)} = \left(u_1,-u_1,0,\frac{1}{2}\right),\quad &&
a_{(1,0,-)} = \left(u_1,-u_1,-\frac{\tau}{2},\frac{1+\tau}{2}\right) \nonumber
\\
&a_{(0,1,+)} = \left(u_1,-u_1,0,\frac{\tau}{2}\right),\quad &&
a_{(0,1,-)} = \left(u_1,-u_1,-\frac{1}{2},\frac{1+\tau}{2}\right) \nonumber
\\
&a_{(1,1,+)} = \left(u_1,-u_1,0,\frac{1+\tau}{2}\right),\quad &&
a_{(1,1,-)} = \left(u_1,-u_1,\frac{1}{2},\frac{\tau}{2}\right) \nonumber
\end{align}
where $ \sum_{i=1}^4 a^i_{k,l,\pm} = \frac{k+l\tau}{2}$.
Contribution of $ a_{(0,0,+)}$ to the elliptic genus is given by
\begin{align}
Z^{O(4),N}_{(0,0,+)}(\tau, z, \xi) = \frac{1}{4} \frac{1}{(2\pi i)^2}\sum_{u_* \,\in\, \mathfrak M^*_\text{sing}} \oint_{u_*} \dd u_1 \dd u_2 ~ Z^{O(4),N}_{(0,0,+), \text{1-loop}}
\end{align}
\begin{align}
&\hspace{-1.5cm}
Z^{O(4),N}_{(0,0,+), \text{1-loop}}=\left(\frac{2\pi \eta(q)^3}{\theta_1(\tau|-z)}\right)^2
\frac{ \theta_1 \big(\tau \big| u_1 \!+\!u_2 \big)}{ \theta_1 \big(\tau \big| \!-\!z \!+\!u_1 \!+\!u_2\big)}
\frac{ \theta_1 \big(\tau \big| u_1 \!-\!u_2 \big)}{ \theta_1 \big(\tau \big| \!-\!z \!+\!u_1 \!-\!u_2\big)}
\frac{ \theta_1 \big(\tau \big| \!-\!u_1 \!+\!u_2 \big)}{ \theta_1 \big(\tau \big| \!-\!z \!-\!u_1 \!+\!u_2\big)}
\frac{ \theta_1 \big(\tau \big| \!-\!u_1 \!-\!u_2 \big)}{ \theta_1 \big(\tau \big| \!-\!z \!-\!u_1 \!-\!u_2\big)} \nonumber
\\
&\hspace{-1.5cm}\qquad\qquad\qquad
\prod_{\alpha=1}^{N} \frac{ \theta_1 \big(\tau \big| -z +\xi_\alpha +u_1 \big)}{ \theta_1 \big(\tau \big| \xi_\alpha +u_1\big)} \frac{ \theta_1 \big(\tau \big| -z +\xi_\alpha -u_1 \big)}{ \theta_1 \big(\tau \big| \xi_\alpha -u_1\big)} \frac{ \theta_1 \big(\tau \big| -z +\xi_\alpha +u_2 \big)}{ \theta_1 \big(\tau \big| \xi_\alpha +u_2\big)} \frac{ \theta_1 \big(\tau \big| -z +\xi_\alpha -u_2 \big)}{ \theta_1 \big(\tau \big| \xi_\alpha -u_2\big)}
\end{align}
where the set $ \mathfrak M^*_\text{sing} $ is determined by JK residue \cite{Benini:2013xpa, Kim:2014dza}. The poles $ u_* = (u_{1*},u_{2*}) $ with non-trivial JK residues are
\begin{align}
&(1)\,:~ -z +u_1 +u_2=0, ~ -z -u_1 +u_2=0 ~ &&\rightarrow (u_{1*},u_{2*}) = \left(\frac{k+l\tau}{2}, z+\frac{k+l\tau}{2}\right)
\\
&(2)\,:~ u_2 + \xi_\alpha, ~ -z +u_1 +u_2=0 ~ &&\rightarrow (u_{1*},u_{2*}) = \left(z+\xi_\alpha, -\xi_\alpha\right)
\\
&(3)\,:~ u_2 + \xi_\alpha, ~ u_1 +\xi_\beta=0 ~ &&\rightarrow (u_{1*},u_{2*}) = \left(-\xi_\beta, -\xi_\alpha\right)
\end{align}
Then $ Z^{O(4),N}_{(k,l,+)}(\tau, z, \xi) $ can be written as
\begin{align}
Z^{O(4),N}_{(0,0,+)}(\tau, z, \xi) = \frac{1}{4} \left( Z^{O(4),N}_{(0,0,+),(1)} +Z^{O(4),N}_{(0,0,+),(2)} +Z^{O(4),N}_{(0,0,+),(3)} \right)
\end{align}
where each contribution of the poles is
\begin{align}
Z^{O(4),N}_{(0,0,+),(1)} =& \frac{1}{2}\sum_{k,l=0}^1 y^{-l} \frac{ \theta_1 \big(\tau \big| -z \big)}{ \theta_1 \big(\tau \big| -2z\big)} \frac{ \theta_1 \big(\tau \big| -z -k-l\tau\big)}{ \theta_1 \big(\tau \big| -2z -k-l\tau \big)}
\\&\prod_{\alpha=1}^{N} \frac{ \theta_1 \big(\tau \big| \!-\!z \!+\!\xi_\alpha \!+\!\frac{k+l\tau}{2} \big)}{ \theta_1 \big(\tau \big| \xi_\alpha \!+\!\frac{k+l\tau}{2}\big)} \frac{ \theta_1 \big(\tau \big| \!-\!z \!+\!\xi_\alpha \!-\!\frac{k+l\tau}{2} \big)}{ \theta_1 \big(\tau \big| \xi_\alpha \!-\!\frac{k+l\tau}{2}\big)} \frac{ \theta_1 \big(\tau \big| \xi_\alpha \!+\!\frac{k+l\tau}{2} \big)}{ \theta_1 \big(\tau \big| z \!+\!\xi_\alpha \!+\!\frac{k+l\tau}{2}\big)} \frac{ \theta_1 \big(\tau \big|\!-\!2z \!+\!\xi_\alpha \!-\!\frac{k+l\tau}{2} \big)}{ \theta_1 \big(\tau \big| \!-\!z \!+\!\xi_\alpha \!-\!\frac{k+l\tau}{2}\big)} \nonumber
\end{align}
\begin{align}
Z^{O(4),N}_{(0,0,+),(2)} =& -\sum_{\alpha=1}^{N} \frac{ \theta_1 \big(\tau \big| -z -2\xi_\alpha \big)}{ \theta_1 \big(\tau \big| -2z -2\xi_\alpha \big)} \frac{ \theta_1 \big(\tau \big| -z\big)}{ \theta_1 \big(\tau \big| -2z\big)} \frac{ \theta_1 \big(\tau \big| z +2\xi_\alpha\big)}{ \theta_1 \big(\tau \big| 2\xi_\alpha\big)}
\\&\hspace{-0.5cm} \left(\prod_{\beta=1, \beta\neq\alpha}^{N} \frac{ \theta_1 \big(\tau \big| \!-\!z \!-\!\xi_\alpha \!+\!\xi_\beta \big)}{ \theta_1 \big(\tau \big| \!-\!\xi_\alpha \!+\!\xi_\beta\big)}\right) \prod_{\gamma=1}^{N} \frac{ \theta_1 \big(\tau \big| \!-\!z \!+\!\xi_\alpha \!+\!\xi_\gamma\big)}{ \theta_1 \big(\tau \big| \xi_\alpha \!+\!\xi_\gamma\big)} \frac{ \theta_1 \big(\tau \big| \!-\!2z \!-\!\xi_\alpha \!+\!\xi_\gamma \big)}{ \theta_1 \big(\tau \big| \!-\!z \!-\!\xi_\alpha \!+\!\xi_\gamma \big)} \frac{ \theta_1 \big(\tau \big|\xi_\alpha \!+\!\xi_\gamma  \big)}{ \theta_1 \big(\tau \big| z \!+\!\xi_\alpha \!+\!\xi_\gamma\big)} \nonumber
\end{align}
\begin{align}
Z^{O(4),N}_{(0,0,+),(3)} =& \sum_{\alpha=1}^{N} \sum_{\beta=1, \beta\neq \alpha}^{N} \frac{ \theta_1 \big(\tau \big| -\xi_\alpha -\xi_\beta\big)}{ \theta_1 \big(\tau \big| -z -\xi_\alpha -\xi_\beta \big)} \frac{ \theta_1 \big(\tau \big| \xi_\alpha +\xi_\beta\big)}{ \theta_1 \big(\tau \big| -z +\xi_\alpha +\xi_\beta\big)}
\\&\hspace{-0.8cm}\left(\prod_{\gamma=1, \gamma\neq\alpha, \beta}^{N} \frac{ \theta_1 \big(\tau \big| \!-\!z \!-\!\xi_\alpha \!+\!\xi_\gamma \big)}{ \theta_1 \big(\tau \big| \!-\!\xi_\alpha \!+\!\xi_\gamma\big)} \frac{ \theta_1 \big(\tau \big| \!-\!z \!-\!\xi_\beta \!+\!\xi_\gamma \big)}{ \theta_1 \big(\tau \big| \!-\!\xi_\beta \!+\!\xi_\gamma\big)}\right) \prod_{\delta=1}^{N} \frac{ \theta_1 \big(\tau \big| \!-\!z \!+\!\xi_\alpha \!+\!\xi_\delta\big)}{ \theta_1 \big(\tau \big|\xi_\alpha \!+\!\xi_\delta\big)} \frac{ \theta_1 \big(\tau \big| \!-\!z \!+\!\xi_\beta \!+\!\xi_\delta \big)}{ \theta_1 \big(\tau \big| \xi_\beta \!+\!\xi_\delta \big)}  \nonumber
\end{align}
Contribution of $ a_{(0,0,-)}$ to the elliptic genus is given by
\begin{align}
&\hspace{-1.5cm}
Z^{O(4),N}_{(0,0,-)}=\frac{1}{8}
\frac{ \theta_1 \big(\tau \big| -\frac{1}{2} \big)}{ \theta_1 \big(\tau \big| -z -\frac{1}{2}\big)}
\frac{ \theta_1 \big(\tau \big| \frac{1+\tau}{2} \big)}{ \theta_1 \big(\tau \big| -z +\frac{1+\tau}{2}\big)}
\frac{ \theta_1 \big(\tau \big| -\frac{\tau}{2} \big)}{ \theta_1 \big(\tau \big| -z -\frac{\tau}{2}\big)}
\frac{ \theta_1 \big(\tau \big| \frac{\tau}{2} \big)}{ \theta_1 \big(\tau \big| -z +\frac{\tau}{2}\big)} \frac{ \theta_1 \big(\tau \big| -\frac{1+\tau}{2} \big)}{ \theta_1 \big(\tau \big| -z -\frac{1+\tau}{2}\big)}
\frac{ \theta_1 \big(\tau \big| \frac{1}{2} \big)}{ \theta_1 \big(\tau \big| -z +\frac{1}{2}\big)} \nonumber
\\
&\hspace{-1.5cm}\qquad\qquad\qquad
\prod_{\alpha=1}^{N} \frac{ \theta_1 \big(\tau \big| -z +\xi_\alpha \big)}{ \theta_1 \big(\tau \big| \xi_\alpha\big)} \frac{ \theta_1 \big(\tau \big| -z +\xi_\alpha -\frac{1}{2} \big)}{ \theta_1 \big(\tau \big| \xi_\alpha -\frac{1}{2}\big)} \frac{ \theta_1 \big(\tau \big| -z +\xi_\alpha +\frac{1+\tau}{2} \big)}{ \theta_1 \big(\tau \big| \xi_\alpha +\frac{1+\tau}{2}\big)} \frac{ \theta_1 \big(\tau \big| -z +\xi_\alpha -\frac{\tau}{2} \big)}{ \theta_1 \big(\tau \big| \xi_\alpha -\frac{\tau}{2}\big)}
\end{align}
Contributions of $ a_{(1,0,\pm)},\, a_{(0,1,\pm)},\, a_{(1,1,\pm)}$ can be written in terms of $ a_{(k,l,\pm)} = (u_1, -u_1, a, b)$,
where $ a $, $ b $ are given in \eqref{O4_Holonomy} corresponding to $ (k,l,\pm) $.
\begin{align}
Z^{O(4),N}_{(k,l,\pm)}=\frac{1}{4} \frac{1}{2\pi i}\sum_{u_* \,\in\, \mathfrak M^*_\text{sing}} \oint_{u_*} \dd u_1 ~ Z^{O(4),N}_{(k,l,\pm), \text{1-loop}}
\end{align}
\begin{align}
Z^{O(4),N}_{(k,l,\pm), \text{1-loop}}=&\left(\frac{2\pi \eta(q)^3}{\theta_1(\tau|-z)}\right)
\frac{ \theta_1 \big(\tau \big| u_1 +a \big)}{ \theta_1 \big(\tau \big| -z +u_1 +a\big)}
\frac{ \theta_1 \big(\tau \big| -u_1 +a \big)}{ \theta_1 \big(\tau \big| -z -u_1 +a\big)}
\\
&
\frac{ \theta_1 \big(\tau \big| u_1 +b \big)}{ \theta_1 \big(\tau \big| -z +u_1 +b\big)}
\frac{ \theta_1 \big(\tau \big| -u_1 +b \big)}{ \theta_1 \big(\tau \big| -z -u_1 +b\big)}
\frac{ \theta_1 \big(\tau \big| a +b \big)}{ \theta_1 \big(\tau \big| -z +a +b\big)} \nonumber
\\
&
\prod_{\alpha=1}^{N} \frac{ \theta_1 \big(\tau \big| -z +\xi_\alpha +u_1 \big)}{ \theta_1 \big(\tau \big| \xi_\alpha +u_1\big)} \frac{ \theta_1 \big(\tau \big| -z +\xi_\alpha -u_1 \big)}{ \theta_1 \big(\tau \big| \xi_\alpha -u_1\big)} \frac{ \theta_1 \big(\tau \big| -z +\xi_\alpha +a \big)}{ \theta_1 \big(\tau \big| \xi_\alpha +a\big)} \frac{ \theta_1 \big(\tau \big| -z +\xi_\alpha +b \big)}{ \theta_1 \big(\tau \big| \xi_\alpha +b\big)} \nonumber
\end{align}
where $ (k,l)=(1,0),\,(0,1),\,(1,1) $ and $ \mathfrak M^*_\text{sing} = \{ z-a,\, z-b,\, -\xi_\alpha\}$. Evaluating JK-residue we obtain
\begin{align}
Z^{O(4),N}_{(k,l,\pm)} = \frac{1}{4} y^{-\frac{N-3}{2}l} \left( Z^{O(4),N}_{(k,l,\pm),(1)} +Z^{O(4),N}_{(k,l,\pm),(2)} +Z^{O(4),N}_{(k,l,\pm),(3)} \right)
\end{align}
where
\begin{align}
Z^{O(4),N}_{(k,l,\pm),(1)} &= -
\frac{ \theta_1 \big(\tau \big| -a +b +z\big)}{ \theta_1 \big(\tau \big| -a +b\big)}
\frac{ \theta_1 \big(\tau \big| -z +2a \big)}{ \theta_1 \big(\tau \big| -2z +2a\big)}
\frac{ \theta_1 \big(\tau \big| -z +a +b \big)}{ \theta_1 \big(\tau \big| -2z +a +b\big)}
\frac{ \theta_1 \big(\tau \big| a +b \big)}{ \theta_1 \big(\tau \big| -z +a +b\big)} \nonumber
\\
&
\prod_{\alpha=1}^{N} \frac{ \theta_1 \big(\tau \big| \xi_\alpha -a \big)}{ \theta_1 \big(\tau \big| \xi_\alpha -a +z\big)} \frac{ \theta_1 \big(\tau \big| \xi_\alpha +a -2z \big)}{ \theta_1 \big(\tau \big| \xi_\alpha +a\big)} \frac{ \theta_1 \big(\tau \big| \xi_\alpha +b -z\big)}{ \theta_1 \big(\tau \big| \xi_\alpha +b\big)} \nonumber
\end{align}

\begin{align}
Z^{O(4),N}_{(k,l,\pm),(2)} &= -
\frac{ \theta_1 \big(\tau \big| -b +a +z\big)}{ \theta_1 \big(\tau \big| -b +a\big)}
\frac{ \theta_1 \big(\tau \big| -z +2b \big)}{ \theta_1 \big(\tau \big| -2z +2b\big)}
\frac{ \theta_1 \big(\tau \big| -z +a +b \big)}{ \theta_1 \big(\tau \big| -2z +a +b\big)}
\frac{ \theta_1 \big(\tau \big| a +b \big)}{ \theta_1 \big(\tau \big| -z +a +b\big)} \nonumber
\\
&
\prod_{\alpha=1}^{N} \frac{ \theta_1 \big(\tau \big| \xi_\alpha -b \big)}{ \theta_1 \big(\tau \big| \xi_\alpha -b +z\big)} \frac{ \theta_1 \big(\tau \big| \xi_\alpha +b -2z \big)}{ \theta_1 \big(\tau \big| \xi_\alpha +b\big)} \frac{ \theta_1 \big(\tau \big| \xi_\alpha +a -z\big)}{ \theta_1 \big(\tau \big| \xi_\alpha +a\big)} \nonumber
\end{align}
\begin{align}
&Z^{O(4),N}_{(k,l,\pm),(3)} =  \sum_{\alpha=1}^N
\frac{ \theta_1 \big(\tau \big| -\xi_\alpha +a \big)}{ \theta_1 \big(\tau \big| -\xi_\alpha +a -z\big)}
\frac{ \theta_1 \big(\tau \big| -\xi_\alpha +b \big)}{ \theta_1 \big(\tau \big| -\xi_\alpha +b -z\big)}
\frac{ \theta_1 \big(\tau \big| a +b \big)}{ \theta_1 \big(\tau \big| -z +a +b\big)}
\frac{ \theta_1 \big(\tau \big| 2\xi_\alpha -z \big)}{ \theta_1 \big(\tau \big| 2\xi_\alpha \big)} \nonumber
\\
&
\qquad\qquad\prod_{\beta=1,\beta\neq\alpha}^{N} \frac{ \theta_1 \big(\tau \big|\xi_\beta -\xi_\alpha -z  \big)}{ \theta_1 \big(\tau \big| \xi_\beta -\xi_\alpha \big)}
\frac{ \theta_1 \big(\tau \big|\xi_\beta +\xi_\alpha -z  \big)}{ \theta_1 \big(\tau \big| \xi_\beta +\xi_\alpha \big)}
\frac{ \theta_1 \big(\tau \big|\xi_\beta +a -z  \big)}{ \theta_1 \big(\tau \big| \xi_\beta +a \big)}
\frac{ \theta_1 \big(\tau \big|\xi_\beta +b -z  \big)}{ \theta_1 \big(\tau \big| \xi_\beta +b \big)}. \nonumber
\end{align}

We fix the mod 2 theta angle and a choice of two orbifolds as in $O(2)$ theories. We define
\begin{align}
Z^{O(4),N}_{(kl)}=Z^{O(4),N}_{(k,l,+)}+(-1)^{N+1}Z^{O(4),N}_{(k,l,-)}
\end{align}
where $ (kl)=(00),\,(10),\,(01),\,(11) $.
Finally we obtain the elliptic genus for the theory A
\begin{align}
Z^{\text{A},\,SO(4),N}(\tau, z, \xi_\alpha)=&
Z^{O(4),N}_{(00)} \label{EG_SO(4)}
\\
Z^{\text{A},\,O_+(4),N}(\tau, z, \xi_\alpha)=&
\frac{1}{2}\left(Z^{O(4),N}_{(00)} +(-1)^{N+1} \left(Z^{O(4),N}_{(10)} + Z^{O(4),N}_{(01)} + Z^{O(4),N}_{(11)} \right)\right)\label{EG_O+(4)}
\\
Z^{\text{A},\,O_-(4),N}(\tau, z, \xi_\alpha)=&
\frac{1}{2}\left(Z^{O(4),N}_{(00)} +(-1)^{N}\left(Z^{O(4),N}_{(10)} +Z^{O(4),N}_{(01)} +Z^{O(4),N}_{(11)} \right)\right)\label{EG_O-(4)}
\end{align}
and for theory B
\begin{align}
Z^{\text{B},\,SO(4),N}(\tau, z, \xi_\alpha)=&
\widetilde Z^{O(4),N}_{(00)} \label{EG_SO(4)_dual}
\\
Z^{\text{B},\,O_+(4),N}(\tau, z, \xi_\alpha)=&
\frac{1}{2}\left(\widetilde Z^{O(4),N}_{(00)} +(-1)^{N+1} \left(\widetilde  Z^{O(4),N}_{(10)} + \widetilde Z^{O(4),N}_{(01)} + \widetilde Z^{O(4),N}_{(11)} \right)\right)\label{EG_O+(4)_dual}
\\
Z^{\text{B},\,O_-(4),N}(\tau, z, \xi_\alpha)=&
\frac{1}{2}\left(\widetilde Z^{O(4),N}_{(00)} +(-1)^{N} \left(\widetilde Z^{O(4),N}_{(10)} + \widetilde Z^{O(4),N}_{(01)} + \widetilde Z^{O(4),N}_{(11)}\right)\right) \label{EG_O-(4)_dual}
\end{align}
where
\begin{align}
\widetilde Z^{O(4),N}_{(kl)}(\tau,z,\xi)&=\left(\prod_{\alpha=1}^{N}\prod_{\beta=\alpha}^{N}\frac{\theta_1(\tau|-z +\xi_\alpha +\xi_\beta)}{\theta_1(\tau|\xi_\alpha +\xi_\beta)}\right)\times Z^{O(4),N}_{(kl)}(\tau,z,-\xi+z/2)
\end{align}

Let's compute the Witten index from the elliptic genus.
\begin{align}
\lim_{y\rightarrow 1}Z^{\text{A},\,SO(4),N}=& \frac{1}{4}\left(N(N-2) +\frac{1+(-1)^{N+1}}{2}\right) =\left\{\begin{array}{ll}
\frac{N(N-2)}{4}&\quad\mbox{for $ N $ even}\\
\frac{(N-1)^2}{4}&\quad\mbox{for $ N $ odd}
\end{array}\right.\label{WI_SO(4)}
\end{align}
\begin{align}
\lim_{y\rightarrow 1}Z^{\text{A},\,O_+(4),N}=&
\frac{1}{8}\left(N(N-2) +\frac{1+(-1)^{N+1}}{2} +(N-1)(1+(-1)^{N+1})\right) +\frac{1}{2}(N-1)\frac{1+(-1)^{N+1}}{2}\nonumber
\\
&=\left\{\begin{array}{ll}
\frac{N(N-2)}{8}&\quad\mbox{for $ N $ even}\\
\frac{N^2-1}{8}+\frac{N-1}{2}&\quad\mbox{for $ N $ odd}
\end{array}\right.\label{WI_O+(4)}
\end{align}
\begin{align}
\lim_{y\rightarrow 1}Z^{\text{A},\,O_-(4),N}=&
\frac{1}{8}\left(N(N-2) +\frac{1+(-1)^{N+1}}{2} -(N-1)(1+(-1)^{N+1})\right) -\frac{1}{2}(N-1)\frac{1+(-1)^{N+1}}{2}\nonumber
\\
&=\left\{\begin{array}{ll}
\frac{N(N-2)}{8}&\quad\mbox{for $ N $ even}\\
\frac{(N-1)(N-3)}{8}-\frac{N-1}{2}&\quad\mbox{for $ N $ odd}
\end{array}\right.\label{WI_O-(4)}
\end{align}
where two terms for odd $ N $ correspond to the number of vacua in the untwisted sector and the twisted sector respectively.
Note that for odd $ N $, the Witten index of the $ O_-(4) $ gauge theory has a relative sign difference between the untwisted sector and the twisted sector.

We have tested dualities for $ N=3,\,4,\,5,\,6,\,7 $.
$N=3$  we checked
\begin{align}
Z^{M,3}=Z^{O(4),3}_{(00)} = Z^{O(4),3}_{(10)} = Z^{O(4),3}_{(01)} = Z^{O(4),3}_{(11)}
\end{align}
analytically at appendix E.
Thus we have
\begin{align}
&Z^{\text{A},\,SO(4),3}=Z^{M,3}
\\
&Z^{\text{A},\,O_+(4),3}=2Z^{M,3}
\\
&Z^{\text{A},\,O_-(4),3}=Z^{M,3}
\end{align}
For $N=5,\,7$ we checked that the elliptic genera satisfy
\begin{align}
\left(\begin{array}{c}
Z^{O(4),N}_{(00)}\\
Z^{O(4),N}_{(10)}\\
Z^{O(4),N}_{(01)}\\
Z^{O(4),N}_{(11)}\\
\end{array}\right)
=
\frac{1}{2}\left(\begin{array}{cccc}
1&1&1&1\\
-1&-1&1&1\\
-1&1&-1&1\\
-1&1&1&-1\\
\end{array}\right)
\left(\begin{array}{c}
\widetilde Z^{O(N-3),N}_{(00)}\\
\widetilde Z^{O(N-3),N}_{(10)}\\
\widetilde Z^{O(N-3),N}_{(01)}\\
\widetilde Z^{O(N-3),N}_{(11)}\\
\end{array}\right)
\end{align}
up to $q^3$ with some specific values of flavor holonomies such as $\xi_\alpha = \frac{\alpha}{N+1}$ for computational simplicity.
For $N=4, 6$ we checked
\begin{align}
\left(\begin{array}{c}
Z^{O(4),N}_{(00)}\\
Z^{O(4),N}_{(10)}\\
Z^{O(4),N}_{(01)}\\
Z^{O(4),N}_{(11)}\\
\end{array}\right)
=
\frac{1}{2}\left(\begin{array}{cccc}
1&1&1&1\\
1&1&-1&-1\\
1&-1&1&-1\\
1&-1&-1&1\\
\end{array}\right)
\left(\begin{array}{c}
\widetilde Z^{O(N-3),N}_{(00)}\\
\widetilde Z^{O(N-3),N}_{(10)}\\
\widetilde Z^{O(N-3),N}_{(01)}\\
\widetilde Z^{O(N-3),N}_{(11)}\\
\end{array}\right)
\end{align}
up to $q^3$ with some specific values of flavor holonomies.

\newpage
\appendix
\section{SU(2) VS SO(3)}
\subsection{$\mathcal N=(2,2)$ pure SU($2$) gauge theory}
Let us consider an $\mathcal N=(2,2)$ pure SU($2$) gauge theory. The elliptic genus is given by \cite{Benini:2013nda}
\begin{align}
Z_(\tau, z, \xi) = \frac{1}{2} \sum_{u_* \,\in\, \mathfrak M^+_\text{sing}}
\frac{i \eta(q)^3}{\theta_1(\tau|-z)}
\oint_{u_*}\dd u\, \frac{\theta_1(\tau|2u)}{\theta_1(\tau| -z +2u)} \,
\frac{\theta_1(\tau|-2u)}{\theta_1(\tau| -z -2u)}
\end{align}
The set $ \mathfrak M^+_\text{sing} $ is
\begin{align}
\mathfrak M^+_\text{sing} =\Big\{  \frac{z+a+b\tau}{2} ~\big|~ a,b=0,1\Big\} \;.
\end{align}
We have
\begin{align}
Z^{SU(2)}(\tau,z,\xi)
&= \frac{1}{4} \sum_{a,b=0}^1 y^{-b} \frac{\theta_1(\tau|-z-a-b\tau)}{\theta_1(\tau|-2z-a-b\tau)}
\\
&= \frac{\theta_1(\tau|-z)}{\theta_1(\tau|-2z)}
\end{align}
This is the same as elliptic genus of a $Y_2=\Tr \Sigma^2$ twisted chiral field which has vector-like R-charge $4$. The Witten index limit $ z\rightarrow 0 $ of the elliptic genus gives $ \lim_{z\rightarrow 0} Z^{SU(2)}(\tau,z,\xi) = \frac{1}{2}$ where non-integer value comes from the bosonic zero mode of $ Y_2 $.

\subsection{$\mathcal N=(2,2)$ pure SO($3$) gauge theory}
Let us consider an $\mathcal N=(2,2)$ pure SO($3$) gauge theory. The elliptic genus is given by
\begin{align}
Z_{T^2}(\tau, z, \xi) =&
- \frac{1}{2}
\frac{i \eta(q)^3}{\theta_1(\tau|-z)}
\oint_{u=z} \dd u\, \frac{\theta_1(\tau|u)}{\theta_1(\tau| -z +u)} \,
\frac{\theta_1(\tau|-u)}{\theta_1(\tau| -z -u)}
\\
&+e^{i\theta} \frac{1}{4} \frac{\theta_1(\tau|+\frac{1}{2})}{\theta_1(\tau|-z +\frac{1}{2})} \frac{\theta_1(\tau|-\frac{1+\tau}{2})}{\theta_1(\tau|-z -\frac{1+\tau}{2})} \frac{\theta_1(\tau|+\frac{\tau}{2})}{\theta_1(\tau|-z +\frac{\tau}{2})} \nonumber
\\
=&-\frac{1}{2} \frac{\theta_1(\tau|-z)}{\theta_1(\tau|-2z)}+ e^{i\theta}\frac{1}{4} \frac{\theta_1(\tau|+\frac{1}{2})}{\theta_1(\tau|-z +\frac{1}{2})} \frac{\theta_1(\tau|-\frac{1+\tau}{2})}{\theta_1(\tau|-z -\frac{1+\tau}{2})} \frac{\theta_1(\tau|+\frac{\tau}{2})}{\theta_1(\tau|-z +\frac{\tau}{2})} \nonumber
\end{align}
where $ \theta $ is a tree level mod 2 theta angle which is a phase difference between holonomies.
We checked that
\begin{align}
\frac{\theta_1(\tau|-z)}{\theta_1(\tau|-2z)} = \frac{1}{2} \frac{\theta_1(\tau|+\frac{1}{2})}{\theta_1(\tau|-z +\frac{1}{2})} \frac{\theta_1(\tau|-\frac{1+\tau}{2})}{\theta_1(\tau|-z -\frac{1+\tau}{2})} \frac{\theta_1(\tau|+\frac{\tau}{2})}{\theta_1(\tau|-z +\frac{\tau}{2})}
\end{align}
so we have
\begin{align}
Z_{T^2}(\tau,z,\xi)= \frac{\theta_1(\tau|-z)}{\theta_1(\tau|-2z)}\left(\frac{-1+e^{i\theta}}{2}\right)
\end{align}
In the convention of \cite{Hori:2011pd}, the pure $SO(3)$ gauge theory is regular for $\theta=0$ and irregular for $ \theta=\pi $. The elliptic genus shows that the pure $SO(3)$ theory has no supersymmetric vacuum for $\theta=0$  while it has a non-compact Coulomb branch for $\theta=\pi$ as in the pure $SU(2)$ theory.  According to \cite{Hori:2011pd}, we expect the supersymmetry is broken for the pure $SO(3)$ gauge theory
whether it is regular or not. The elliptic genus computation of the regular theory is consistent with it, but for the irregular theory it gives
the unexpected answer. We interpret such discrepancy is due to the existence of the noncompact branch \cite{Aharony:2016jki}.

\section{Elliptic genus formula}
The elliptic genus of a gauge theory with gauge group $G$ of rank one is given by \cite{Benini:2013nda}
\begin{align}
Z_{T^2}(\tau,z,\xi)
= - \frac{1}{|W|} \sum_{u_j \,\in\, \mathfrak M^+_\text{sing}} \oint_{u=u_j} \hspace{-1em} \dd u\, \frac{i\eta(q)^3}{\theta_1(q, y^{-1})} \prod_{\alpha \,\in\, G} \frac{\theta_1(q, x^{\alpha})}{\theta_1(q, y^{-1} x^\alpha)} \times\phantom{,} \nonumber\\
\phantom{,}\times \prod_{\Phi_i} \prod_{\rho\,\in\,\mathfrak R_i} \frac{\theta_1 \big(q \,,\, y^{R_i/2-1} \, x^{\rho} \, e^{2\pi i P_i(\xi)} \big)}{\theta_1 \big( q \,,\, y^{R_i/2} \, x^\rho \, e^{2\pi i P_i(\xi)} \big)}
\end{align}
where $|W|$ is the order of the Weyl group. The integrand in the first line is a contribution of gauge multiplet and the factors in the second line come from matter fields. Elliptic genus of a chiral multiplet is given by
\begin{align}
Z_{\Phi,R, Q}(\tau,z,u)
&=\frac{\theta_1\left(\tau|(\frac{R}{2}-1)z+Qu\right)}{\theta_1\left(\tau|\frac{R}{2}z+Qu\right)}
\\
&=-y^{\frac{R}{2}-\frac{1}{2}}x^{Q}
\frac{1-y^{1-\frac{R}{2}}x^{-Q}}{1-y^{\frac{R}{2}}x^{Q}}
\prod_{n=1}^\infty \frac{\left(1-y^{\frac{R}{2}-1}x^{Q} q^n\right) \left(1-y^{-\left(\frac{R}{2}-1\right)}x^{-Q}q^{n} \right) }{\left(1-y^{\frac{R}{2}}x^{Q} q^n\right) \left(1-y^{-\frac{R}{2}}x^{-Q}q^{n} \right)}~.
\end{align}

\section{Eta and theta functions}
The Dedekind eta function and the Jacobi theta function are
\begin{align}
&\eta(q)=q^{1/24}\prod_{n = 1}^\infty (1-q^n)
\\
&\theta_1(\tau | z)= -i q^{1/8} y^{1/2} \prod_{k=1}^\infty (1-q^k) (1-y q^k) (1-y^{-1}q^{k-1})
\\
&\theta_3(\tau | z)= \prod_{n=1}^\infty(1-q^n)(1+yq^{n-\frac{1}{2}})(1+y^{-1}q^{n-\frac{1}{2}})
\end{align}
where $q = e^{2\pi i \tau}$, $y = e^{2\pi i z}$ and $\im\tau > 0$. Two Jacobi theta functions are related as
\begin{align}
\theta_1 \left(\tau \big| z \right) = e^{\frac{\pi i}{4}\tau +\frac{\pi i}{2} (2z -1)}\theta_3 \left(\tau \big| z +\frac{1+\tau}{2}\right)
\end{align}
The modular properties are
\begin{align}
&\theta_1 \left(\tau +1 \big| z \right) =e^{\frac{\pi i}{4}} \theta_1 \left(\tau \big| z \right),\quad
\theta_1 \left(-\frac{1}{\tau} \big| \frac{z}{\tau} \right) =-i \sqrt{-i\tau} e^{\frac{\pi i}{\tau}z^2}\theta_1 \left(\tau \big| z \right)
\\
&\theta_3 \left(\tau +1 \big| z \right) = \theta_3 \left(\tau \big| z+\frac{1}{2} \right),\quad
\theta_3 \left(-\frac{1}{\tau} \big| \frac{z}{\tau} \right) = \sqrt{-i\tau} e^{\frac{\pi i}{\tau}z^2}\theta_3 \left(\tau \big| z \right)
\end{align}
Under shifts of $z$ the Jacobi theta functions transform as
\begin{align}
&\theta_1 \left(\tau \big| z + a+ b\tau \right) = (-1)^{a+b} e^{-2\pi i b z - \pi i b^2 \tau} \theta_1 \left(\tau \big| z \right)
\\
&\theta_3 \left(\tau \big| z + a+ b\tau \right) = e^{-2\pi i b z - \pi i  b^2 \tau} \theta_3 \left(\tau \big| z \right)
\end{align}

Useful integral formula for poles associated with a chiral multiplet and a twisted chiral multiplet are
\begin{align}
&\frac{1}{2\pi i}\sum_{a,b=0}^{Q-1}\oint_{u \,=\, \frac{1}{Q}(-rz +a +b\tau)} \hspace{-.5em} \dd u \; \frac{\theta_1(\tau|(r-1)z+Qu)}{\theta_1(\tau|rz+Qu)}f(u)
= \frac{\theta_1(\tau|-z)}{2\pi \eta(q)^3} \frac{1}{Q}\sum_{a,b=0}^{Q-1} y^{b} f\left(\frac{1}{Q}(-rz +a +b\tau)\right) \;.\nonumber
\\
&\frac{1}{2\pi i}\sum_{a,b=0}^{Q-1}\oint_{u \,=\, \frac{1}{Q}(rz +a +b\tau)} \hspace{-.5em} \dd u \; \frac{\theta_1(\tau|(1-r)z+Qu)}{\theta_1(\tau|-rz+Qu)}f(u)
= -\frac{\theta_1(\tau|-z)}{2\pi \eta(q)^3} \frac{1}{Q}\sum_{a,b=0}^{Q-1} y^{-b} f\left(\frac{1}{Q}(rz +a +b\tau)\right) \;.
\end{align}
where $f(u)$ is analytic at $ u =\frac{rz +a +b\tau}{Q} $.

\section{Superconformal index}
In this section we would like to compute the superconformal index for $ O_\pm(1) $ and $ O_\pm(2) $, $ SO(2) $ theories (including free and Landau-Ginzburg theories). We also compute $ (c,c) $ ring elements and check their matching between dual theories.
To obtain the chiral ring elements we need the superconformal index, which imposes the NS boundary condition for fermions \cite{Gadde:2013ftv}.
This can be obtained from the elliptic genus using the spectral flow. Thus the main object to discuss in this section is the superconformal
index.
\subsection{$ O(1) $ gauge theories}
The superconformal index of a free chiral multiplet is given by
\begin{align}
\mathcal I^{\Phi} (q,y,a)=\Delta\left(q, y, a \right)  = \prod_{i=1}^{\infty}\frac{\left(1-ay^{-1}q^{i-\frac{1}{2}}\right)\left(1-a^{-1}yq^{i-\frac{1}{2}}\right)}{\left(1-aq^{i-1}\right)\left(1-a^{-1}q^{i}\right)}
= \frac{\theta_3 \left(\tau \big| -z +\xi -\frac{1}{2}\right)}{\theta_3 \left(\tau \big| \xi -\frac{1+\tau}{2} \right)}
\end{align}
where $ q=e^{2\pi i \tau} $, $ y=e^{2\pi i z} $, $ a=e^{2\pi i \xi} $ \cite{Gadde:2013ftv}. The superconformal index of a chiral multiplet with a left-moving R-charge $r$ is $ \mathcal I^{\Phi_r}(q,y,a)= \Delta(q,y,a(q^{\frac{1}{2}}y)^r) $. We shall sometimes simply write $ \Delta\left(a \right) $.

Let's compute the superconformal index for $ O_\pm(1) $ and $ SO(1) $ (free) theory A. It is given by
\begin{align}
\mathcal I^{A, SO(1), N}(q,y,a)&= \mathcal I^{O(1), N}_{(00)}
\\
\mathcal I^{A, O_+(1), N}(q,y,a)&= \frac{1}{2}\left(\mathcal I^{O(1), N}_{(00)} +\mathcal I^{O(1), N}_{(10)} +(-1)^{N} \mathcal I^{O(1), N}_{(01)} + \mathcal I^{O(1), N}_{(11)}\right)
\\
\mathcal I^{A, O_-(1), N}(q,y,a)&= \frac{1}{2}\left(\mathcal I^{O(1), N}_{(00)} +\mathcal I^{O(1), N}_{(10)} +(-1)^{N+1} \mathcal I^{O(1), N}_{(01)} + \mathcal I^{O(1), N}_{(11)}\right)
\end{align}
where
\begin{align}
&\CalI^{O(1), N}_{(kl)}(q,y,a) = \left(q^{\frac{1}{4}}y^{-\frac{1}{2}}\right)^{Nl} \prod_{\alpha=1}^N\Delta((-1)^k q^{\frac{1}{2}l} a_\alpha)
\end{align}
The phase factors in front of the $ (01) $ component come from a choice of two orbifolds where $ O_+(1) $ (resp. $ O_-(1) $) theory is a standard $ \mathbf Z_2 $ (resp. non-standard $ \mathbf Z_2 (-1)^{F_s}$) orbifold for even $ N $. For odd $ N $, the role of $ O_+(1) $ and $ O_-(1) $ are reversed.
The effects of $(-1)^{F_s}$ on RR and NSNS sectors are \cite{Hori:2011pd}
\begin{align}
(-1)^{F_s}=
\left\{\begin{array}{ll}
1&\quad\mbox{in untwisted NSNS and twisted RR}\\
-1&\quad\mbox{in twisted NSNS and untwisted RR.}
\end{array}\right.
\end{align}
The superconformal indices are consistent with a complex mass deformation where a massive chiral field contribute to the index given by
\begin{align}
\CalI^{O(1), 1}_{(kl),\textrm{massive}}(q,y,a)=(q^{\frac{1}{4}}y^{-\frac{1}{2}})^l \Delta(q,y,(-1)^k q^{\frac{1}{2}l} (q^\frac{1}{2}y)^{\frac{1}{2}}) = \left\{\begin{array}{ll}
1&\mbox{~for $(kl)=(00),\,(10),\,(11)$}\\
-1&\mbox{~for $(kl)=(01)$}.
\end{array}\right.
\end{align}

The $ O_\pm(1) $ and $ SO(1) $ (Landau-Ginzburg) theory B have the superconformal index
\begin{align}
\mathcal I^{B, SO(1), N}(q,y,a)&= \widetilde \CalI^{O(1), N}_{(00)}
\\
\mathcal I^{B, O_+(1), N}(q,y,a)&= \frac{1}{2}\left(\widetilde \CalI^{O(1), N}_{(00)} +\widetilde \CalI^{O(1), N}_{(10)} +(-1)^{N}\widetilde \CalI^{O(1), N}_{(01)} + \widetilde \CalI^{O(1), N}_{(11)}\right)
\\
\mathcal I^{B, O_-(1), N}(q,y,a)&= \frac{1}{2}\left(\widetilde \CalI^{O(1), N}_{(00)} +\widetilde \CalI^{O(1), N}_{(10)} +(-1)^{N+1}\widetilde \CalI^{O(1), N}_{(01)} + \widetilde \CalI^{O(1), N}_{(11)}\right)
\end{align}
where
\begin{align}
\widetilde \CalI^{O(1), N}_{(kl)}(q,y,a) = \CalI^{M, N}(q,y,a) \CalI^{O(1), N}_{(kl)}(q,y,a^{-1} (q^{\frac{1}{2}}y)^{\frac{1}{2}})
\end{align}
$ \CalI^{M, N} $ is the contribution of $ \frac{N(N+1)}{2} $ singlets $ M_{\alpha\beta} $,
\begin{align}
\CalI^{M, N}(q,y,a)=\prod_{\alpha=1}^{N}\prod_{\beta=\alpha}^{N}\Delta\left(q, y, a_\alpha a_\alpha \right)
\end{align}

We would like to check the operator matching for the dualities. We focus on $ (c,c) $ ring elements whose left-moving energy and R-charge satisfy $ h = \frac{j}{2} $. They can be obtained by
\begin{align}
P(x,a)=\lim_{t\rightarrow 0}\CalI(qt,yt^{-\frac{1}{2}},a)
\end{align}
where the right-hand side is a function of $x=q^{\frac{1}{2}}y$ and $a$, which means only states satisfying $ h = \frac{j}{2} $ survive in the limit. For theories A, chiral ring contributions are
\begin{align}
P^{A, SO(1), N}(q,y,a)&= P^{O(1), N}_{(00)}
\\
P^{A, O_+(1), N}(q,y,a)&= \frac{1}{2}\left(P^{O(1), N}_{(00)} +P^{O(1), N}_{(10)} +(-1)^{N} P^{O(1), N}_{(01)} + P^{O(1), N}_{(11)}\right)
\\
P^{A, O_-(1), N}(q,y,a)&= \frac{1}{2}\left(P^{O(1), N}_{(00)} +P^{O(1), N}_{(10)} +(-1)^{N+1} P^{O(1), N}_{(01)} + P^{O(1), N}_{(11)}\right)
\end{align}
where
\begin{align}
&P^{O(1), N}_{(00)} = \prod_{\alpha=1}^N \frac{1-a_\alpha^{-1}x}{1-a_\alpha} ,
&&P^{O(1), N}_{(10)} = \prod_{\alpha=1}^N \frac{1+a_\alpha^{-1}x}{1+a_\alpha}
\\
&P^{O(1), N}_{(01)} = (-1)^N x^{\frac{N}{2}} \prod_{\alpha=1}^N a_\alpha^{-1} ,
&&P^{O(1), N}_{(11)} = x^{\frac{N}{2}} \prod_{\alpha=1}^N a_\alpha^{-1}.
\end{align}
For theories B, chiral ring contributions are
\begin{align}
P^{B, SO(1), N}(q,y,a)&= \widetilde P^{O(1), N}_{(00)}
\\
P^{B, O_+(1), N}(q,y,a)&= \frac{1}{2}\left(\widetilde P^{O(1), N}_{(00)} +\widetilde P^{O(1), N}_{(10)} +(-1)^{N}\widetilde P^{O(1), N}_{(01)} + \widetilde P^{O(1), N}_{(11)}\right) \label{SCI_B_O(1)+_N}
\\
P^{B, O_-(1), N}(q,y,a)&= \frac{1}{2}\left(\widetilde P^{O(1), N}_{(00)} +\widetilde P^{O(1), N}_{(10)} +(-1)^{N+1}\widetilde P^{O(1), N}_{(01)} + \widetilde P^{O(1), N}_{(11)}\right)
\end{align}
where
\begin{align}
&\widetilde P^{O(1), N}_{(kl)}(x, a_\alpha) = P^{M, N} \times P^{O(1),N}_{(kl)}(x,a^{-1}_{\alpha}x^{\frac{1}{2}})  ,\\
&P^{M,N} = \prod_{\alpha=1}^N \prod_{\beta=\alpha}^N \frac{1-a_\alpha^{-1} a_\beta^{-1} x}{1-a_\alpha a_\beta}
\end{align}
%
%
and $P^{M,N}$ is the contribution of the singlets $M_{\alpha \beta}$.

We have checked the simplest dualities
\begin{align}
\CalI^{\text{A},\,SO(1),1}&= \CalI^{\text{B},\,O_+(1),1}
\\
\CalI^{\text{A},\,O_+(1),1}&= \CalI^{\text{B},\,SO(1),1}
\\
\CalI^{\text{A},\,O_-(1),1}&= \CalI^{\text{B},\,O_-(1),1}
\end{align}
and in terms of contributions of four components,
\begin{align}
&\CalI^{O(1),1}_{(00)}=\frac{1}{2}\left(\widetilde \CalI^{O(1),1}_{(00)} +\widetilde \CalI^{O(1),1}_{(10)} -\widetilde \CalI^{O(1),1}_{(01)} +\widetilde \CalI^{O(1),1}_{(11)}\right) \label{SCI:SO(1)N=1/O+(1)}
\\
&\widetilde \CalI^{O(1),1}_{(00)} =\frac{1}{2}\left(\CalI^{O(1),1}_{(00)} +\CalI^{O(1),1}_{(10)} -\CalI^{O(1),1}_{(01)} +\CalI^{O(1),1}_{(11)}\right)  \label{SCI:O+(1)N=1/SO(1)}
\\
&\CalI^{O(1),1}_{(00)} +\CalI^{O(1),1}_{(10)} +\CalI^{O(1),1}_{(01)} +\CalI^{O(1),1}_{(11)}
=\widetilde \CalI^{O(1),1}_{(00)} +\widetilde \CalI^{O(1),1}_{(10)} +\widetilde \CalI^{O(1),1}_{(01)} +\widetilde \CalI^{O(1),1}_{(11)} \label{SCI:O-(1)N=1/O-(1)}
\end{align}
For dualities between $ O_-(1)/O_-(1) $ theories we also have checked that the untwisted (resp. twisted) NSNS sector of the $ O_-(1) $ theory A is the same as the untwisted (resp. twisted) NSNS sector of the $ O_-(1) $ theory B as derived in \cite{Hori:2011pd}. This is also true for $ O_-(k)/O_-(N-k+1) $ duality in general. Thus the four components of the superconformal index of dual theories are related as
\begin{align}
\left(\begin{array}{c}
\CalI^{O(1),1}_{(00)}\\
\CalI^{O(1),1}_{(10)}\\
\CalI^{O(1),1}_{(01)}\\
\CalI^{O(1),1}_{(11)}\\
\end{array}\right)
=
\frac{1}{2}\left(\begin{array}{cccc}
1&1&-1&1\\
1&1&1&-1\\
-1&1&1&1\\
1&-1&1&1\\
\end{array}\right)
\left(\begin{array}{c}
\widetilde \CalI^{O(1),1}_{(00)}\\
\widetilde \CalI^{O(1),1}_{(10)}\\
\widetilde \CalI^{O(1),1}_{(01)}\\
\widetilde \CalI^{O(1),1}_{(11)}\\
\end{array}\right)
\end{align}

Let's consider $ SO(1) $ theory A (free theory) with $ N=1 $ and its dual description $ O_+(1) $ theory B. The $ (c,c) $ ring contribution of $ SO(1) $ theory A is given by
\begin{align}
P^{A, SO(1),1}(q,y,a) = P^{O(1), 1}_{(00)} =\frac{1-a^{-1}x}{1-a}
\end{align}
The chiral ring of the $ SO(1) $ theory A is parameterized by free scalar field $ Q $ whose contribution is $ \frac{1}{1-a} $. $ -a^{-1}x $ is the contribution of a left-moving fermion $ \bar \psi_Q $.
In the $ O_+(1) $ theory B, the untwisted NSNS sector has contributions
\begin{align}
\frac{1}{2}\left(\widetilde P^{O(1), 1}_{(00)} +\widetilde P^{O(1), 1}_{(10)}\right) = \frac{1-x}{1-a^2} ~.\label{SCI:B_O1+_untiwsted}
\end{align}
with identifications $\frac{1}{1-a^2} \rightarrow M $ and $-x \rightarrow q\bar\psi_q $.
The twisted NSNS sector has contributions
\begin{align}
\frac{1}{2}\left(-\widetilde P^{O(1), 1}_{(01)} +\widetilde P^{O(1), 1}_{(11)}\right) = \frac{1-a^{-2}x}{1-a^2}\times a~.
\end{align}
where $ \frac{1}{1-a^2} \rightarrow M $, $ -a^{-2}x \rightarrow \bar \psi_M $, $ a \rightarrow \bar \psi_q $ where $ \bar \psi_q$ is a zero mode of the fermion in the twisted NSNS sector. Twisted $ \bar \psi_q$ operator in the twisted NSNS sector is anti-invariant under the standard $ \mathbf Z_2 $ but it is invariant under non-standard $ \mathbf Z_2(-1)^{F_s} $.
Operator map is
\begin{align}
Q^{2n} \leftrightarrow M^n,\quad Q^{2n-1} \leftrightarrow M^{n-1} \bar \psi_q,\quad \psi_Q \leftrightarrow \bar\psi_M \bar\psi_q
\end{align}
Note that the operator $ Q $ corresponds to the twist operator $ \bar \psi_q $ in the dual theory.

Let's consider $ (c,c) $ ring for the $ O_+(1) $ gauge theory A with $ N=1 $ and its dual, $ SO(1) $ (LG) theory B. $ SO(1) $ theory B is a theory of two chiral multiplet with a superpotential $ W = Mq^2 $ so we have classical relations
\begin{align}
Mq = q^2 =0~.
\end{align}
$ (c,c) $ ring contribution is given by
\begin{align}
\widetilde P^{O(1), 1}_{(00)} = \frac{1-a^{-2}x}{1-a^2}\times \frac{1-ax^{\frac{1}{2}}}{1-a^{-1}x^{\frac{1}{2}}} = \frac{1-x}{1-a^2}+a^{-1}x^{\frac{1}{2}},\
\end{align}
where $ \frac{1}{1-a^2} \rightarrow M $, $ \frac{1}{1-a^{-1}x^{\frac{1}{2}}} \rightarrow q$, $ -a^{-2}x \rightarrow \bar \psi_M $, $ -a x^{\frac{1}{2}} \rightarrow \bar\psi_q$ in the first equality. On the right-hand side of the second equality, the remaining elements are $ \frac{1}{1-a^2}\rightarrow M $, $ -x\rightarrow q\bar\psi_q $, $ a^{-1}x^{\frac{1}{2}}\rightarrow q $. Thus the contribution of $ Mq $ and $ q^2 $ are canceled out with $ \bar \psi_q $ and $ \bar \psi_M $ respectively.
The $ O_+(1) $ gauge theory A with $ N=1 $ is a non-standard $ \mathbf Z_2 $ orbifold. The untwisted NSNS sector has the contribution from $ M=QQ $, which is
\begin{align}
\frac{1}{2}\left(P^{O(1), 1}_{(00)} +P^{O(1), 1}_{(10)}\right) = \frac{1-x}{1-a^2} \label{SCI:A_O1+_untiwsted}
\end{align}
where $ \frac{1}{1-a^2} \rightarrow Q^2 $, $ -x \rightarrow Q\bar\psi_Q $.
The twisted NSNS sector has the contribution from the zero mode of the fermion $\bar \psi $ which is bosonic,
\begin{align}
\frac{1}{2}\left(- P^{O(1), 1}_{(01)} +P^{O(1), 1}_{(11)}\right) = a^{-1}x^{\frac{1}{2}}
\end{align}
so we have identification between the twist operator $ \bar \psi_Q $ of the $ O_+(1) $ theory A and $q$ of the $ SO(1) $ theory B.

Now let's consider the $ O_{-}(1) $ theory A with $ N=1 $, which is a standard $ \mathbf Z_2$ orbifold, and its dual description, the $ O_{-}(1) $ theory B with $ N=1 $. For $ O_-(k)/O_-(N-k+1) $ dualities, the untwisted (resp. the twisted) NSNS sector of theory A is mapped to that of theory B. Furthermore, the untwisted NSNS sector of $ O_-(k) $ theory is the same as that of $ O_+(k) $ theory because $ (-1)^{F_s} $ operator has no effect on the untwisted NSNS sector. Thus we have seen the untwisted NSNS sectors of the theories in \eqref{SCI:B_O1+_untiwsted} and \eqref{SCI:A_O1+_untiwsted}, which are the same. The twisted NSNS sector does not have $ (c,c) $ ring elements.

\subsection{$ O(2) $ gauge theories}
The superconformal index of $ O(2) $ gauge theories is computed by summing over all flat connections. We already identified seven components of the moduli space of the flat connection in the computation of the elliptic genus. Each contribution of components to the elliptic genus and the superconformal index has the same form because both of them have the same pole structure. However, relative phase factors are different, which are related to the action of $ (-1)^{F_s} $ operator, the mod 2 theta angle and zero point energy and R-charge. Let us first write down the superconformal index and explain about the phase factors.

The superconformal index of the theory A is given by
\begin{align}
\mathcal I^{A, SO(2), N}(q,y,a)&= \mathcal I^{O(2), N}_{(00)}
\\
\mathcal I^{A, O_+(2), N}(q,y,a)&= \frac{1}{2}\left(\mathcal I^{O(2), N}_{(00)} +\mathcal I^{O(2), N}_{(10)} + (-1)^{N+1}\mathcal I^{O(2), N}_{(01)} + \mathcal I^{O(2), N}_{(11)}\right)
\\
\mathcal I^{A, O_-(2), N}(q,y,a)&= \frac{1}{2}\left(\mathcal I^{O(2), N}_{(00)} +\mathcal I^{O(2), N}_{(10)} +(-1)^{N} \mathcal I^{O(2), N}_{(01)} + \mathcal I^{O(2), N}_{(11)}\right)
\end{align}
where
\begin{align}
\mathcal I^{O(2), N}_{(00)} =& \sum_{\alpha=1}^{N}\left(\prod_{\beta=1,\beta\neq\alpha}^{N}\Delta\left(a_\alpha^{-1}a_\beta\right)\right) \left(\prod_{\gamma=1}^{N}\Delta\left(a_\alpha a_\gamma\right)\right)
\\
\mathcal I^{O(2), N}_{(10)} =&\Delta\left(-q^{\frac{1}{2}}y\right) \left( \left(\prod_{\alpha=1}^{N} \Delta\left(a_\alpha\right) \Delta\left(-a_\alpha\right)\right) +(-1)^{N+1} \left(\prod_{\alpha=1}^{N} \Delta\left(q^{-\frac{1}{2}}a_\alpha\right) \Delta\left(-q^{\frac{1}{2}}a_\alpha\right)\right)\right)
\\
\mathcal I^{O(2), N}_{(01)} =&\left(y^{-\frac{1}{2}}q^{\frac{1}{4}}\right)^{\!N-1} \!\!\!\!\Delta\left(q^{-\frac{1}{2}}(q^{\frac{1}{2}}y)\right) \left(\!\! \left(\prod_{\alpha=1}^{N} \Delta\left(a_\alpha\right) \Delta\left(q^{\frac{1}{2}}a_\alpha\right)\!\!\right) \!+\!(-1)^{N+1} \left(\prod_{\alpha=1}^{N} \Delta\left(-a_\alpha\right) \Delta\left(-q^{\frac{1}{2}}a_\alpha\right)\!\!\right)\!\!\right)
\\
\mathcal I^{O(2), N}_{(11)} =&\left(y^{-\frac{1}{2}}q^{\frac{1}{4}}\right)^{\!N-1}  \!\!\!\!\Delta\left(-q^{-\frac{1}{2}}(q^{\frac{1}{2}}y)\right) \left(\!\! \left(\prod_{\alpha=1}^{N} \Delta\left(a_\alpha\right) \Delta\left(-q^{\frac{1}{2}}a_\alpha\right)\!\!\right) \!+\! (-1)^{N+1}\left(\prod_{\alpha=1}^{N} \Delta\left(-a_\alpha\right) \Delta\left(q^{\frac{1}{2}}a_\alpha\right)\!\!\right)\!\!\right)~.
\end{align}
The sign factors $(-1)^{N+1}$ and $(-1)^{N}$ in front of $\mathcal I^{O(2), N}_{(01)}$ are the choice of $(-1)^{F_s} $ action on the twisted NSNS sector. $ (-1)^{N+1} $ factors inside of $\mathcal I^{O(2), N}_{(kl)}$ come from the mod 2 theta angle and are determined by introducing a massive doublet with a complex mass and demanding that the theory is a regular theory.
In the twisted sector ($ (01),\,(11) $ contributions) we have additional factor $ \left(y^{-\frac{1}{2}}q^{\frac{1}{4}}\right)^{N-1} $ which shift zero point energy and R-charge. This factor is compatible with matching of operators between dual theories.

The superconformal index of the $ O_\pm(2) $ and $ SO(2) $ theory B can be written as
\begin{align}
\mathcal I^{B, SO(2), N}(q,y,a)&= \widetilde \CalI^{O(2), N}_{(00)}
\\
\mathcal I^{B, O_+(2), N}(q,y,a)&= \frac{1}{2}\left(\widetilde \CalI^{O(2), N}_{(00)} +\widetilde \CalI^{O(2), N}_{(10)} +(-1)^{N+1}\widetilde \CalI^{O(2), N}_{(01)} + \widetilde \CalI^{O(2), N}_{(11)}\right)
\\
\mathcal I^{B, O_-(2), N}(q,y,a)&= \frac{1}{2}\left(\widetilde \CalI^{O(2), N}_{(00)} +\widetilde \CalI^{O(2), N}_{(10)} +(-1)^{N}\widetilde \CalI^{O(2), N}_{(01)} + \widetilde \CalI^{O(2), N}_{(11)}\right)
\end{align}
where
\begin{align}
\widetilde \CalI^{O(2), N}_{(kl)}(q,y,a) = \CalI^{M, N}(q,y,a) \CalI^{O(2), N}_{(kl)}(q,y,a^{-1} (q^{\frac{1}{2}}y)^{\frac{1}{2}})
\end{align}

We have checked agreements of superconformal indices for the dualities, $ O_\pm(2) $ and $ SO(2) $ theory A with $ N=1,2,3 $. When the theory A has one flavor $ N=1 $ it is dual to a free theory. We have
\begin{align}
\CalI^{M, 1}=\CalI^{O(2), 1}_{(00)}=\CalI^{O(2), 1}_{(10)}=\CalI^{O(2), 1}_{(01)}=\CalI^{O(2), 1}_{(11)} =\frac{1}{1-a^2}\left(\prod_{i=1}^{\infty}\frac{\left(1-a^2y^{-1}q^{i-\frac{1}{2}}\right)\left(1-a^{-2}yq^{i-\frac{1}{2}}\right)}{\left(1-a^2q^{i}\right)\left(1-a^{-2}q^{i}\right)}\right)
\end{align}
Therefore the $ SO(2)  $ and $ O_-(2) $ theory A with $ N=1 $ is dual to the free theory of a meson $ M $ and the $ O_+(2) $ theory A with $ N=1 $ is dual to double copies of the free theory.
For $ N=2, 3 $ we have following relation,
\begin{align}
\left(\begin{array}{c}
\CalI^{O(2),N}_{(00)}\\
\CalI^{O(2),N}_{(10)}\\
(-1)^{N+1}\CalI^{O(2),N}_{(01)}\\
\CalI^{O(2),N}_{(11)}\\
\end{array}\right)
=
\frac{1}{2}\left(\begin{array}{cccc}
1&1&1&1\\
1&1&-1&-1\\
1&-1&1&-1\\
1&-1&-1&1\\
\end{array}\right)
\left(\begin{array}{c}
\widetilde \CalI^{O(N-1),N}_{(00)}\\
\widetilde \CalI^{O(N-1),N}_{(10)}\\
\widetilde \CalI^{O(N-1),N}_{(01)}\\
\widetilde \CalI^{O(N-1),N}_{(11)}\\
\end{array}\right)~,
\end{align}
which shows agreement of superconformal index for the dualities. We expect this relation persists for the the dualities with any $ N $.

We would like to identify contributions of chiral primary operators in the superconformal index. We obtain $ (c,c) $ ring elements from the superconformal index,
\begin{align}
P^{O(2), N}_{(00)} =& \sum_{\alpha=1}^{N}\left(\prod_{\beta=1,\beta\neq\alpha}^{N}\frac{1-a_\alpha a_\beta^{-1}x}{1-a_\alpha^{-1}a_\beta}\right) \left(\prod_{\gamma=1}^{N}\frac{1-a_\alpha^{-1} a_\gamma^{-1}x}{1-a_\alpha a_\gamma}\right)
\\
P^{O(2), N}_{(10)} =&\frac{2}{1+x} \left( \left(\prod_{\alpha=1}^{N} \frac{1-a_\alpha^{-1}x}{1-a_\alpha}\frac{1+a_\alpha^{-1}x}{1+a_\alpha}\right) -x^N\prod_{\alpha=1}^{N}a_\alpha^{-2}\right)
\\
P^{O(2), N}_{(01)} =& (-1)^{N} x^{\frac{N-1}{2}}  \left( \left(\prod_{\alpha=1}^{N} \frac{1-a_\alpha^{-1}x}{1-a_\alpha} a_\alpha^{-1}\right) - \left(\prod_{\alpha=1}^{N} \frac{1+a_\alpha^{-1}x}{1+a_\alpha} a_\alpha^{-1}\right)\right)
\\
P^{O(2), N}_{(11)} =& x^{\frac{N-1}{2}}  \left( \left(\prod_{\alpha=1}^{N} \frac{1-a_\alpha^{-1}x}{1-a_\alpha} a_\alpha^{-1}\right) - \left(\prod_{\alpha=1}^{N} \frac{1+a_\alpha^{-1}x}{1+a_\alpha} a_\alpha^{-1}\right)\right).
\end{align}
First example is $ SO(2) $ theory A and $ O_+(1) $ theory B with $ N=2 $. $ SO(2) $ theory A has $ (c,c) $ ring generators, three mesons and one baryon,
\begin{align}
M_{\alpha \beta}=(Q_\alpha Q_\beta),\quad b_{12} = [Q_1 Q_2]
\end{align}
where $ (\cdots) $ and $ [\cdots] $ are symmetric and anti-symmetric in indices respectively. The $ (c,c) $ ring generators are subject to one relation
\begin{align}
J_2: \quad b_{12}^2 = \det (M_{\alpha\beta})
\end{align}
where the notation follows \cite{Hori:2011pd}. Thus all operators $ b_{12}^n $, $ n\geq 2 $ are not independent operators. The superconformal index of the theory is given by
\begin{align}
\mathcal I^{A, SO(2), N}(q,y,a) = \frac{1+a_1a_2}{(1-a_1^2)(1-a_1a_2)(1-a_2^2)} + O(q^{\frac{1}{2}})
\end{align}
where $ a_1a_2 $ in the numerator corresponds to $ b_{12} $ and the three factors in the denominator correspond to three mesons. The superconformal index of the $ O_+(1) $ theory B can be seen from \eqref{SCI_B_O(1)+_N}. Contributions from the untwisted and the twisted NSNS sectors are
\begin{align}
\frac{1}{2}\left(\widetilde P^{O(1), 2}_{(00)} +\widetilde P^{O(1), 2}_{(10)} \right) =&\frac{1}{2}\left(\prod_{\alpha=1}^{2}\prod_{\beta=\alpha}^{2}\frac{1-a_\alpha^{-1} a_\beta^{-1}x}{1-a_\alpha a_\beta}\right)\left(\prod_{\alpha=1}^{2} \frac{1-a_\alpha x^{\frac{1}{2}}}{1-a_\alpha^{-1}x^{\frac{1}{2}}} +\prod_{\alpha=1}^{2} \frac{1+a_\alpha x^{\frac{1}{2}}}{1+a_\alpha^{-1}x^{\frac{1}{2}}}\right)\nonumber
\\
=& \frac{1}{(1-a_1^2)(1-a_1a_2)(1-a_2^2)} + O(q^{\frac{1}{2}})
\\
\frac{1}{2}\left((-1)^2\widetilde P^{O(1), 2}_{(01)} + \widetilde P^{O(1), 2}_{(11)}\right) =&
\left(\prod_{\alpha=1}^{2}\prod_{\beta=\alpha}^{2}\frac{1-a_\alpha^{-1} a_\beta^{-1}x}{1-a_\alpha a_\beta}\right)\prod_{\alpha=1}^{2}a_\alpha\nonumber
\\
=& \frac{1}{(1-a_1^2)(1-a_1a_2)(1-a_2^2)}\times a_1a_2 + O(q^{\frac{1}{2}})
\end{align}
where $ \frac{1}{(1-a_1^2)(1-a_1a_2)(1-a_2^2)} $ comes from singlets $ M_{\alpha\beta} $. The twisted sector has contribution $ a_1a_2 $ which comes from zero modes of fermions in the twisted NSNS sector, $ \bar\psi_{q_1}\bar\psi_{q_2} $ so we have $ (\bar\psi_{q_1}\bar\psi_{q_2})^2=0 $. Therefore, we confirm the identification that the baryon operator in the $ SO(2) $ gauge theory A corresponds to the twist operator in the $ O_+(1) $ theory B.

Let's consider the $ O_+(1) $ theory A with $ N=2 $ and the $ SO(2) $ theory B with $ N=2 $. In the untwisted sector of the $ O_+(1) $ theory the $ (c,c) $ ring is generated by three mesons,
\begin{align}
M_{11},~ M_{12},~ M_{22}
\end{align}
and they are subject to a relation, $ \det (M_{\alpha\beta})=0 $ and explicitly
\begin{align}
(M_{12})^2=M_{11}M_{22}~. \label{CCRelationO+1_N2}
\end{align}
Thus $ (M_{12})^2 $ operator and its higher order product are not independent operators. The superconformal index of the untwisted sector has the form of
\begin{align}
\frac{1}{2}\left( P^{O(1), 2}_{(00)} +P^{O(1), 2}_{(10)} \right) =&\frac{1}{2}\left(\prod_{\alpha=1}^{N}\frac{1-a_\alpha^{-1}x}{1-a_\alpha} +\prod_{\alpha=1}^{N}\frac{1+a_\alpha^{-1}x}{1+a_\alpha}\right)
\\
=& \frac{1+a_1a_2}{(1-a_1^2)(1-a_2^2)} + O(q^1)
\end{align}
where $ a_1a_2 \rightarrow M_{12}$ and $ \frac{1}{(1-a_\alpha^2)} \rightarrow M_{\alpha\alpha}$.
The $ O_+(1) $ theory also have $ (c,c) $ ring elements from the twisted sector.
\begin{align}
\frac{1}{2}\left( (-1)^2P^{O(1), 2}_{(01)} +P^{O(1), 2}_{(11)} \right) = a_1^{-1}a_2^{-1} x~,
\end{align}
which comes from twist operator $ \bar \psi_{Q_1} \bar \psi_{Q_2}$.
In the $ SO(2) $ theory B the relations in the $ (c,c) $ ring of theory A are not visible in the classical theory but appear in the infra-red limit of the quantum theory. \eqref{CCRelationO+1_N2} can be seen as integrating out $ (q_1q_2) $ operator \cite{Hori:2011pd}. The superconformal index of the $ SO(2) $ theory B is given by
\begin{align}
\CalI^{SO(2), 2}=\frac{1}{(1-a_1^2)(1-a_1a_2)(1-a_2^2)}\times \left(\frac{(1-a_1a_2)(1-a_1^2)}{1-\frac{a_1}{a_2}}+\frac{(1-a_1a_2)(1-a_2^2)}{1-\frac{a_2}{a_1}}\right) +O(q^{\frac{1}{2}})
\end{align}
one can see the leading order is the same as $ \frac{1+a_1a_2}{(1-a_1^2)(1-a_2^2)} $. One can also see that the baryon operator $ [q_1q_2] $ contributes $ a_1^{-1}a_2^{-1} x $ to the superconformal index and corresponds to the twist operator of the $ O_+(1) $ theory A.

\section{Analytic proof of the equality of the elliptic genus of several dual pairs}
\subsection{Jacobi Theta Functions}
 In this appendix, we prove the equality of the elliptic genus of dual pairs analytically for simple cases. We need some addition rules between the Jacobi theta functions.  We use the notation  $\theta_{i}(z)\equiv \theta_{i}(\tau|z)$ for $i=1,2,3,4$ for simplicity.
 The relation between $\theta_i(z)$ is given by
\begin{eqnarray}
\theta_{1}(z+1/2) &=& \theta_{2}(z) \label{1}\\
\theta_{1}(z+\tau/2) &=& i q^{-1/8}y^{-1/2}\theta_{4}(z) \label{2} \\
\theta_{1}(z+1/2+\tau/2) &=& q^{-1/8}y^{-1/2}\theta_{3}(z) \label{3}
\end{eqnarray}
The addition rules are as follows.
\begin{eqnarray}
\theta_1(y\pm z)\theta_2(y\mp z)\theta_3\theta_4
 &=&\theta_1(y)\theta_2(y)\theta_3(z)\theta_4(z)\pm\theta_3(y)\theta_4(y)\theta_1(z)\theta_2(z) \label{4} \\
\theta_1(y\pm z)\theta_3(y\mp z)\theta_2\theta_4
 &=&\theta_1(y)\theta_3(y)\theta_2(z)\theta_4(z)\pm \theta_2(y)\theta_4(y)\theta_1(z)\theta_3(z) \label{5} \\
\theta_1(y\pm z)\theta_4(y\mp z)\theta_2\theta_3
 &=&\theta_1(y)\theta_4(y)\theta_2(z)\theta_3(z)\pm \theta_2(y)\theta_3(y)\theta_1(z)\theta_4(z) \label{6} \\
\theta_2(y\pm z)\theta_3(y\mp z)\theta_2\theta_3
 &=&\theta_2(y)\theta_3(y)\theta_2(z)\theta_3(z)\mp \theta_1(y)\theta_4(y)\theta_1(z)\theta_4(z) \label{7} \\
\theta_2(y\pm z)\theta_4(y\mp z)\theta_2\theta_4
 &=&\theta_2(y)\theta_4(y)\theta_2(z)\theta_4(z)\mp \theta_1(y)\theta_3(y)\theta_1(z)\theta_3(z) \label{8} \\
\theta_3(y\pm z)\theta_4(y\mp z)\theta_3\theta_4
 &=&\theta_3(y)\theta_4(y)\theta_3(z)\theta_4(z)\mp \theta_1(y)\theta_2(y)\theta_1(z)\theta_2(z) \label{9}
\end{eqnarray}
Where $\theta_i \equiv \theta_i(0)$. From the eq.(\ref{4}), with $y=z$, we  get
\begin{eqnarray}
\theta_1(2z)\theta_2\theta_3\theta_4=2\theta_1(z)\theta_2(z)\theta_3(z)\theta_4(z) \label{10}.
\end{eqnarray}
Also, the periodicity of the Jacobi theta functions is given by
\begin{eqnarray}
\theta_1(z+a+b\tau)&=&(-1)^{a+b}y^{-b}q^{-\frac{b^2}{2}}\theta_1(z) \\
\theta_2(z+a+b\tau)&=&(-1)^{a}y^{-b}q^{-\frac{b^2}{2}}\theta_2(z) \\
\theta_3(z+a+b\tau)&=&y^{-b}q^{-\frac{b^2}{2}}\theta_3(z) \\
\theta_4(z+a+b\tau)&=&(-1)^{b}y^{-b}q^{-\frac{b^2}{2}}\theta_4(z),
\end{eqnarray}
where $y=e^{2\pi i z}$.
From the fact that $\theta_1(a+b\tau)=0$ for $a,b \in \mathbb{Z}$, we can deduce that
\begin{eqnarray}
\theta_2(1/2)=\theta_3(1/2+\tau/2)=\theta_4(\tau/2)=0.
\end{eqnarray}

Also, the other relations between the Jacobi theta functions are
\begin{eqnarray}
&&\theta_2(z+\frac{1}{2})=-\theta_1(z)\nonumber\\
&&\theta_2(z+\frac{\tau}{2})=q^{-\frac{1}{8}}y^{-\frac{1}{2}}\theta_3(z)\nonumber\\
&&\theta_2(z+\frac{1+\tau}{2})=-i q^{-\frac{1}{8}}y^{-\frac{1}{2}}\theta_4(z)\nonumber\\
\nonumber\\
&&\theta_3(z+\frac{1}{2})=\theta_4(z)\nonumber\\
&&\theta_3(z+\frac{\tau}{2})=q^{-\frac{1}{8}}y^{-\frac{1}{2}}\theta_2(z)\nonumber\\
&&\theta_3(z+\frac{1+\tau}{2})=i q^{-\frac{1}{8}}y^{-\frac{1}{2}}\theta_1(z)\nonumber\\
\nonumber\\
&&\theta_4(z+\frac{1}{2})=\theta_3(z)\nonumber\\
&&\theta_4(z+\frac{\tau}{2})=i q^{-\frac{1}{8}}y^{-\frac{1}{2}}\theta_1(z)\nonumber\\
&&\theta_4(z+\frac{1+\tau}{2})=q^{-\frac{1}{8}}y^{-\frac{1}{2}}\theta_2(z).
\end{eqnarray}

For small $z$ with $(a,b,i)=(0,0,1),(1,0,2),(0,1,4),(1,1,3)$, we have
\begin{eqnarray}
\theta_i(z+\frac{a+b\tau}{2})\sim z^1.
\end{eqnarray}

\subsection{Elliptic Genus of $O(1), O(2)$ theories}
We start by writing down the elliptic genus of the $O(1) ,O(2)$ gauge groups with $N$ fundamentals and the mesonic variables $Z^{M,N}$. The elliptic genus of the $SO(1), O_{\pm}(1)$ gauge theory with $N$ fundamentals is
\begin{eqnarray}
Z^{A,SO(1),N}&=&\prod_{\alpha=1}^{N}\frac{\theta_1(\xi_\alpha -z)}{\theta_1(\xi_\alpha)}\\
Z^{A,O_{+}(1),N}&=&\frac{1}{2}\Bigg( \!\!(-1)^{N}\!\!\prod_{\alpha=1}^{N}\frac{\theta_1(\xi_\alpha\!\!-z)}{\theta_1(\xi_\alpha)}+\prod_{\alpha=1}^{N}\frac{\theta_2(\xi_\alpha\!\!-z)}{\theta_2(\xi_\alpha)} +\prod_{\alpha=1}^{N}\frac{\theta_3(\xi_\alpha \!\!-z)}{\theta_3(\xi_\alpha)} +\prod_{\alpha=1}^{N}\frac{\theta_4(\xi_\alpha\!\!-z)}{\theta_4(\xi_\alpha)}\Bigg)\nonumber\\
\end{eqnarray}
\begin{eqnarray}
Z^{A,O_{-}(1),N}&=&\frac{1}{2}\Bigg( \!\!(-1)^{N+1}\!\!\prod_{\alpha=1}^{N}\frac{\theta_1(\xi_\alpha\!\!-z)}{\theta_1(\xi_\alpha)}+\prod_{\alpha=1}^{N}\frac{\theta_2(\xi_\alpha\!\!-z)}{\theta_2(\xi_\alpha)} +\prod_{\alpha=1}^{N}\frac{\theta_3(\xi_\alpha \!\!-z)}{\theta_3(\xi_\alpha)} +\prod_{\alpha=1}^{N}\frac{\theta_4(\xi_\alpha\!\!-z)}{\theta_4(\xi_\alpha)}\Bigg).\nonumber\\
\end{eqnarray}
The elliptic genus of the $SO(2), O_{\pm}(2)$ gauge theory with $N$ fundamentals is
\begin{eqnarray}
Z^{A,SO(2),N}&=&Z_{(00)}^{O(2),N}\label{18}\\
Z^{A,O_{+}(2),N}&=&\frac{1}{2}\Big(Z_{(00)}^{O(2),N}+(-1)^{N+1}(Z_{(01)}^{O(2),N}+Z_{(10)}^{O(2),N}+Z_{(11)}^{O(2),N}) \Big)\label{19}\\
Z^{A,O_{-}(2),N}&=&\frac{1}{2}\Big(Z_{(00)}^{O(2),N}+(-1)^{N}(Z_{(01)}^{O(2),N}+Z_{(10)}^{O(2),N}+Z_{(11)}^{O(2),N}) \Big)\label{20}
\end{eqnarray}
where
\begin{eqnarray}
Z_{(00)}^{O(2),N}&=&\sum_{\beta=1}^{N}\Bigg(\prod_{\alpha\neq\beta}\frac{\theta_1(\xi_{\alpha}-\xi_{\beta}-z)}{\theta_1(\xi_{\alpha}-\xi_{\beta})}\Bigg)\Bigg(\prod_{\alpha=1}\frac{\theta_1(\xi_{\alpha}+\xi_{\beta}-z)}{\theta_1(\xi_{\alpha}+\xi_{\beta})}\Bigg)\\
Z_{(01)}^{O(2),N}&=&\frac{1}{2}\frac{\theta_{4}(0)}{\theta_4(z)}\Bigg(\prod_{\alpha=1}^{N}\frac{\theta_1(\xi_{\alpha}\!\!-\!\!z)}{\theta_1(\xi_\alpha)}\frac{\theta_4(\xi_{\alpha}\!\!-\!\!z)}{\theta_4(\xi_\alpha)} +\!(-1)^{N+1}\!\prod_{\alpha=1}^{N}\frac{\theta_2(\xi_{\alpha}\!\!-\!\!z)}{\theta_2(\xi_\alpha)}\frac{\theta_3(\xi_{\alpha}\!\!-\!\!z)}{\theta_3(\xi_\alpha)}\Bigg)\\
Z_{(10)}^{O(2),N}&=&\frac{1}{2}\frac{\theta_{2}(0)}{\theta_2(z)}\Bigg(\prod_{\alpha=1}^{N}\frac{\theta_1(\xi_{\alpha}\!\!-\!\!z)}{\theta_1(\xi_\alpha)}\frac{\theta_2(\xi_{\alpha}\!\!-\!\!z)}{\theta_2(\xi_\alpha)} +\!(-1)^{N+1}\!\prod_{\alpha=1}^{N}\frac{\theta_3(\xi_{\alpha}\!\!-\!\!z)}{\theta_3(\xi_\alpha)}\frac{\theta_4(\xi_{\alpha}\!\!-\!\!z)}{\theta_4(\xi_\alpha)}\Bigg)\\
Z_{(11)}^{O(2),N}&=&\frac{1}{2}\frac{\theta_{3}(0)}{\theta_3(z)}\Bigg(\prod_{\alpha=1}^{N}\frac{\theta_1(\xi_{\alpha}\!\!-\!\!z)}{\theta_1(\xi_\alpha)}\frac{\theta_3(\xi_{\alpha}\!\!-\!\!z)}{\theta_3(\xi_\alpha)} +\!(-1)^{N+1}\!\prod_{\alpha=1}^{N}\frac{\theta_2(\xi_{\alpha}\!\!-\!\!z)}{\theta_2(\xi_\alpha)}\frac{\theta_4(\xi_{\alpha}\!\!-\!\!z)}{\theta_4(\xi_\alpha)}\Bigg).
\end{eqnarray}
Finally the elliptic genus of $N(N+1)/2$ mesonic variables is
\begin{eqnarray}
Z^{M,N}=\prod_{\alpha=1}^{N}\prod_{\beta=\alpha}^{N} \frac{\theta_1(\xi_{\alpha}+\xi_{\beta}-z)}{\theta_1(\xi_{\alpha}+\xi_{\beta})}.
\end{eqnarray}

\subsection{Dual pairs of consideration}
In this section, we enumerate dual pairs we are interested in. Before writing them, let us define some notations. We  denote the elliptic genus of the gauge group $G$ with $N$ fundamentals of the A theory as
\begin{eqnarray}
  Z^{A,G,N} \equiv Z^{A,G,N}(\tau,z,\xi)
\end{eqnarray}
and the elliptic genus of $N(N+1)/2$ mesonic variables is denoted by
\begin{eqnarray}
Z^{M,N} \equiv Z^{M,N}(\tau,z,\xi).
\end{eqnarray}
Also, we denote the elliptic genus of substitution $\xi \rightarrow -\xi+\frac{z}{2}$ multiplied by $Z^{M,N}$, which becomes the dual B theory, using tilde notation as
\begin{eqnarray}
\widetilde{Z}^{A,G,N}&\equiv& Z^{A,G,N}(\tau,z,-\xi+\frac{z}{2})\times Z^{M,N}\nonumber\\
&=&Z^{B,G,N}.
\end{eqnarray}
It is easy to see that the elliptic genus of the dual of the dual theory is that of the original theory.
\begin{eqnarray}
\widetilde{Z}^{B,G,N}&=&\widetilde{\widetilde{Z}}^{A,G,N}\nonumber\\
&=&Z^{A,G,N}(\tau,z,-(-\xi+\frac{z}{2})+\frac{z}{2})\times Z^{M,N}(\tau,z,-\xi+\frac{z}{2})\times Z^{M,N}\nonumber\\
&=&Z^{A,G,N}(\tau,z,\xi)\nonumber\\
&=&Z^{A,G,N}\label{pd}
\end{eqnarray}
where we use
\begin{eqnarray}
Z^{M,N}(\tau,z,-\xi+\frac{z}{2})&=&\prod_{\alpha=1}^{N}\prod_{\beta=\alpha}^{N} \frac{\theta_1(-\xi_{\alpha}-\xi_{\beta})}{\theta_1(-\xi_{\alpha}-\xi_{\beta}+z)}\nonumber\\
&=&\prod_{\alpha=1}^{N}\prod_{\beta=\alpha}^{N} \frac{\theta_1(\xi_{\alpha}+\xi_{\beta})}{\theta_1(\xi_{\alpha}+\xi_{\beta}-z)}\nonumber\\
&=&(Z^{M,N})^{-1}.
\end{eqnarray}
If we have a dual pair between two different gauge groups $G_1$ in the A theory and $G_2$ in the B theory, we must have the same
 elliptic genus. Using the above notations we should have
\begin{eqnarray}
Z^{A,G_1,N}&=&Z^{B,G_2,N}.\label{dn}
\end{eqnarray}
Also, if we check the eq.(\ref{dn}), the equality of the elliptic genus of $G_2$ in the A theory and $G_1$ in the B theory automatically follows since
\begin{eqnarray}
Z^{A,G_2,N}&=&\widetilde{Z}^{B,G_2,N}\nonumber\\
&=&\widetilde{Z}^{A,G_1,N}\nonumber\\
&=&Z^{B,G_1,N}
\label{dnn}\end{eqnarray}
where we use the eq.(\ref{pd}) in the first line and the eq.(\ref{dn}) in the second line.
Now, we list the possible dual pairs that we have to check analytically
\begin{enumerate}
\item $O(1),1$ $\longleftrightarrow$ $O(1),1$
\begin{eqnarray}
Z^{A,O_{+}(1),1}&=&Z^{B,SO(1),1}\\
Z^{A,O_{-}(1),1}&=&Z^{B,O_{-}(1),1}
\end{eqnarray}
\item $O(2),1$ $\longleftrightarrow$ $O(0),1$
\begin{eqnarray}
Z^{A,O_{+}(2),1}&=&2Z^{M,1}\\
Z^{A,SO(2),1}&=& Z^{M,1}\\
Z^{A,O_{-}(2),1}&=&- Z^{M,1}
\end{eqnarray}
\item $O(2),2$ $\longleftrightarrow$ $O(1),2$
\begin{eqnarray}
Z^{A,O_{+}(2),2}&=&Z^{B,SO(1),2}\\
Z^{A,SO(2),2}&=&Z^{B,O_{+}(1),2}\\
Z^{A,O_{-}(2),2}&=&Z^{B,O_{-}(1),2}.
\end{eqnarray}
Later we will also consider
\item $O(3),2$ $\longleftrightarrow$ $O(0),2$
\begin{eqnarray}
Z^{A,SO(3),2}&=&Z_{(00)}^{O(3),2}=Z^{M,2}\\
Z^{A,O_{+}(3),2}&=&2Z^{M,2}\\
Z^{A,O_{-}(3),2}&=&Z^{M,2}.
\end{eqnarray}
\item $O(4),3$ $\longleftrightarrow$ $O(0),3$
\begin{eqnarray}
Z^{A,SO(4),3}&=&Z_{(00)}^{O(4),3}=Z^{M,3}
\\
Z^{A,O_{+}(4),3}&=&2Z^{M,3}
\\
Z^{A,O_{-}(4),3}&=&-Z^{M,3}.
\end{eqnarray}
\end{enumerate}

\subsection{One Fundamental cases}
\subsubsection{$SO(1), O_{\pm}(1)$ theories with 1 fundamental}
From the elliptic genus expressions and the relations of eq.(\ref{1})-eq.(\ref{3}), we have
\begin{eqnarray}
Z^{A,SO(1),1}&=& \frac{\theta_1(\xi-z)}{\theta_1(\xi)} \label{11} \\
Z^{A,O_{+}(1),1}&=&\frac{1}{2}\Bigg(-\frac{\theta_1(\xi-z)}{\theta_1(\xi)}+\frac{\theta_2(\xi-z)}{\theta_2(\xi)}+\frac{\theta_3(\xi-z)}{\theta_3(\xi)}+\frac{\theta_4(\xi-z)}{\theta_4(\xi)} \Bigg) \label{12}\\
Z^{A,O_{-}(1),1}&=&\frac{1}{2}\Bigg(\frac{\theta_1(\xi-z)}{\theta_1(\xi)}+\frac{\theta_2(\xi-z)}{\theta_2(\xi)}+\frac{\theta_3(\xi-z)}{\theta_3(\xi)}+\frac{\theta_4(\xi-z)}{\theta_4(\xi)} \label{13}\Bigg).
\end{eqnarray}
We can simplify them as
\begin{eqnarray}
Z^{A,O_{+}(1),1}&=&\frac{\{\theta_1(\xi)\theta_2(\xi \!\!-\!\!z)\!\!-\!\!\theta_1(\xi\!\!-\!\!z)\theta_2(\xi) \}\theta_3(\xi)\theta_4(\xi)\!\!+ \!\!\theta_1(\xi)\theta_2(\xi)\{\theta_3(\xi\!\!-\!\!z)\theta_4(\xi)\!\!+\!\!\theta_3(\xi)\theta_4(\xi\!\!-\!\!z) \}}{2\theta_1(\xi)\theta_2(\xi)\theta_3(\xi)\theta_4(\xi)} {}\nonumber\\
&=&{}\frac{\theta_1(\frac{z}{2})\theta_2(\frac{z}{2})\theta_3(\xi\!\!-\!\!\frac{z}{2})\theta_4(\xi\!\!-\!\!\frac{z}{2})\theta_3(\xi)\theta_4(\xi)\!\!+\!\!\theta_3(\frac{z}{2})\theta_4(\frac{z}{2})\theta_3(\xi\!\!-\!\!\frac{z}{2})\theta_4(\xi\!\!-\!\!\frac{z}{2})\theta_1(\xi)\theta_2(\xi)}{\theta_1(\xi)\theta_2(\xi)\theta_3(\xi)\theta_4(\xi)\theta_3(0)\theta_4(0)} {}\nonumber\\
&=&{}\frac{\theta_3(\xi\!\!-\!\!\frac{z}{2})\theta_4(\xi\!\!-\!\!\frac{z}{2})}{\theta_1(\xi)\theta_2(\xi)\theta_3(\xi)\theta_4(\xi)\theta_3(0)\theta_4(0)}\Big(\theta_1(\frac{z}{2})\theta_2(\frac{z}{2})\theta_3(\xi)\theta_4(\xi)\!+\!\theta_1(\xi)\theta_2(\xi)\theta_3(\frac{z}{2})\theta_4(\frac{z}{2}) \Big){}\nonumber\\
&=&{}\frac{\theta_1(\xi+\frac{z}{2})\theta_2(\xi-\frac{z}{2})\theta_3(\xi-\frac{z}{2})\theta_4(\xi-\frac{z}{2})}{\theta_1(\xi)\theta_2(\xi)\theta_3(\xi)\theta_4(\xi)}{}\nonumber\\
&=&{}\frac{\theta_1(\xi+\frac{z}{2})}{\theta_1(\xi-\frac{z}{2})}\frac{\theta_1(\xi-\frac{z}{2})\theta_2(\xi-\frac{z}{2})\theta_3(\xi-\frac{z}{2})\theta_4(\xi-\frac{z}{2})}{\theta_1(\xi)\theta_2(\xi)\theta_3(\xi)\theta_4(\xi)}{}\nonumber\\
&=&{}\frac{\theta_1(\xi+\frac{z}{2})}{\theta_1(\xi-\frac{z}{2})}\frac{\theta_1(2\xi-z)}{\theta_1(2\xi)}\label{14}.
\end{eqnarray}
At the first line, we rewrite eq.(\ref{12}) to have a common denominator. At the second and third line, we use the addition rules of eq.(\ref{4}), eq.(\ref{9}). At the last line, we use eq.(\ref{10}). The duality between $Z^{A,SO(1),1}$ can be checked by changing the holonomy at $Z^{A,O_{+}(1),1}$ as $\xi\rightarrow -\xi+\frac{z}{2}$ and multiplying $Z^{M,1}=\frac{\theta_1(2\xi-z)}{\theta_1(2\xi)}$.
\begin{eqnarray}
Z^{B,O_{+}(1),1}&=&Z^{A,O_{+}(1),1}(\xi\rightarrow -\xi+\frac{z}{2})\times Z^{M,1}\nonumber\\
&=&\frac{\theta_1(-\xi+z)}{\theta_1(-\xi)}\frac{\theta_1(-2\xi)}{\theta_1(-2\xi+z)}\times \frac{\theta_1(2\xi-z)}{\theta_1(2\xi)}{}\nonumber\\
&=&{}\frac{\theta_1(\xi-z)}{\theta_1(\xi)}\frac{\theta_1(2\xi)}{\theta_1(2\xi-z)}\frac{\theta_1(2\xi-z)}{\theta_1(2\xi)}{}\nonumber\\
&=&{}\frac{\theta_1(\xi-z)}{\theta_1(\xi)}=Z^{A,SO(1),1}.
\end{eqnarray}
For $Z^{O_{-}(1),1}$, we have
\begin{eqnarray}
Z^{A,O_{-}(1),1}&=&\frac{\{\theta_1(\xi)\theta_2(\xi \!\!-\!\!z)\!\!+\!\!\theta_1(\xi\!\!-\!\!z)\theta_2(\xi) \}\theta_3(\xi)\theta_4(\xi)\!\!+ \!\!\theta_1(\xi)\theta_2(\xi)\{\theta_3(\xi\!\!-\!\!z)\theta_4(\xi)\!\!+\!\!\theta_3(\xi)\theta_4(\xi\!\!-\!\!z) \}}{2\theta_1(\xi)\theta_2(\xi)\theta_3(\xi)\theta_4(\xi)} {}\nonumber\\
&=&{}\frac{\theta_3(\frac{z}{2})\theta_4(\frac{z}{2})}{\theta_1(\xi)\theta_2(\xi)\theta_3(\xi)\theta_4(\xi)\theta_3(0)\theta_4(0)}\Big(\theta_1(\xi\!\!-\!\!\frac{z}{2})\theta_2(\xi\!\!-\!\!\frac{z}{2})\theta_3(\xi)\theta_4(\xi)\!+\!\theta_1(\xi)\theta_2(\xi)\theta_3(\xi\!\!-\!\!\frac{z}{2})\theta_4(\xi\!\!-\!\!\frac{z}{2}) \Big){}\nonumber\\
&=&{}\frac{\theta_1(2\xi-\frac{z}{2})\theta_2(\frac{z}{2})\theta_3(\frac{z}{2})\theta_4(\frac{z}{2})}{\theta_1(\xi)\theta_2(\xi)\theta_3(\xi)\theta_4(\xi)}{}\nonumber\\
&=&{}\frac{\theta_1(2\xi-\frac{z}{2})}{\theta_1(\frac{z}{2})}\frac{\theta_1(\frac{z}{2})\theta_2(\frac{z}{2})\theta_3(\frac{z}{2})\theta_4(\frac{z}{2})}{\theta_1(\xi)\theta_2(\xi)\theta_3(\xi)\theta_4(\xi)}{}\nonumber\\
&=&{}\frac{\theta_1(2\xi-\frac{z}{2})}{\theta_1(2\xi)}\frac{\theta_1(z)}{\theta_1(\frac{z}{2})}. \label{16}
\end{eqnarray}
At the first line, we rewrite eq.(\ref{12}) to have a common denominator. At the second line, we use the addition rules of eq.(\ref{4}), eq.(\ref{9}). At the last line, we use eq.(\ref{10}). From this we can obtain
\begin{eqnarray}
Z^{B,O_{-}(1),1}&=&
Z^{A,O_{-}(1),1}(\xi\rightarrow -\xi+\frac{z}{2})\times Z^{M,1}\nonumber\\
&=&\frac{\theta_1(-2\xi+\frac{z}{2})}{\theta_1(-2\xi+z)}\frac{\theta_1(z)}{\theta_1(\frac{z}{2})}\times \frac{\theta_1(2\xi-z)}{\theta_1(2\xi)}{}\nonumber\\
&=&{}\frac{\theta_1(2\xi-\frac{z}{2})}{\theta_1(2\xi-z)}\frac{\theta_1(z)}{\theta_1(\frac{z}{2})}\frac{\theta_1(2\xi-z)}{\theta_1(2\xi)}{}\nonumber\\
&=&{}\frac{\theta_1(2\xi-\frac{z}{2})}{\theta_1(2\xi)}\frac{\theta_1(z)}{\theta_1(\frac{z}{2})}=Z^{A,O_{-}(1),1}.
\end{eqnarray}

We also check the following identities appearing in the main text
\begin{eqnarray}
 Z_{(00)}^{O(1),1}+Z_{(10)}^{O(1),1}&=&\widetilde{Z}_{(01)}^{O(1),1}+\widetilde{Z}_{(11)}^{O(1),1}\label{pp1}\\
  Z_{(01)}^{O(1),1}+Z_{(11)}^{O(1),1}&=&\widetilde{Z}_{(00)}^{O(1),1}+\widetilde{Z}_{(10)}^{O(1),1}\label{pp2}\\
   Z_{(01)}^{O(1),1}-Z_{(11)}^{O(1),1}&=&\widetilde{Z}_{(11)}^{O(1),1}-\widetilde{Z}_{(01)}^{O(1),1}\label{pp3}.
\end{eqnarray}
Actually,  eq.(\ref{pp1}) and  eq.(\ref{pp2}) are equivalent since eq.(\ref{pp2}) is obtained from   eq.(\ref{pp1}) by
taking the duality transformation, which adds the tilde in both sides and $\widetilde{\widetilde{Z}}=Z$ for involved expressions. So, we only need to check eq.(\ref{pp1}) and eq.(\ref{pp3}).
\begin{eqnarray}
Z_{(00)}^{O(1),1}+Z_{(10)}^{O(1),1}&=&\frac{\theta_1(\xi-z)}{\theta_1(\xi)}+\frac{\theta_2(\xi-z)}{\theta_2(\xi)}\nonumber\\
&=&\frac{\theta_1(\xi\!\!-\!\!z)\theta_2(\xi)\theta_3(\xi)\theta_4(\xi)+\theta_1(\xi)\theta_2(\xi\!\!-\!\!z)\theta_3(\xi)\theta_4(\xi)}{\theta_1(\xi)\theta_2(\xi)\theta_3(\xi)\theta_4(\xi)}\nonumber\\
&=&\frac{\{\theta_1(\xi\!\!-\!\!z)\theta_2(\xi)+\theta_1(\xi)\theta_2(\xi\!\!-\!\!z)\}\theta_3\theta_4\frac{\theta_3(\xi)\theta_4(\xi)}{\theta_3\theta_4}}{\theta_1(\xi)\theta_2(\xi)\theta_3(\xi)\theta_4(\xi)}\nonumber\\
&=&2\frac{\theta_1(\xi\!\!-\!\!\frac{z}{2})\theta_2(\xi\!\!-\!\!\frac{z}{2})\theta_3(\xi)\theta_4(\xi)\theta_3(\frac{z}{2})\theta_4(\frac{z}{2})}{\theta_1(\xi)\theta_2(\xi)\theta_3(\xi)\theta_4(\xi)\theta_3\theta_4}\nonumber\\
&=&\frac{\theta_1(\xi\!\!-\!\!\frac{z}{2})\theta_2(\xi\!\!-\!\!\frac{z}{2})\{\theta_3(\xi\!\!+\!\!\frac{z}{2})\theta_4(\xi\!\!-\!\!\frac{z}{2})+\theta_3(\xi\!\!-\!\!\frac{z}{2})\theta_4(\xi\!\!+\!\!\frac{z}{2})\}\theta_3\theta_4}{\theta_1(\xi)\theta_2(\xi)\theta_3(\xi)\theta_4(\xi)\theta_3\theta_4}\nonumber\\
&=&\frac{\theta_1(\xi\!\!-\!\!\frac{z}{2})\theta_2(\xi\!\!-\!\!\frac{z}{2})\theta_3(\xi\!\!-\!\!\frac{z}{2})\theta_4(\xi\!\!-\!\!\frac{z}{2})}{\theta_1(\xi)\theta_2(\xi)\theta_3(\xi)\theta_4(\xi)}\Bigg(\frac{\theta_3(\xi\!\!+\!\!\frac{z}{2})}{\theta_3(\xi\!\!-\!\!\frac{z}{2})}+\frac{\theta_4(\xi\!\!+\!\!\frac{z}{2})}{\theta_4(\xi\!\!-\!\!\frac{z}{2})}\Bigg)\nonumber\\
&=&\frac{\theta_1(2\xi-z)}{\theta_1(2\xi)}\Bigg(\frac{\theta_3(\xi\!\!+\!\!\frac{z}{2})}{\theta_3(\xi\!\!-\!\!\frac{z}{2})}+\frac{\theta_4(\xi\!\!+\!\!\frac{z}{2})}{\theta_4(\xi\!\!-\!\!\frac{z}{2})}\Bigg)\nonumber\\
&=&\widetilde{Z}_{(01)}^{O(1),1}+\widetilde{Z}_{(11)}^{O(1),1}.
\end{eqnarray}
Similarly,
\begin{eqnarray}
Z_{(01)}^{O(1),1}-Z_{(11)}^{O(1),1}&=&\frac{\theta_4(\xi-z)}{\theta_4(\xi)}-\frac{\theta_3(\xi-z)}{\theta_3(\xi)}\nonumber\\
&=&\frac{\theta_3(\xi)\theta_4(\xi-z)-\theta_3(\xi-z)\theta_4(\xi)}{\theta_3(\xi)\theta_4(\xi)}\nonumber\\
&=&\frac{-2}{\theta_3\theta_4}\frac{\theta_1(\xi)\theta_2(\xi)\theta_1(\frac{z}{2})\theta_2(\frac{z}{2})}{\theta_1(\xi)\theta_2(\xi)\theta_3(\xi)\theta_4(\xi)}\theta_1(\xi\!\!-\!\!\frac{z}{2})\theta_2(\xi\!\!-\!\!\frac{z}{2})\nonumber\\
&=&\frac{-2}{\theta_3\theta_4}\frac{\theta_1(\xi\!\!-\!\!\frac{z}{2})\theta_2(\xi\!\!-\!\!\frac{z}{2})}{\theta_1(\xi)\theta_2(\xi)\theta_3(\xi)\theta_4(\xi)}\frac{\theta_3\theta_4}{2}\Big(\theta_3(\xi\!\!-\!\!\frac{z}{2})\theta_4(\xi\!\!+\!\!\frac{z}{2})\!\!-\!\!\theta_3(\xi\!\!+\!\!\frac{z}{2})\theta_4(\xi\!\!-\!\!\frac{z}{2}) \Big)\nonumber\\
&=&\frac{\theta_1(\xi\!\!-\!\!\frac{z}{2})\theta_2(\xi\!\!-\!\!\frac{z}{2})\theta_3(\xi\!\!-\!\!\frac{z}{2})\theta_4(\xi\!\!-\!\!\frac{z}{2})}{\theta_1(\xi)\theta_2(\xi)\theta_3(\xi)\theta_4(\xi)}\Bigg(\frac{\theta_3(\xi\!\!+\!\!\frac{z}{2})}{\theta_3(\xi\!\!-\!\!\frac{z}{2})}-\frac{\theta_4(\xi\!\!+\!\!\frac{z}{2})}{\theta_4(\xi\!\!-\!\!\frac{z}{2})} \Bigg)\nonumber\\
&=&\frac{\theta_1(2\xi-z)}{\theta_1(2\xi)}\Bigg(\frac{\theta_3(\xi\!\!+\!\!\frac{z}{2})}{\theta_3(\xi\!\!-\!\!\frac{z}{2})}-\frac{\theta_4(\xi\!\!+\!\!\frac{z}{2})}{\theta_4(\xi\!\!-\!\!\frac{z}{2})} \Bigg)\nonumber\\
&=&\widetilde{Z}_{(11)}^{O(1),1}-\widetilde{Z}_{(01)}^{O(1),1}.
\end{eqnarray}

\subsubsection{$SO(2), O_{\pm}(2)$ theories with 1 fundamental}
Let us show the case of one fundamental with $SO(2), O_{\pm}(2)$ gauge theories. The elliptic genus expressions of the  $SO(2), O_{\pm}(2)$ theories with 1 fundamental are given by
\begin{eqnarray}
Z_{(10)}^{O(2),1}&=&\frac{1}{2}\frac{\theta_2(0)}{\theta_2(z)}\Bigg(\frac{\theta_1(\xi\!\!-\!\!z)\theta_2(\xi\!\!-\!\!z)}{\theta_1(\xi)\theta_2(\xi)}+\frac{\theta_3(\xi\!\!-\!\!z)\theta_4(\xi\!\!-\!\!z)}{\theta_3(\xi)\theta_4(\xi)}\Bigg) ,\\
Z_{(01)}^{O(2),1}&=&\frac{1}{2}\frac{\theta_4(0)}{\theta_4(z)}\Bigg(\frac{\theta_1(\xi\!\!-\!\!z)\theta_4(\xi\!\!-\!\!z)}{\theta_1(\xi)\theta_4(\xi)}+\frac{\theta_2(\xi\!\!-\!\!z)\theta_3(\xi\!\!-\!\!z)}{\theta_2(\xi)\theta_3(\xi)}\Bigg) ,\\
Z_{(11)}^{O(2),1}&=&\frac{1}{2}\frac{\theta_3(0)}{\theta_3(z)}\Bigg(\frac{\theta_1(\xi\!\!-\!\!z)\theta_3(\xi\!\!-\!\!z)}
{\theta_1(\xi)\theta_3(\xi)}+\frac{\theta_2(\xi\!\!-\!\!z)\theta_4(\xi\!\!-\!\!z)}{\theta_2(\xi)\theta_4(\xi)}\Bigg).
\end{eqnarray}
Using the theta functions addition rules, we can get
\begin{eqnarray}
Z_{(00)}^{O(2),1}&=&\frac{\theta_1(2\xi-z)}{\theta_1(2\xi)}\\
Z_{(10)}^{O(2),1}&=&\frac{1}{2}\frac{\theta_2(0)}{\theta_2(z)}\Bigg(\frac{\theta_1(\xi\!\!-\!\!z)\theta_2(\xi\!\!-\!\!z)\theta_3(\xi)\theta_4(\xi)+\theta_1(\xi)\theta_2(\xi)\theta_3(\xi\!\!-\!\!z)\theta_4(\xi\!\!-\!\!z)}{\theta_1(\xi)\theta_2(\xi)\theta_3(\xi)\theta_4(\xi)}\Bigg){}\nonumber\\
&=&{}\frac{1}{2}\frac{\theta_2(0)}{\theta_2(z)}\frac{\theta_1(2\xi-z)\theta_2(-z)\theta_3(0)\theta_4(0)}{\theta_1(\xi)\theta_2(\xi)\theta_3(\xi)\theta_4(\xi)}{}\nonumber\\
&=&{}\frac{\theta_1(2\xi-z)}{\theta_1(2\xi)}\\
Z_{(01)}^{O(2),1}&=&\frac{1}{2}\frac{\theta_4(0)}{\theta_4(z)}\Bigg(\frac{\theta_1(\xi\!\!-\!\!z)\theta_4(\xi\!\!-\!\!z)\theta_2(\xi)\theta_3(\xi)+\theta_1(\xi)\theta_4(\xi)\theta_2(\xi\!\!-\!\!z)\theta_3(\xi\!\!-\!\!z)}{\theta_1(\xi)\theta_2(\xi)\theta_3(\xi)\theta_4(\xi)}\Bigg){}\nonumber\\
&=&{}\frac{1}{2}\frac{\theta_4(0)}{\theta_4(z)}\frac{\theta_1(2\xi-z)\theta_4(-z)\theta_2(0)\theta_3(0)}{\theta_1(\xi)\theta_2(\xi)\theta_3(\xi)\theta_4(\xi)}{}\nonumber\\
&=&{}\frac{\theta_1(2\xi-z)}{\theta_1(2\xi)}\\
Z_{(11)}^{O(2),1}&=&\frac{1}{2}\frac{\theta_3(0)}{\theta_3(z)}\Bigg(\frac{\theta_1(\xi\!\!-\!\!z)\theta_3(\xi\!\!-\!\!z)\theta_2(\xi)\theta_4(\xi)+\theta_1(\xi)\theta_3(\xi)\theta_2(\xi\!\!-\!\!z)\theta_4(\xi\!\!-\!\!z)}{\theta_1(\xi)\theta_2(\xi)\theta_3(\xi)\theta_4(\xi)}\Bigg){}\nonumber\\
&=&{}\frac{1}{2}\frac{\theta_3(0)}{\theta_3(z)}\frac{\theta_1(2\xi-z)\theta_3(-z)\theta_2(0)\theta_4(0)}{\theta_1(\xi)\theta_2(\xi)\theta_3(\xi)\theta_4(\xi)}{}\nonumber\\
&=&{}\frac{\theta_1(2\xi-z)}{\theta_1(2\xi)}.
\end{eqnarray}
We use eq.(\ref{10}) at each of the last line. Hence, we have,
\begin{eqnarray}
Z_{(00)}^{O(2),1}=Z_{(01)}^{O(2),1}=Z_{(10)}^{O(2),1}=Z_{(11)}^{O(2),1}=\frac{\theta_1(2\xi-z)}{\theta_1(2\xi)}=Z^{M,1}
\end{eqnarray}
and with eq.(\ref{18})-eq.(\ref{20}) we get
\begin{eqnarray}
Z^{A,SO(2),1}&=&Z^{M,1}\\
Z^{A,O_{+}(2),1}&=&2Z^{M,1}\\
Z^{A,O_{-}(2),1}&=&-Z^{M,1}
\end{eqnarray}
as desired.

\subsection{$SO(2), O_{\pm}(2)$ theories with 2 fundamentals}
\subsubsection{Duality between $O_{+}(2),2$ $\leftrightarrow$ $SO(1),2$}\label{o2}
We consider the duality between the $O_{+}(2)$ gauge theory with 2 fundamentals and the SO(1) gauge theory with 2 fundamentals.
We should have the equality of the elliptic genus
\begin{eqnarray}
Z^{A,SO(1),2}=Z^{B,O_{+}(2),2}.  \label{eqso1}
\end{eqnarray}
Once we prove this  we can automatically have
\begin{eqnarray}
Z^{A,O_{+}(2),2}=Z^{B,SO(1),2}.
\end{eqnarray}
It turns out that it's more convenient to prove the  eq. (\ref{eqso1}). The elliptic genus is given by
\begin{eqnarray}
Z^{A,SO(1),2}=\prod_{\alpha=1}^{2}\frac{\theta_1(\xi_{\alpha}-z)}{\theta_1(\xi_{\alpha})}
\end{eqnarray}
and
\begin{eqnarray}
Z^{A,O_{+}(2),2}=\frac{1}{2}\Big(Z_{(00)}^{O(2),2}-(Z_{(01)}^{O(2),2}+Z_{(10)}^{O(2),2}+Z_{(11)}^{O(2),2}) \Big)
\end{eqnarray}
where
\begin{eqnarray}
Z_{(00)}^{O(2),2}&=&\Bigg(\frac{\theta_1(\xi_2\!\!-\!\!\xi_1\!\!-\!\!z)}{\theta_1(\xi_2\!\!-\!\!\xi_1)}\frac{\theta_1(2\xi_1\!\!-\!\!z)}{\theta_1(2\xi_1)}\!\!+\!\!\frac{\theta_1(\xi_1\!\!-\!\!\xi_2\!\!-\!\!z)}{\theta_1(\xi_1\!\!-\!\!\xi_2)}\frac{\theta_1(2\xi_2\!\!-\!\!z)}{\theta_1(2\xi_2)}\Bigg)\frac{\theta_1(\xi_1\!\!+\!\!\xi_2\!\!-\!\!z)}{\theta_1(\xi_1\!\!+\!\!\xi_2)}\\
Z_{(01)}^{O(2),2}&=&\frac{1}{2}\frac{\theta_{4}(0)}{\theta_4(z)}\Bigg(\prod_{\alpha=1}^{2}\frac{\theta_1(\xi_{\alpha}\!\!-\!\!z)}{\theta_1(\xi_\alpha)}\frac{\theta_4(\xi_{\alpha}\!\!-\!\!z)}{\theta_4(\xi_\alpha)} -\prod_{\alpha=1}^{2}\frac{\theta_2(\xi_{\alpha}\!\!-\!\!z)}{\theta_2(\xi_\alpha)}\frac{\theta_3(\xi_{\alpha}\!\!-\!\!z)}{\theta_3(\xi_\alpha)}\Bigg)\\
Z_{(10)}^{O(2),2}&=&\frac{1}{2}\frac{\theta_{2}(0)}{\theta_2(z)}\Bigg(\prod_{\alpha=1}^{2}\frac{\theta_1(\xi_{\alpha}\!\!-\!\!z)}{\theta_1(\xi_\alpha)}\frac{\theta_2(\xi_{\alpha}\!\!-\!\!z)}{\theta_2(\xi_\alpha)} -\prod_{\alpha=1}^{2}\frac{\theta_3(\xi_{\alpha}\!\!-\!\!z)}{\theta_3(\xi_\alpha)}\frac{\theta_4(\xi_{\alpha}\!\!-\!\!z)}{\theta_4(\xi_\alpha)}\Bigg)\\
Z_{(11)}^{O(2),2}&=&\frac{1}{2}\frac{\theta_{3}(0)}{\theta_3(z)}\Bigg(\prod_{\alpha=1}^{2}\frac{\theta_1(\xi_{\alpha}\!\!-\!\!z)}{\theta_1(\xi_\alpha)}\frac{\theta_3(\xi_{\alpha}\!\!-\!\!z)}{\theta_3(\xi_\alpha)} -\prod_{\alpha=1}^{2}\frac{\theta_2(\xi_{\alpha}\!\!-\!\!z)}{\theta_2(\xi_\alpha)}\frac{\theta_4(\xi_{\alpha}\!\!-\!\!z)}{\theta_4(\xi_\alpha)}\Bigg).
\end{eqnarray}
We first simplify $Z^{A,O_{+}(2),2}$  using the Jacobi theta function addition rules.
 Consider first $Z_{(01)}^{O(2),2}$. The $\xi_1$ part of the first term can be written as
\begin{eqnarray}
\frac{\theta_1(\xi_1\!\!-\!\!z)}{\theta_1(\xi_1)}\frac{\theta_4(\xi_1\!\!-\!\!z)}{\theta_4(\xi_1)}&=&\frac{\theta_1(\xi_1\!\!-\!\!z)\theta_4(\xi_1\!\!-\!\!z)\theta_2(\xi_1)\theta_3(\xi_1)}{\theta_1(\xi_1)\theta_4(\xi_1)\theta_2(\xi_1)\theta_3(\xi_1)}\nonumber\\
&=&\frac{\frac{\theta_2\theta_3}{2}(\theta_1(2\xi_1\!\!-\!\!z)\theta_4(z)-\theta_1(z)\theta_4(2\xi_1\!\!-\!\!z))}{\frac{1}{2}\theta_1(2\xi_1)\theta_2\theta_3\theta_4}\nonumber\\
&=&\frac{\theta_1(2\xi_1\!\!-\!\!z)\theta_4(z)}{\theta_1(2\xi_1)\theta_4}-\frac{\theta_1(z)\theta_4(2\xi_1\!\!-\!\!z)}{\theta_1(2\xi_1)\theta_4}
\end{eqnarray}
and the $\xi_1$ part of the second term can be written as
\begin{eqnarray}
\frac{\theta_2(\xi_1\!\!-\!\!z)}{\theta_2(\xi_1)}\frac{\theta_3(\xi_1\!\!-\!\!z)}{\theta_3(\xi_1)}&=&\frac{\theta_2(\xi_1\!\!-\!\!z)\theta_3(\xi_1\!\!-\!\!z)\theta_1(\xi_1)\theta_4(\xi_1)}{\theta_2(\xi_1)\theta_3(\xi_1)\theta_1(\xi_1)\theta_4(\xi_1)}\nonumber\\
&=&\frac{\theta_1(2\xi_1\!\!-\!\!z)\theta_4(z)}{\theta_1(2\xi_1)\theta_4}+\frac{\theta_1(z)\theta_4(2\xi_1\!\!-\!\!z)}{\theta_1(2\xi_1)\theta_4}.
\end{eqnarray}
Combining these results, we get
\begin{eqnarray}
Z_{(01)}^{O(2),2}=\frac{1}{2}\frac{\theta_1(2\xi_1\!\!-\!\!z)}{\theta_1(2\xi_1)}\Big(\Delta_1\Delta_4-\Delta_2\Delta_3\Big)
\!\!-\!\!\frac{1}{2}\frac{\theta_1(z)\theta_4(2\xi_1\!\!-\!\!z)}{\theta_1(2\xi_1)\theta_4(z)}\Big(\Delta_1\Delta_4+\Delta_2\Delta_3\Big),
\end{eqnarray}
where $\Delta_i\equiv \frac{\theta_i(\xi_2-z)}{\theta_i(\xi_2)}$.
Also we can evaluate
\begin{eqnarray}
\Delta_1\Delta_4\!\!+\!\!\Delta_2\Delta_3&=&\frac{\theta_1(\xi_2\!\!-\!\!z)\theta_4(\xi_2\!\!-\!\!z)\theta_2(\xi_2)\theta_3(\xi_2)\!\!+\!\!\theta_2(\xi_2\!\!-\!\!z)\theta_3(\xi_2\!\!-\!\!z)\theta_1(\xi_2)\theta_4(\xi_2)}{\theta_1(\xi_2)\theta_2(\xi_2)\theta_3(\xi_2)\theta_4(\xi_2)}\nonumber\\
&=&\frac{\theta_2\theta_3\theta_1(\xi_2-z+\xi_2)\theta_4(\xi_2-z-\xi_2)}{\frac{1}{2}\theta_1(2\xi_2)\theta_2\theta_3\theta_4}\nonumber\\
&=&2\frac{\theta_1(2\xi_2-z)\theta_4(z)}{\theta_1(2\xi_2)\theta_4}
\end{eqnarray}
and similarly
\begin{eqnarray}
\Delta_1\Delta_4\!\!-\!\!\Delta_2\Delta_3=-2\frac{\theta_1(z)\theta_4(2\xi_2-z)}{\theta_1(2\xi_2)\theta_4}
\end{eqnarray}
so that
\begin{eqnarray}
Z_{(01)}^{O(2),2}&=&-\frac{\theta_1(2\xi_1\!\!-\!\!z)}{\theta_1(2\xi_1)}\frac{\theta_1(z)\theta_4(2\xi_2\!\!-\!\!z)}{\theta_1(2\xi_2)\theta_4}-\frac{\theta_1(z)\theta_4(2\xi_1\!\!-\!\!z)}{\theta_1(2\xi_1)\theta_4(z)}\frac{\theta_1(2\xi_2\!\!-\!\!z)\theta_4(z)}{\theta_1(2\xi_2)\theta_4}\nonumber\\
&=&-\frac{\theta_1(z)}{\theta_1(2\xi_1)\theta_2(2\xi_2)\theta_4}\Big(\theta_1(2\xi_1\!\!-\!\!z)\theta_4(2\xi_2\!\!-\!\!z)+\theta_1(2\xi_2\!\!-\!\!z)\theta_4(2\xi_1\!\!-\!\!z)\Big)\nonumber\\
&=&-2\frac{\theta_1(z)\theta_1(\xi_1\!\!+\!\!\xi_2\!\!-\!\!z)\theta_4(\xi_1\!\!+\!\!\xi_2\!\!-\!\!z)\theta_2(\xi_1\!\!-\!\!\xi_2)\theta_3(\xi_1\!\!-\!\!\xi_2)}{\theta_1(2\xi_1)\theta_1(2\xi_2)\theta_2\theta_3\theta_4}.\label{Z01}
\end{eqnarray}
We can follow similar procedures for the $Z_{(10)}^{O(2),2}$ and $Z_{(11)}^{O(2),2}$. The results are given by
\begin{eqnarray}
Z_{(10)}^{O(2),2}=-2\frac{\theta_1(z)\theta_1(\xi_1\!\!+\!\!\xi_2\!\!-\!\!z)\theta_2(\xi_1\!\!+\!\!\xi_2\!\!-\!\!z)\theta_3(\xi_1\!\!-\!\!\xi_2)\theta_4(\xi_1\!\!-\!\!\xi_2)}{\theta_1(2\xi_1)\theta_1(2\xi_2)\theta_2\theta_3\theta_4}.\label{Z10}
\end{eqnarray}
\begin{eqnarray}
Z_{(11)}^{O(2),2}=-2\frac{\theta_1(z)\theta_1(\xi_1\!\!+\!\!\xi_2\!\!-\!\!z)\theta_3(\xi_1\!\!+\!\!\xi_2\!\!-\!\!z)\theta_2(\xi_1\!\!-\!\!\xi_2)\theta_4(\xi_1\!\!-\!\!\xi_2)}{\theta_1(2\xi_1)\theta_1(2\xi_2)\theta_2\theta_3\theta_4}.\label{Z11}
\end{eqnarray}
Now consider $Z_{(01)}^{O(2),2}+Z_{(10)}^{O(2),2}+Z_{(11)}^{O(2),2}\equiv Z^{O(2),2}$.
\begin{eqnarray}
Z^{O(2),2}&=&-\frac{\theta_1(\xi_1\!\!+\!\!\xi_2\!\!-\!\!z)\theta_1(z)}{\theta_1(2\xi_1)\theta_1(2\xi_2)\theta_2\theta_3\theta_4}\Big(2\theta_2(\xi_1\!\!+\!\!\xi_2\!\!-\!\!z)\theta_3(\xi_1\!\!-\!\!\xi_2)\theta_4(\xi_1\!\!-\!\!\xi_2)\nonumber\\
&&+2\theta_3(\xi_1\!\!+\!\!\xi_2\!\!-\!\!z)\theta_2(\xi_1\!\!-\!\!\xi_2)\theta_4(\xi_1\!\!-\!\!\xi_2)+2\theta_4(\xi_1\!\!+\!\!\xi_2\!\!-\!\!z)\theta_2(\xi_1\!\!-\!\!\xi_2)\theta_3(\xi_1\!\!-\!\!\xi_2) \Big)\nonumber\\
&=&-\frac{\theta_1(\xi_1\!\!+\!\!\xi_2\!\!-\!\!z)\theta_1(z)}{\theta_1(2\xi_1)\theta_1(2\xi_2)\theta_2\theta_3\theta_4}\Bigg(\theta_2(\xi_1\!\!-\!\!\xi_2)\Big(\theta_4(\xi_1\!\!+\!\!\xi_2\!\!-\!\!z)\theta_3(\xi_1\!\!-\!\!\xi_2)+\theta_3(\xi_1\!\!+\!\!\xi_2\!\!-\!\!z)\theta_4(\xi_1\!\!-\!\!\xi_2)\Big)\nonumber\\
&&+\theta_3(\xi_1\!\!-\!\!\xi_2)\Big(\theta_4(\xi_1\!\!+\!\!\xi_2\!\!-\!\!z)\theta_2(\xi_1\!\!-\!\!\xi_2)+\theta_2(\xi_1\!\!+\!\!\xi_2\!\!-\!\!z)\theta_4(\xi_1\!\!-\!\!\xi_2)\Big)\nonumber\\
&&+\theta_4(\xi_1\!\!-\!\!\xi_2)\Big(\theta_2(\xi_1\!\!+\!\!\xi_2\!\!-\!\!z)\theta_3(\xi_1\!\!-\!\!\xi_2)+\theta_3(\xi_1\!\!+\!\!\xi_2\!\!-\!\!z)\theta_2(\xi_1\!\!-\!\!\xi_2)\Big)\Bigg)\nonumber\\
&=&-\frac{2\theta_1(\xi_1\!\!+\!\!\xi_2\!\!-\!\!z)\theta_1(z)}{\theta_1(2\xi_1)\theta_1(2\xi_2)\theta_2\theta_3\theta_4}\Bigg(
\frac{\theta_2(\xi_1\!\!-\!\!\xi_2)}{\theta_3\theta_4}\theta_3(\xi_1\!\!-\!\!\frac{z}{2})\theta_4(\xi_1\!\!-\!\!\frac{z}{2})\theta_3(\xi_2\!\!-\!\!\frac{z}{2})\theta_4(\xi_2\!\!-\!\!\frac{z}{2})\nonumber\\
&&+\frac{\theta_3(\xi_1\!\!-\!\!\xi_2)}{\theta_2\theta_4}\theta_2(\xi_1\!\!-\!\!\frac{z}{2})\theta_4(\xi_1\!\!-\!\!\frac{z}{2})\theta_2(\xi_2\!\!-\!\!\frac{z}{2})\theta_4(\xi_2\!\!-\!\!\frac{z}{2})\nonumber\\
&&+\frac{\theta_4(\xi_1\!\!-\!\!\xi_2)}{\theta_2\theta_3}\theta_2(\xi_1\!\!-\!\!\frac{z}{2})\theta_3(\xi_1\!\!-\!\!\frac{z}{2})\theta_2(\xi_2\!\!-\!\!\frac{z}{2})\theta_3(\xi_2\!\!-\!\!\frac{z}{2})\Bigg)
\end{eqnarray}
So we get
\begin{eqnarray}
Z^{A,O_+(2),2}&=&\frac{1}{2}\Big(Z_{(00)}^{O(2),2}-Z^{O(2),2}\Big)\nonumber\\
&=&\frac{1}{2}\Bigg(\frac{\theta_1(\xi_2\!\!-\!\!\xi_1\!\!-\!\!z)}{\theta_1(\xi_2\!\!-\!\!\xi_1)}\frac{\theta_1(2\xi_1\!\!-\!\!z)}{\theta_1(2\xi_1)}\!\!+\!\!\frac{\theta_1(\xi_1\!\!-\!\!\xi_2\!\!-\!\!z)}{\theta_1(\xi_1\!\!-\!\!\xi_2)}\frac{\theta_1(2\xi_2\!\!-\!\!z)}{\theta_1(2\xi_2)}\Bigg)\frac{\theta_1(\xi_1\!\!+\!\!\xi_2\!\!-\!\!z)}{\theta_1(\xi_1\!\!+\!\!\xi_2)}+\frac{\theta_1(z)\theta_1(\xi_1\!\!+\!\!\xi_2\!\!-\!\!z)}{\theta_1(2\xi_1)\theta_1(2\xi_2)}\nonumber\\
&&\times\Bigg(\frac{\theta_2(\xi_1\!\!-\!\!\xi_2)}{\theta_2}\frac{\theta_3(\xi_1\!\!-\!\!\frac{z}{2})\theta_4(\xi_1\!\!-\!\!\frac{z}{2})\theta_3(\xi_2\!\!-\!\!\frac{z}{2})\theta_4(\xi_2\!\!-\!\!\frac{z}{2})}{\theta_3\theta_4\theta_3\theta_4}+(2\leftrightarrow 3)+(2\leftrightarrow 4)\Bigg).
\end{eqnarray}
We should change the holonomies as $\xi \rightarrow -\xi +\frac{z}{2}$ and multiply $Z^{M,2}$ to compare with the dual theory $Z^{A,SO(1),2}$
\begin{eqnarray}
Z^{B,O_{+}(2),2}&=&\widetilde{Z}^{A,O_{+}(2),2}\nonumber\\ &=&\frac{1}{2}\Bigg(\frac{\theta_1(\xi_2\!\!-\!\!\xi_1\!\!-\!\!z)}{\theta_1(\xi_2\!\!-\!\!\xi_1)}\frac{\theta_1(2\xi_1\!\!-\!\!z)}{\theta_1(2\xi_1)}\!\!+\!\!\frac{\theta_1(\xi_1\!\!-\!\!\xi_2\!\!-\!\!z)}{\theta_1(\xi_1\!\!-\!\!\xi_2)}\frac{\theta_1(2\xi_2\!\!-\!\!z)}{\theta_1(2\xi_2)}\Bigg)-\frac{\theta_1(z)\theta_1(\xi_1\!\!+\!\!\xi_2\!\!-\!\!z)}{\theta_1(2\xi_1)\theta_1(2\xi_2)}\nonumber\\
&&\times\Bigg[\frac{\theta_2(\xi_1\!\!-\!\!\xi_2)}{\theta_2}\frac{\theta_3(\xi_1)\theta_4(\xi_1)\theta_3(\xi_2)\theta_4(\xi_2)}{\theta_3\theta_4\theta_3\theta_4}+(2\leftrightarrow 3)+(2\leftrightarrow 4)\Bigg].
\end{eqnarray}
We can have further simplifications. Let's consider the square bracket $[\cdots]$,
\begin{eqnarray}
[\cdots]&=&\frac{\theta_1(\xi_1\!\!-\!\!\xi_2)}{\theta_1(\xi_1\!\!-\!\!\xi_2)}[\cdots]\nonumber\\
&=&\frac{1}{2\theta_1(\xi_1\!\!-\!\!\xi_2)}\Bigg[\Bigg(\frac{\theta_1(\xi_1\!\!-\!\!\xi_2)\theta_2(\xi_1\!\!-\!\!\xi_2)\theta_3(\xi_1)\theta_4(\xi_1)}{\theta_2\theta_3\theta_4}\frac{\theta_3(\xi_2)\theta_4(\xi_2)}{\theta_3\theta_4}\nonumber\\
&&\qquad\qquad+\frac{\theta_1(\xi_1\!\!-\!\!\xi_2)\theta_3(\xi_1\!\!-\!\!\xi_2)\theta_2(\xi_1)\theta_4(\xi_1)}{\theta_2\theta_3\theta_4}\frac{\theta_2(\xi_2)\theta_4(\xi_2)}{\theta_2\theta_4}\nonumber\\
&&\qquad\qquad+\frac{\theta_1(\xi_1\!\!-\!\!\xi_2)\theta_4(\xi_1\!\!-\!\!\xi_2)\theta_2(\xi_1)\theta_3(\xi_1)}{\theta_2\theta_3\theta_4}\frac{\theta_2(\xi_2)\theta_3(\xi_2)}{\theta_2\theta_3}\Bigg)-(\xi_1\leftrightarrow\xi_2)\Bigg].
\end{eqnarray}
Using the relations below,
\begin{eqnarray}
\theta_1(\xi_1\!\!-\!\!\xi_2)\theta_2(\xi_1\!\!-\!\!\xi_2)\theta_3(\xi_1)\theta_4(\xi_1)&=&\frac{\theta_3\theta_4}{2}\Big(\theta_1(2\xi_1\!\!-\!\!\xi_2)\theta_2(\xi_2)\!\!-\!\!\theta_1(\xi_2)\theta_2(2\xi_1\!\!-\!\!\xi_2)\Big)\nonumber\\
\theta_1(\xi_1\!\!-\!\!\xi_2)\theta_3(\xi_1\!\!-\!\!\xi_2)\theta_2(\xi_1)\theta_4(\xi_1)&=&\frac{\theta_2\theta_4}{2}\Big(\theta_1(2\xi_1\!\!-\!\!\xi_2)\theta_3(\xi_2)\!\!-\!\!\theta_1(\xi_2)\theta_3(2\xi_1\!\!-\!\!\xi_2)\Big)\nonumber\\
\theta_1(\xi_1\!\!-\!\!\xi_2)\theta_4(\xi_1\!\!-\!\!\xi_2)\theta_2(\xi_1)\theta_3(\xi_1)&=&\frac{\theta_2\theta_3}{2}\Big(\theta_1(2\xi_1\!\!-\!\!\xi_2)\theta_4(\xi_2)\!\!-\!\!\theta_1(\xi_2)\theta_4(2\xi_1\!\!-\!\!\xi_2)\Big)
\end{eqnarray}
we get
\begin{eqnarray}
[\cdots]&=&\frac{1}{8\theta_1(\xi_1\!\!-\!\!\xi_2)}\Bigg[\Bigg(3\frac{\theta_1(2\xi_1\!\!-\!\!\xi_2)}{\theta_1(\xi_2)}-\frac{\theta_2(2\xi_1\!\!-\!\!\xi_2)}{\theta_2(\xi_2)}-\frac{\theta_3(2\xi_1\!\!-\!\!\xi_2)}{\theta_3(\xi_2)}-\frac{\theta_4(2\xi_1\!\!-\!\!\xi_2)}{\theta_4(\xi_2)}\Bigg)\theta_1(2\xi_2) \nonumber\\
&&\qquad\qquad+\Bigg(\frac{\theta_2(2\xi_2\!\!-\!\!\xi_1)}{\theta_2(\xi_1)}+\frac{\theta_3(2\xi_2\!\!-\!\!\xi_1)}{\theta_3(\xi_1)}+\frac{\theta_4(2\xi_2\!\!-\!\!\xi_1)}{\theta_4(\xi_1)}-3\frac{\theta_1(2\xi_2\!\!-\!\!\xi_1)}{\theta_1(\xi_1)}\Bigg)\theta_1(2\xi_1) \Bigg]\nonumber\\
&=&\frac{\theta_1(2\xi_2)\theta_1(2\xi_1\!\!-\!\!\xi_2)}{2\theta_1(\xi_2)\theta_1(\xi_1\!\!-\!\!\xi_2)}\!\!-\!\!\frac{\theta_1(2\xi_2)}{4\theta_1(\xi_1\!\!-\!\!\xi_2)}\frac{1}{2}\Bigg(\!\!\frac{\theta_1(2\xi_1\!\!-\!\!\xi_2)}{\theta_1(\xi_2)}\!\!+\!\!\frac{\theta_2(2\xi_1\!\!-\!\!\xi_2)}{\theta_2(\xi_2)}\!\!+\!\!\frac{\theta_3(2\xi_1\!\!-\!\!\xi_2)}{\theta_3(\xi_2)}\!\!+\!\!\frac{\theta_4(2\xi_1\!\!-\!\!\xi_2)}{\theta_4(\xi_2)}\!\!\Bigg)\nonumber\\
&+&\frac{\theta_1(2\xi_1)\theta_1(2\xi_2\!\!-\!\!\xi_1)}{2\theta_1(\xi_1)\theta_1(\xi_2\!\!-\!\!\xi_1)}\!\!-\!\!\frac{\theta_1(2\xi_1)}{4\theta_1(\xi_2\!\!-\!\!\xi_1)}\frac{1}{2}\Bigg(\!\!\frac{\theta_1(2\xi_2\!\!-\!\!\xi_1)}{\theta_1(\xi_1)}\!\!+\!\!\frac{\theta_2(2\xi_2\!\!-\!\!\xi_1)}{\theta_2(\xi_1)}\!\!+\!\!\frac{\theta_3(2\xi_2\!\!-\!\!\xi_1)}{\theta_3(\xi_1)}\!\!+\!\!\frac{\theta_4(2\xi_2\!\!-\!\!\xi_1)}{\theta_4(\xi_1)}\!\!\Bigg).
\nonumber\\
\end{eqnarray}
Using  eq.(\ref{16}) for $\xi=\xi_2(\xi_1)$, $z=2\xi_2-2\xi_1(2\xi_1-2\xi_2)$ we can simplify the expressions in the two brackets as
\begin{eqnarray}
\frac{1}{2}\Bigg(\!\!\frac{\theta_1(2\xi_1\!\!-\!\!\xi_2)}{\theta_1(\xi_2)}\!\!+\!\!\frac{\theta_2(2\xi_1\!\!-\!\!\xi_2)}{\theta_2(\xi_2)}\!\!+\!\!\frac{\theta_3(2\xi_1\!\!-\!\!\xi_2)}{\theta_3(\xi_2)}\!\!+\!\!\frac{\theta_4(2\xi_1\!\!-\!\!\xi_2)}{\theta_4(\xi_2)}\!\!\Bigg)&=&\frac{\theta_1(\xi_1\!\!+\!\!\xi_2)\theta_1(2\xi_2\!\!-\!\!2\xi_1)}{\theta_1(2\xi_2)\theta_1(\xi_2\!\!-\!\!\xi_1)}\nonumber\\
\frac{1}{2}\Bigg(\!\!\frac{\theta_1(2\xi_2\!\!-\!\!\xi_1)}{\theta_1(\xi_1)}\!\!+\!\!\frac{\theta_2(2\xi_2\!\!-\!\!\xi_1)}{\theta_2(\xi_1)}\!\!+\!\!\frac{\theta_3(2\xi_2\!\!-\!\!\xi_1)}{\theta_3(\xi_1)}\!\!+\!\!\frac{\theta_4(2\xi_2\!\!-\!\!\xi_1)}{\theta_4(\xi_1)}\!\!\Bigg)&=&\frac{\theta_1(\xi_1\!\!+\!\!\xi_2)\theta_1(2\xi_1\!\!-\!\!2\xi_2)}{\theta_1(2\xi_1)\theta_1(\xi_1\!\!-\!\!\xi_2)}
\end{eqnarray}
so that
\begin{eqnarray}
[\cdots]&=&\frac{1}{2}\frac{\theta_1(2\xi_2)\theta_1(2\xi_1\!\!-\!\!\xi_2)}{\theta_1(\xi_2)\theta_1(\xi_1\!\!-\!\!\xi_2)}-\frac{1}{4}\frac{\theta_1(2\xi_2)}{\theta_1(\xi_1\!\!-\!\!\xi_2)}\frac{\theta_1(\xi_1\!\!+\!\!\xi_2)\theta_1(2\xi_2\!\!-\!\!2\xi_1)}{\theta_1(2\xi_2)\theta_1(\xi_2\!\!-\!\!\xi_1)}\nonumber\\
&+&\frac{1}{2}\frac{\theta_1(2\xi_1)\theta_1(2\xi_2\!\!-\!\!\xi_1)}{\theta_1(\xi_1)\theta_1(\xi_2\!\!-\!\!\xi_1)}-\frac{1}{4}\frac{\theta_1(2\xi_1)}{\theta_1(\xi_2\!\!-\!\!\xi_1)}\frac{\theta_1(\xi_1\!\!+\!\!\xi_2)\theta_1(2\xi_1\!\!-\!\!2\xi_2)}{\theta_1(2\xi_1)\theta_1(\xi_1\!\!-\!\!\xi_2)}\nonumber\\
&=&\frac{1}{2}\frac{\theta_1(2\xi_2)\theta_1(2\xi_1\!\!-\!\!\xi_2)}{\theta_1(\xi_2)\theta_1(\xi_1\!\!-\!\!\xi_2)}-\frac{1}{4}\frac{\theta_1(\xi_1\!\!+\!\!\xi_2)\theta_1(2\xi_1\!\!-\!\!2\xi_2)}{\{\theta_1(\xi_1\!\!-\!\!\xi_2)\}^2}\nonumber\\
&+&\frac{1}{2}\frac{\theta_1(2\xi_1)\theta_1(2\xi_2\!\!-\!\!\xi_1)}{\theta_1(\xi_1)\theta_1(\xi_2\!\!-\!\!\xi_1)}+\frac{1}{4}\frac{\theta_1(\xi_1\!\!+\!\!\xi_2)\theta_1(2\xi_1\!\!-\!\!2\xi_2)}{\{\theta_1(\xi_1\!\!-\!\!\xi_2)\}^2}\nonumber\\
&=&\frac{1}{2}\frac{\theta_1(2\xi_2)\theta_1(2\xi_1\!\!-\!\!\xi_2)}{\theta_1(\xi_2)\theta_1(\xi_1\!\!-\!\!\xi_2)}+\frac{1}{2}\frac{\theta_1(2\xi_1)\theta_1(2\xi_2\!\!-\!\!\xi_1)}{\theta_1(\xi_1)\theta_1(\xi_2\!\!-\!\!\xi_1)}.
\end{eqnarray}
Thus we finally get
\begin{eqnarray}
Z^{B,O_{+}(2),2} &=&
\frac{1}{2}\frac{\theta_1(\xi_2\!\!-\!\!\xi_1\!\!-\!\!z)\theta_1(2\xi_1\!\!-\!\!z)}{\theta_1(\xi_2\!\!-\!\!\xi_1)\theta_1(2\xi_1)}\!\!+\!\!\frac{1}{2}\frac{\theta_1(\xi_1\!\!+\!\!\xi_2\!\!-\!\!z)\theta_1(2\xi_1\!\!-\!\!\xi_2)\theta_1(z)}{\theta_1(\xi_2\!\!-\!\!\xi_1)\theta_1(2\xi_1)\theta_1(\xi_2)}\nonumber\\
&+&\frac{1}{2}\frac{\theta_1(\xi_1\!\!-\!\!\xi_2\!\!-\!\!z)\theta_1(2\xi_2\!\!-\!\!z)}{\theta_1(\xi_1\!\!-\!\!\xi_2)\theta_1(2\xi_2)}\!\!+\!\!\frac{1}{2}\frac{\theta_1(\xi_1\!\!+\!\!\xi_2\!\!-\!\!z)\theta_1(2\xi_2\!\!-\!\!\xi_1)\theta_1(z)}{\theta_1(\xi_1\!\!-\!\!\xi_2)\theta_1(2\xi_2)\theta_1(\xi_1)}.
\end{eqnarray}
Now we show the equality of $Z^{A,SO(1),2}=Z^{B,O_{+}(2),2}$ using the fact that {\it doubly periodic entire function is a constant}.\footnote{We thank a referee of the previous version of the draft for suggesting this method.} Consider  $Q$ defined as
\begin{eqnarray}
Q &\equiv& \frac{Z^{B,O_{+}(2),2}}{Z^{A,SO(1),2}}\nonumber\\
&=&\frac{1}{2}\Bigg[\frac{\theta_1(\xi_2\!\!-\!\!\xi_1\!\!-\!\!z)\theta_1(2\xi_1\!\!-\!\!z)\theta_1(\xi_1)\theta_1(\xi_2)}{\theta_1(\xi_2\!\!-\!\!\xi_1)\theta_1(2\xi_1)\theta_1(\xi_1\!\!-\!\!z)\theta_1(\xi_2\!\!-\!\!z)}\!\!+\!\!\frac{\theta_1(\xi_1\!\!-\!\!\xi_2\!\!-\!\!z)\theta_1(2\xi_2\!\!-\!\!z)\theta_1(\xi_1)\theta_1(\xi_2)}{\theta_1(\xi_1\!\!-\!\!\xi_2)\theta_1(2\xi_2)\theta_1(\xi_1\!\!-\!\!z)\theta_1(\xi_2\!\!-\!\!z)}\nonumber\\
&&+\frac{\theta_1(\xi_1\!\!+\!\!\xi_2\!\!-\!\!z)\theta_1(2\xi_1\!\!-\!\!\xi_2)\theta_1(z)\theta_1(\xi_1)}{\theta_1(\xi_2\!\!-\!\!\xi_1)\theta_1(2\xi_1)\theta_1(\xi_1\!\!-\!\!z)\theta_1(\xi_2\!\!-\!\!z)}\!\!+\!\!\frac{\theta_1(\xi_1\!\!+\!\!\xi_2\!\!-\!\!z)\theta_1(2\xi_2\!\!-\!\!\xi_1)\theta_1(z)\theta_1(\xi_2)}{\theta_1(\xi_1\!\!-\!\!\xi_2)\theta_1(2\xi_2)\theta_1(\xi_1\!\!-\!\!z)\theta_1(\xi_2\!\!-\!\!z)}\Bigg]\label{Qd}.
\end{eqnarray}
If we think $Q$ as a function of $z$ for fixed $\xi_1$ and $\xi_2$, we can easily check that $Q$ is invariant under the changes $z\rightarrow z+1$ and $z \rightarrow z+\tau$ which means $Q$ is a doubly periodic function of $z$. Also we should check that  $Q$ has no poles. Since  the Jacobi theta functions have no poles in $z$, there are only two points where they have the possibility of poles: $z=\xi_1$, $\xi_2$ (we don't consider the points $z=\xi_{\alpha}+a+b\tau$ with $a,b$ nonzero integers since $Q$ is doubly periodic). $Q$ is symmetric under the change $\xi_1 \leftrightarrow \xi_2$ so that we just need to check a point $z=\xi_1$. The fact that $Z^{B,O_{+}(2),2}=0$ at $z=\xi_1$ gives
\begin{eqnarray}
Z^{B,O_{+}(2),2}\sim\mathcal{O}((z-\xi_1)^n)
\end{eqnarray}
as $z\rightarrow \xi_1$ for some $n\geq 1$. Meanwhile $Z^{A,SO(1),2}$ becomes zero at $z=\xi_1$ and $\theta_1(\xi_1-z)\sim\mathcal{O}((z-\xi_1)^1)$ as $z\rightarrow \xi_1$ so that
\begin{eqnarray}
Z^{A,SO(1),2}\sim\mathcal{O}((z-\xi_1)^1)
\end{eqnarray}
as $z\rightarrow \xi_1$. Thus we can deduce that $Q=\frac{Z^{B,O_{+}(2),2}}{Z^{A,SO(1),2}}$ converges to a finite value and it tells us that the $Q$ has no pole at $z=\xi_1$. From the fact that doubly periodic entire function is a constant, we can easily find the constant as
\begin{eqnarray}
Q=\lim_{z\rightarrow 0}Q = \frac{1}{2}(1+1+0+0)=1,
\end{eqnarray}
which implies
\begin{eqnarray}
Z^{A,SO(1),2}=Z^{B,O_{+}(2),2}.
\end{eqnarray}
 One might worry that Q can have a pole at $\xi_1\rightarrow\xi_2$, $\xi_1 \rightarrow 0$ and $\xi_2 \rightarrow 0$ where Q seems to diverge. If we take a limit $\xi_1\rightarrow\xi_2$, the divergent parts of the first and the second term of the eq.(\ref{Qd}) are cancelled by the third and the fourth term respectively so that $Q$ becomes finite. Also for the limit $\xi_1 \rightarrow 0$(we don't consider the $\xi_2 \rightarrow 0$ case since $Q$ is symmetric under $\xi_1 \leftrightarrow \xi_2$), one can easily check that
\begin{eqnarray}
\lim_{\xi_1\rightarrow 0}Q = \frac{1}{2}\Big(\frac{1}{2}+0+\frac{(-1)^2}{2}+(-1)^2\Big)=1.
\end{eqnarray}

\subsubsection{Dualities between $O(2),2$ $\leftrightarrow$ $O(1),2$}
We can also check the following equalities
\begin{eqnarray}
Z^{A,SO(2),2}&=&Z^{B,O_{+}(1),2}\\
Z^{A,O_{-}(2),2}&=&Z^{B,O_{-}(1),2},
\end{eqnarray}
from which we have the following relations
\begin{eqnarray}
Z^{A,O_{+}(1),2}&=&Z^{B,SO(2),2}\\
Z^{A,O_{-}(1),2}&=&Z^{B,O_{-}(2),2}.
\end{eqnarray}
We prove these equalities by showing relations below
\begin{eqnarray}
\widetilde{Z}_{(01)}^{O(2),2}-\widetilde{Z}_{(10)}^{O(2),2}&=&Z_{(10)}^{O(1),2}-Z_{(01)}^{O(1),2}\label{k1}\\
\widetilde{Z}_{(11)}^{O(2),2}-\widetilde{Z}_{(10)}^{O(2),2}&=&Z_{(10)}^{O(1),2}-Z_{(11)}^{O(1),2}\label{k2}\\
\widetilde{Z}_{(01)}^{O(2),2}+\widetilde{Z}_{(11)}^{O(2),2}&=&Z_{(10)}^{O(1),2}-Z_{(00)}^{O(1),2}\label{k3}.
\end{eqnarray}

Let's first show the relation eq.(\ref{k1}). From the eq.(\ref{Z01}) and eq.(\ref{Z10}), $Z_{(01)}^{O(2),2}-Z_{(10)}^{O(2),2}$ becomes
\begin{eqnarray}
Z_{(01)}^{O(2),2}-Z_{(10)}^{O(2),2}&=&2\frac{\theta_1(z)\theta_1(\xi_1\!\!+\!\!\xi_2\!\!-\!\!z)\theta_3(\xi_1\!\!-\!\!\xi_2)}{\theta_1(2\xi_1)\theta_1(2\xi_2)\theta_2\theta_3\theta_4}\Big\{\theta_2(\xi_1\!\!+\!\!\xi_2\!\!-\!\!z)\theta_4(\xi_1\!\!-\!\!\xi_2)\!\!-\!\!\theta_2(\xi_1\!\!-\!\!\xi_2)\theta_4(\xi_1\!\!+\!\!\xi_2\!\!-\!\!z)\Big\}\nonumber\\
&=&-4\frac{\theta_1(z)\theta_1(\xi_1\!\!+\!\!\xi_2\!\!-\!\!z)\theta_3(\xi_1\!\!-\!\!\xi_2)}{\theta_1(2\xi_1)\theta_1(2\xi_2)\theta_2\theta_3\theta_4\theta_2\theta_4}\theta_1(\xi_1\!\!-\!\!\frac{z}{2})\theta_3(\xi_1\!\!-\!\!\frac{z}{2})\theta_1(\xi_2\!\!-\!\!\frac{z}{2})\theta_3(\xi_2\!\!-\!\!\frac{z}{2})\nonumber\\
\end{eqnarray}
Now changing the holonomies $\xi_{i}\rightarrow -\xi_{i}+\frac{z}{2}$ and multiplying $Z^{M,2}$, we get
\begin{eqnarray}
\widetilde{Z}_{(01)}^{O(2),2}-\widetilde{Z}_{(10)}^{O(2),2}&=&4\frac{\theta_1(z)\theta_1(\xi_1\!\!+\!\!\xi_2\!\!-\!\!z)\theta_3(\xi_1\!\!-\!\!\xi_2)}{\theta_1(2\xi_1)\theta_1(2\xi_2)\theta_2\theta_3\theta_4\theta_2\theta_4}\theta_1(\xi_1)\theta_3(\xi_1)\theta_1(\xi_2)\theta_3(\xi_2)\nonumber\\
&=&\frac{\theta_1(z)\theta_3\theta_1(\xi_1\!\!+\!\!\xi_2\!\!-\!\!z)\theta_3(\xi_1\!\!-\!\!\xi_2)}{\theta_1(\xi_1)\theta_2(\xi_1)\theta_3(\xi_1)\theta_4(\xi_1)\theta_1(\xi_2)\theta_2(\xi_2)\theta_3(\xi_2)\theta_4(\xi_2)}\theta_1(\xi_1)\theta_3(\xi_1)\theta_1(\xi_2)\theta_3(\xi_2)\nonumber\\
&=&\frac{\theta_1(z)\theta_3\theta_1(\xi_1\!\!+\!\!\xi_2\!\!-\!\!z)\theta_3(\xi_1\!\!-\!\!\xi_2)}{\theta_2(\xi_1)\theta_4(\xi_1)\theta_2(\xi_2)\theta_4(\xi_2)}\nonumber\\
&=&\frac{2}{\theta_2^2\theta_4^2}\frac{1}{\theta_2(\xi_1)\theta_4(\xi_1)\theta_2(\xi_2)\theta_4(\xi_2)}\nonumber\\
&&\times\Big\{\theta_1(\xi_1\!\!-\!\!\frac{z}{2})\theta_3(\xi_1\!\!-\!\!\frac{z}{2})\theta_1(\frac{z}{2})\theta_3(\frac{z}{2})\cdot\theta_2(\xi_2\!\!-\!\!\frac{z}{2})\theta_4(\xi_2\!\!-\!\!\frac{z}{2})\theta_2(\frac{z}{2})\theta_4(\frac{z}{2})\nonumber\\
&&\qquad+\theta_1(\xi_2\!\!-\!\!\frac{z}{2})\theta_3(\xi_2\!\!-\!\!\frac{z}{2})\theta_1(\frac{z}{2})\theta_3(\frac{z}{2})\cdot\theta_2(\xi_1\!\!-\!\!\frac{z}{2})\theta_4(\xi_1\!\!-\!\!\frac{z}{2})\theta_2(\frac{z}{2})\theta_4(\frac{z}{2})\Big\}\nonumber\\
&=&\frac{\theta_2(\xi_1\!\!-\!\!z)\theta_2(\xi_2\!\!-\!\!z)\theta_4(\xi_1)\theta_4(\xi_2)-\theta_2(\xi_1)\theta_2(\xi_2)\theta_4(\xi_1\!\!-\!\!z)\theta_4(\xi_2\!\!-\!\!z)}{\theta_2(\xi_1)\theta_4(\xi_1)\theta_2(\xi_2)\theta_4(\xi_2)}\nonumber\\
&=&\frac{\theta_2(\xi_1\!\!-\!\!z)\theta_2(\xi_2\!\!-\!\!z)}{\theta_2(\xi_1)\theta_2(\xi_2)}-\frac{\theta_4(\xi_1\!\!-\!\!z)\theta_4(\xi_2\!\!-\!\!z)}{\theta_4(\xi_1)\theta_4(\xi_2)}\nonumber\\
&=&Z_{(10)}^{O(1),2}-Z_{(01)}^{O(1),2},
\end{eqnarray}
which shows the eq.(\ref{k1}).

Secondly from the eq.(\ref{Z11}) and eq.(\ref{Z10}), $Z_{(11)}^{O(2),2}-Z_{(10)}^{O(2),2}$ becomes
\begin{eqnarray}
Z_{(11)}^{O(2),2}-Z_{(10)}^{O(2),2}&=&2\frac{\theta_1(z)\theta_1(\xi_1\!\!+\!\!\xi_2\!\!-\!\!z)\theta_4(\xi_1\!\!-\!\!\xi_2)}{\theta_1(2\xi_1)\theta_1(2\xi_2)\theta_2\theta_3\theta_4}\Big\{\theta_2(\xi_1\!\!+\!\!\xi_2\!\!-\!\!z)\theta_3(\xi_1\!\!-\!\!\xi_2)\!\!-\!\!\theta_2(\xi_1\!\!-\!\!\xi_2)\theta_3(\xi_1\!\!+\!\!\xi_2\!\!-\!\!z)\Big\}\nonumber\\
&=&-4\frac{\theta_1(z)\theta_1(\xi_1\!\!+\!\!\xi_2\!\!-\!\!z)\theta_4(\xi_1\!\!-\!\!\xi_2)}{\theta_1(2\xi_1)\theta_1(2\xi_2)\theta_2\theta_3\theta_4\theta_2\theta_3}\theta_1(\xi_1\!\!-\!\!\frac{z}{2})\theta_4(\xi_1\!\!-\!\!\frac{z}{2})\theta_1(\xi_2\!\!-\!\!\frac{z}{2})\theta_4(\xi_2\!\!-\!\!\frac{z}{2})\nonumber\\
\end{eqnarray}
Now changing the holonomies $\xi_{i}\rightarrow -\xi_{i}+\frac{z}{2}$ and multiplying $Z^{M,2}$, we get
\begin{eqnarray}
\widetilde{Z}_{(11)}^{O(2),2}-\widetilde{Z}_{(10)}^{O(2),2}&=&4\frac{\theta_1(z)\theta_1(\xi_1\!\!+\!\!\xi_2\!\!-\!\!z)\theta_4(\xi_1\!\!-\!\!\xi_2)}{\theta_1(2\xi_1)\theta_1(2\xi_2)\theta_2\theta_3\theta_4\theta_2\theta_3}\theta_1(\xi_1)\theta_4(\xi_1)\theta_1(\xi_2)\theta_4(\xi_2)\nonumber\\
&=&\frac{\theta_1(z)\theta_4\theta_1(\xi_1\!\!+\!\!\xi_2\!\!-\!\!z)\theta_4(\xi_1\!\!-\!\!\xi_2)}{\theta_2(\xi_1)\theta_3(\xi_1)\theta_2(\xi_2)\theta_3(\xi_2)}\nonumber\\
&=&\frac{2}{\theta_2^2\theta_3^2}\frac{1}{\theta_2(\xi_1)\theta_3(\xi_1)\theta_2(\xi_2)\theta_3(\xi_2)}\nonumber\\
&&\times\Big\{\theta_1(\xi_1\!\!-\!\!\frac{z}{2})\theta_4(\xi_1\!\!-\!\!\frac{z}{2})\theta_1(\frac{z}{2})\theta_4(\frac{z}{2})\cdot\theta_2(\xi_2\!\!-\!\!\frac{z}{2})\theta_3(\xi_2\!\!-\!\!\frac{z}{2})\theta_2(\frac{z}{2})\theta_3(\frac{z}{2})\nonumber\\
&&\qquad+\theta_1(\xi_2\!\!-\!\!\frac{z}{2})\theta_4(\xi_2\!\!-\!\!\frac{z}{2})\theta_1(\frac{z}{2})\theta_4(\frac{z}{2})\cdot\theta_2(\xi_1\!\!-\!\!\frac{z}{2})\theta_3(\xi_1\!\!-\!\!\frac{z}{2})\theta_2(\frac{z}{2})\theta_3(\frac{z}{2})\Big\}\nonumber\\
&=&\frac{\theta_2(\xi_1\!\!-\!\!z)\theta_2(\xi_2\!\!-\!\!z)\theta_3(\xi_1)\theta_3(\xi_2)-\theta_2(\xi_1)\theta_2(\xi_2)\theta_3(\xi_1\!\!-\!\!z)\theta_3(\xi_2\!\!-\!\!z)}{\theta_2(\xi_1)\theta_3(\xi_1)\theta_2(\xi_2)\theta_3(\xi_2)}\nonumber\\
&=&\frac{\theta_2(\xi_1\!\!-\!\!z)\theta_2(\xi_2\!\!-\!\!z)}{\theta_2(\xi_1)\theta_2(\xi_2)}-\frac{\theta_3(\xi_1\!\!-\!\!z)\theta_3(\xi_2\!\!-\!\!z)}{\theta_3(\xi_1)\theta_3(\xi_2)}\nonumber\\
&=&Z_{(10)}^{O(1),2}-Z_{(11)}^{O(1),2},
\end{eqnarray}
which shows the eq.(\ref{k2}).

Finally from the eq.(\ref{Z01}) and eq.(\ref{Z11}), $Z_{(01)}^{O(2),2}+Z_{(11)}^{O(2),2}$ becomes
\begin{eqnarray}
Z_{(01)}^{O(2),2}+Z_{(11)}^{O(2),2}&=&-2\frac{\theta_1(z)\theta_1(\xi_1\!\!+\!\!\xi_2\!\!-\!\!z)\theta_2(\xi_1\!\!-\!\!\xi_2)}{\theta_1(2\xi_1)\theta_1(2\xi_2)\theta_2\theta_3\theta_4}\Big\{\theta_3(\xi_1\!\!+\!\!\xi_2\!\!-\!\!z)\theta_4(\xi_1\!\!-\!\!\xi_2)\!\!+\!\!\theta_3(\xi_1\!\!-\!\!\xi_2)\theta_4(\xi_1\!\!+\!\!\xi_2\!\!-\!\!z)\Big\}\nonumber\\
&=&-4\frac{\theta_1(z)\theta_1(\xi_1\!\!+\!\!\xi_2\!\!-\!\!z)\theta_2(\xi_1\!\!-\!\!\xi_2)}{\theta_1(2\xi_1)\theta_1(2\xi_2)\theta_2\theta_3\theta_4\theta_3\theta_4}\theta_3(\xi_1\!\!-\!\!\frac{z}{2})\theta_4(\xi_1\!\!-\!\!\frac{z}{2})\theta_3(\xi_2\!\!-\!\!\frac{z}{2})\theta_4(\xi_2\!\!-\!\!\frac{z}{2})\nonumber\\
\end{eqnarray}
Now changing the holonomies $\xi_{i}\rightarrow -\xi_{i}+\frac{z}{2}$ and multiplying $Z^{M,2}$, we get
\begin{eqnarray}
\widetilde{Z}_{(01)}^{O(2),2}+\widetilde{Z}_{(11)}^{O(2),2}&=&4\frac{\theta_1(z)\theta_1(\xi_1\!\!+\!\!\xi_2\!\!-\!\!z)\theta_2(\xi_1\!\!-\!\!\xi_2)}{\theta_1(2\xi_1)\theta_1(2\xi_2)\theta_2\theta_3\theta_4\theta_3\theta_4}\theta_3(\xi_1)\theta_4(\xi_1)\theta_3(\xi_2)\theta_4(\xi_2)\nonumber\\
&=&\frac{\theta_1(z)\theta_2\theta_1(\xi_1\!\!+\!\!\xi_2\!\!-\!\!z)\theta_2(\xi_1\!\!-\!\!\xi_2)}{\theta_1(\xi_1)\theta_2(\xi_1)\theta_1(\xi_2)\theta_2(\xi_2)}\nonumber\\
&=&\frac{2}{\theta_3^2\theta_4^2}\frac{1}{\theta_1(\xi_1)\theta_2(\xi_1)\theta_1(\xi_2)\theta_2(\xi_2)}\nonumber\\
&&\times\Big\{\theta_1(\xi_1\!\!-\!\!\frac{z}{2})\theta_2(\xi_1\!\!-\!\!\frac{z}{2})\theta_3(\frac{z}{2})\theta_4(\frac{z}{2})\cdot\theta_3(\xi_2\!\!-\!\!\frac{z}{2})\theta_4(\xi_2\!\!-\!\!\frac{z}{2})\theta_1(\frac{z}{2})\theta_2(\frac{z}{2})\nonumber\\
&&\qquad+\theta_1(\xi_2\!\!-\!\!\frac{z}{2})\theta_2(\xi_2\!\!-\!\!\frac{z}{2})\theta_3(\frac{z}{2})\theta_4(\frac{z}{2})\cdot\theta_3(\xi_1\!\!-\!\!\frac{z}{2})\theta_4(\xi_1\!\!-\!\!\frac{z}{2})\theta_1(\frac{z}{2})\theta_2(\frac{z}{2})\Big\}\nonumber\\
&=&\frac{\theta_2(\xi_1\!\!-\!\!z)\theta_2(\xi_2\!\!-\!\!z)\theta_1(\xi_1)\theta_1(\xi_2)-\theta_2(\xi_1)\theta_2(\xi_2)\theta_1(\xi_1\!\!-\!\!z)\theta_1(\xi_2\!\!-\!\!z)}{\theta_1(\xi_1)\theta_2(\xi_1)\theta_1(\xi_2)\theta_2(\xi_2)}\nonumber\\
&=&\frac{\theta_2(\xi_1\!\!-\!\!z)\theta_2(\xi_2\!\!-\!\!z)}{\theta_2(\xi_1)\theta_2(\xi_2)}-\frac{\theta_1(\xi_1\!\!-\!\!z)\theta_1(\xi_2\!\!-\!\!z)}{\theta_1(\xi_1)\theta_1(\xi_2)}\nonumber\\
&=&Z_{(10)}^{O(1),2}-Z_{(00)}^{O(1),2},
\end{eqnarray}
which shows the eq.(\ref{k3}).

 Thus we have shown the eq.(\ref{k1})-eq.(\ref{k3}). By subtracting the eq.(\ref{k2}) from the eq.(\ref{k1}), we can get
\begin{eqnarray}
\widetilde{Z}_{(01)}^{O(2),2}-\widetilde{Z}_{(11)}^{O(2),2}=Z_{(11)}^{O(1),2}-Z_{(01)}^{O(1),2}.\label{k4}
\end{eqnarray}
Also  we can use the proof of the last subsection
\begin{eqnarray}
Z_{(00)}^{O(1),2}=\frac{1}{2}\Big(\widetilde{Z}_{(00)}^{O(2),2}-\widetilde{Z}_{(01)}^{O(2),2}-\widetilde{Z}_{(10)}^{O(2),2}-\widetilde{Z}_{(11)}^{O(2),2}\Big).\label{st}
\end{eqnarray}
Adding the eq.(\ref{k3}) to the eq.(\ref{k4}), we can get
\begin{eqnarray}
\widetilde{Z}_{(11)}^{O(2),2}=\frac{1}{2}\Big(-Z_{(00)}^{O(1),2}+Z_{(10)}^{O(1),2}+Z_{(01)}^{O(1),2}-Z_{(11)}^{O(1),2}\Big).\label{k5}
\end{eqnarray}
Subtracting the eq.(\ref{k3}) from the eq.(\ref{k4}), we can get
\begin{eqnarray}
\widetilde{Z}_{(01)}^{O(2),2}=\frac{1}{2}\Big(-Z_{(00)}^{O(1),2}+Z_{(10)}^{O(1),2}-Z_{(01)}^{O(1),2}+Z_{(11)}^{O(1),2}\Big).\label{k6}
\end{eqnarray}
Using the eq.(\ref{k6}), the eq.(\ref{k1}) gives
\begin{eqnarray}
\widetilde{Z}_{(10)}^{O(2),2}=\frac{1}{2}\Big(-Z_{(00)}^{O(1),2}-Z_{(10)}^{O(1),2}+Z_{(01)}^{O(1),2}+Z_{(11)}^{O(1),2}\Big).\label{k7}
\end{eqnarray}
Again using the eq.(\ref{k5})-eq.(\ref{k7}), the eq.(\ref{st}) gives
\begin{eqnarray}
\widetilde{Z}_{(00)}^{O(2),2}=\frac{1}{2}\Big(Z_{(00)}^{O(1),2}+Z_{(10)}^{O(1),2}+Z_{(01)}^{O(1),2}+Z_{(11)}^{O(1),2}\Big).\label{k8}
\end{eqnarray}
So from the eq.(\ref{k5})-eq.(\ref{k8}), we get relations given by
\begin{eqnarray}
\left( \begin{array}{c}
 \widetilde{Z}_{(00)}^{O(2),2}\\
 \widetilde{Z}_{(10)}^{O(2),2}\\
 \widetilde{Z}_{(01)}^{O(2),2}\\
 \widetilde{Z}_{(11)}^{O(2),2}
\end{array} \right)=
\frac{1}{2}
\left( \begin{array}{cccc}
1&1&1&1\\
-1&-1&1&1\\
-1&1&-1&1\\
-1&1&1&-1
\end{array} \right)
\left( \begin{array}{c}
Z_{(00)}^{O(1),2}\\
Z_{(10)}^{O(1),2}\\
Z_{(01)}^{O(1),2}\\
Z_{(11)}^{O(1),2}
\end{array} \right),
\end{eqnarray}
from which it follows that
\begin{eqnarray}
\left( \begin{array}{c}
Z_{(00)}^{O(2),2}\\
Z_{(10)}^{O(2),2}\\
Z_{(01)}^{O(2),2}\\
Z_{(11)}^{O(2),2}
\end{array} \right)=
\frac{1}{2}
\left( \begin{array}{cccc}
1&1&1&1\\
-1&-1&1&1\\
-1&1&-1&1\\
-1&1&1&-1
\end{array} \right)
\left( \begin{array}{c}
\widetilde{Z}_{(00)}^{O(1),2}\\
\widetilde{Z}_{(10)}^{O(1),2}\\
\widetilde{Z}_{(01)}^{O(1),2}\\
\widetilde{Z}_{(11)}^{O(1),2}
\end{array} \right),
\end{eqnarray}
where we use the fact that $\widetilde{\widetilde{Z}}_{(ab)}^{G,N}=Z_{(ab)}^{G,N}$ from the eq.(\ref{pd}).{\tiny {\tiny }}

This tells us that
\begin{eqnarray}
Z^{A,SO(2),2}&=&Z_{(00)}^{O(2),2}\nonumber\\
&=&\frac{1}{2}\Big(\widetilde{Z}_{(00)}^{O(1),2}+\widetilde{Z}_{(10)}^{O(1),2}+\widetilde{Z}_{(01)}^{O(1),2}+\widetilde{Z}_{(11)}^{O(1),2}\Big)\nonumber\\
&=&Z^{B,O_{+}(1),2}
\end{eqnarray}
and
\begin{eqnarray}
Z^{A,O_{-}(2),2}&=&\frac{1}{2}\Big(Z_{(00)}^{O(2),2}+Z_{(10)}^{O(2),2}+Z_{(01)}^{O(2),2}+Z_{(11)}^{O(2),2}\Big)\nonumber\\
&=&\frac{1}{2}\Big(-\widetilde{Z}_{(00)}^{O(1),2}+\widetilde{Z}_{(10)}^{O(1),2}+\widetilde{Z}_{(01)}^{O(1),2}+\widetilde{Z}_{(11)}^{O(1),2}\Big)\nonumber\\
&=&Z^{B,O_{-}(1),2}.
\end{eqnarray}
So we check the equalities $Z^{A,SO(2),2}=Z^{B,O_{+}(1),2}$ (or equivalently $Z^{A,O_{+}(1),2}=Z^{B,SO(2),2}$) and $Z^{A,O_{-}(2),2}=Z^{B,O_{-}(1),2}$ (or equivalently $Z^{A,O_{-}(1),2}=Z^{B,O_{-}(2),2}$) analytically.

\subsection{$SO(3), O_{\pm}(3)$ theories with 2 fundamentals}

Using the same method as adopted at the subsection E.5.1 one can prove the duality between  $SO(3), O_{\pm}(3)$ theories with 2 fundamentals
and the theory of 3 free mesons. Even though the length of the computation is long, the argument is straightforward.
And the same method applies to the the duality between  $SO(4), O_{\pm}(4)$ theories with 3 fundamentals and the theory of 6 free mesons.

In the below we evaluate $\lim_{z\rightarrow x}\theta_i(z-x+\frac{a+b\tau}{2})f(z)$ instead of $\lim_{z\rightarrow x}(z-x)f(z)$ when we evaluate residue of $f(z)$ at $z=x$ since $\theta_i(z-x+\frac{a+b\tau}{2})$ has zero at $z=x$.
Also recall that for small $z$ with $(a,b,i)=(0,0,1),(1,0,2),(0,1,4),(1,1,3)$, we have
\begin{equation}
\theta_i(z+\frac{a+b\tau}{2})\sim z.
\end{equation}

Using the Jacobi theta function's relations we can get

\begin{eqnarray}
Z_{(0,0,+)}^{O(3),N}&=&-\frac{1}{2}\frac{\theta_1(z)}{\theta_1(2z)}
\prod_{\alpha=1}^{N}
\frac{\theta_1(\xi_{\alpha}\!\!-\!\!2z)}{\theta_1(\xi_{\alpha}\!\!+\!\!z)}
\!\!+\!\!\frac{1}{2}\sum_{\alpha=1}^{N}
\frac{\theta_1(2\xi_{\alpha}\!\!-\!\!z)}{\theta_1(2\xi_{\alpha})}
\frac{\theta_1(\xi_{\alpha})}{\theta_1(\xi_{\alpha}\!\!+\!\!z)}
\prod_{\beta\neq\alpha}
\frac{\theta_1(\xi_{\alpha}\!\!+\!\!\xi_{\beta}\!\!-\!\!z)}{\theta_1(\xi_{\alpha}\!\!+\!\!\xi_{\beta})}
\frac{\theta_1(\xi_{\beta}\!\!-\!\!\xi_{\alpha}\!\!-\!\!z)}{\theta_1(\xi_{\beta}\!\!-\!\!\xi_{\alpha})}
\frac{\theta_1(\xi_{\beta}\!\!-\!\!z)}{\theta_1(\xi_{\beta})}
\nonumber\\
Z_{(1,0,+)}^{O(3),N}&=&-\frac{1}{2}\frac{\theta_1(z)}{\theta_1(2z)}
\prod_{\alpha=1}^{N}
\frac{\theta_2(\xi_{\alpha}\!\!-\!\!2z)}{\theta_2(\xi_{\alpha}\!\!+\!\!z)}
\!\!+\!\!\frac{1}{2}\sum_{\alpha=1}^{N}
\frac{\theta_1(2\xi_{\alpha}\!\!-\!\!z)}{\theta_1(2\xi_{\alpha})}
\frac{\theta_2(\xi_{\alpha})}{\theta_2(\xi_{\alpha}\!\!+\!\!z)}
\prod_{\beta\neq\alpha}
\frac{\theta_1(\xi_{\alpha}\!\!+\!\!\xi_{\beta}\!\!-\!\!z)}{\theta_1(\xi_{\alpha}\!\!+\!\!\xi_{\beta})}
\frac{\theta_1(\xi_{\beta}\!\!-\!\!\xi_{\alpha}\!\!-\!\!z)}{\theta_1(\xi_{\beta}\!\!-\!\!\xi_{\alpha})}
\frac{\theta_2(\xi_{\beta}\!\!-\!\!z)}{\theta_2(\xi_{\beta})}
\nonumber\\
Z_{(0,1,+)}^{O(3),N}&=&-\frac{1}{2}\frac{\theta_1(z)}{\theta_1(2z)}
\prod_{\alpha=1}^{N}
\frac{\theta_4(\xi_{\alpha}\!\!-\!\!2z)}{\theta_4(\xi_{\alpha}\!\!+\!\!z)}
\!\!+\!\!\frac{1}{2}\sum_{\alpha=1}^{N}
\frac{\theta_1(2\xi_{\alpha}\!\!-\!\!z)}{\theta_1(2\xi_{\alpha})}
\frac{\theta_4(\xi_{\alpha})}{\theta_4(\xi_{\alpha}\!\!+\!\!z)}
\prod_{\beta\neq\alpha}
\frac{\theta_1(\xi_{\alpha}\!\!+\!\!\xi_{\beta}\!\!-\!\!z)}{\theta_1(\xi_{\alpha}\!\!+\!\!\xi_{\beta})}
\frac{\theta_1(\xi_{\beta}\!\!-\!\!\xi_{\alpha}\!\!-\!\!z)}{\theta_1(\xi_{\beta}\!\!-\!\!\xi_{\alpha})}
\frac{\theta_4(\xi_{\beta}\!\!-\!\!z)}{\theta_4(\xi_{\beta})}
\nonumber\\
Z_{(1,1,+)}^{O(3),N}&=&-\frac{1}{2}\frac{\theta_1(z)}{\theta_1(2z)}
\prod_{\alpha=1}^{N}
\frac{\theta_3(\xi_{\alpha}\!\!-\!\!2z)}{\theta_3(\xi_{\alpha}\!\!+\!\!z)}
\!\!+\!\!\frac{1}{2}\sum_{\alpha=1}^{N}
\frac{\theta_1(2\xi_{\alpha}\!\!-\!\!z)}{\theta_1(2\xi_{\alpha})}
\frac{\theta_3(\xi_{\alpha})}{\theta_3(\xi_{\alpha}\!\!+\!\!z)}
\prod_{\beta\neq\alpha}
\frac{\theta_1(\xi_{\alpha}\!\!+\!\!\xi_{\beta}\!\!-\!\!z)}{\theta_1(\xi_{\alpha}\!\!+\!\!\xi_{\beta})}
\frac{\theta_1(\xi_{\beta}\!\!-\!\!\xi_{\alpha}\!\!-\!\!z)}{\theta_1(\xi_{\beta}\!\!-\!\!\xi_{\alpha})}
\frac{\theta_3(\xi_{\beta}\!\!-\!\!z)}{\theta_3(\xi_{\beta})}
\nonumber\\
\end{eqnarray}
and
\begin{eqnarray}
Z_{(0,0,-)}^{O(3),N}&=&\frac{1}{2}
\frac{\theta_1(z)}{\theta_1(2z)}
\prod_{\alpha=1}^{N}
\frac{\theta_1(\xi_{\alpha})}{\theta_1(\xi_{\alpha}-z)}
\frac{\theta_1(2\xi_{\alpha}-2z)}{\theta_1(2\xi_{\alpha})}\nonumber\\
Z_{(1,0,-)}^{O(3),N}&=&\frac{1}{2}
\frac{\theta_1(z)}{\theta_1(2z)}
\prod_{\alpha=1}^{N}
\frac{\theta_2(\xi_{\alpha})}{\theta_2(\xi_{\alpha}-z)}
\frac{\theta_1(2\xi_{\alpha}-2z)}{\theta_1(2\xi_{\alpha})}\nonumber\\
Z_{(0,1,-)}^{O(3),N}&=&\frac{1}{2}
\frac{\theta_1(z)}{\theta_1(2z)}
\prod_{\alpha=1}^{N}
\frac{\theta_4(\xi_{\alpha})}{\theta_4(\xi_{\alpha}-z)}
\frac{\theta_1(2\xi_{\alpha}-2z)}{\theta_1(2\xi_{\alpha})}\nonumber\\
Z_{(1,1,-)}^{O(3),N}&=&\frac{1}{2}
\frac{\theta_1(z)}{\theta_1(2z)}
\prod_{\alpha=1}^{N}
\frac{\theta_3(\xi_{\alpha})}{\theta_3(\xi_{\alpha}-z)}
\frac{\theta_1(2\xi_{\alpha}-2z)}{\theta_1(2\xi_{\alpha})}\nonumber\\\label{Zn}
\end{eqnarray}
so that $Z_{(kl)}^{O(3),N}=Z_{(k,l,+)}^{O(3),N}+(-1)^{N}Z_{(k,l,-)}^{O(3),N}$ can be evaluated. Note that in eq.(\ref{Zn}), $\frac{\theta_1(2\xi_{\alpha}-2z)}{\theta_1(2\xi_{\alpha})}$ can also be written as,
\begin{eqnarray}
\frac{\theta_1(2\xi_{\alpha}-2z)}{\theta_1(2\xi_{\alpha})}
=
\frac{\theta_1(\xi_{\alpha}-z)\theta_2(\xi_{\alpha}-z)\theta_3(\xi_{\alpha}-z)\theta_4(\xi_{\alpha}-z)}{\theta_1(\xi_{\alpha})\theta_2(\xi_{\alpha})\theta_3(\xi_{\alpha})\theta_4(\xi_{\alpha})}.
\end{eqnarray}

 The expected elliptic genus matching that we want to check is
\begin{eqnarray}
Z^{M,2}&=&Z_{(00)}^{O(3),2}=Z_{(10)}^{O(3),2}=Z_{(01)}^{O(3),2}=Z_{(11)}^{O(3),2}\nonumber\\
&=&\frac{\theta_1(2\xi_1-z)}{\theta_1(2\xi_1)}
\frac{\theta_1(\xi_1+\xi_2-z)}{\theta_1(\xi_1+\xi_2)}
\frac{\theta_1(2\xi_2-z)}{\theta_1(2\xi_2)}
\end{eqnarray}
where
\begin{eqnarray}
Z_{(00)}^{O(3),2}&=&\frac{1}{2}\frac{\theta_1(z)}{\theta_1(2z)}
\Bigg(
\prod_{\alpha=1}^{2}\frac{\theta_2(\xi_{\alpha}\!\!-\!\!z)\theta_3(\xi_{\alpha}\!\!-\!\!z)\theta_4(\xi_{\alpha}\!\!-\!\!z)}{\theta_2(\xi_{\alpha})\theta_3(\xi_{\alpha})\theta_4(\xi_{\alpha})}
-
\prod_{\alpha=1}^{2}\frac{\theta_1(\xi_{\alpha}\!\!-\!\!2z)}{\theta_1(\xi_{\alpha}\!\!+\!\!z)}
\Bigg)\nonumber\\
&&+\frac{1}{2}\sum_{\beta\neq\alpha}^{2}
\frac{\theta_1(2\xi_{\alpha}\!\!-\!\!z)}{\theta_1(2\xi_{\alpha})}
\frac{\theta_1(\xi_{\alpha}\!\!+\!\!\xi_{\beta}\!\!-\!\!z)}{\theta_1(\xi_{\alpha}\!\!+\!\!\xi_{\beta})}
\frac{\theta_1(\xi_{\beta}\!\!-\!\!\xi_{\alpha}\!\!-\!\!z)}{\theta_1(\xi_{\beta}\!\!-\!\!\xi_{\alpha})}
\frac{\theta_1(\xi_{\alpha})}{\theta_1(\xi_{\alpha}\!\!+\!\!z)}
\frac{\theta_1(\xi_{\beta}\!\!-\!\!z)}{\theta_1(\xi_{\beta})}
\nonumber\\
Z_{(10)}^{O(3),2}&=&\frac{1}{2}\frac{\theta_1(z)}{\theta_1(2z)}
\Bigg(
\prod_{\alpha=1}^{2}\frac{\theta_1(\xi_{\alpha}\!\!-\!\!z)\theta_3(\xi_{\alpha}\!\!-\!\!z)\theta_4(\xi_{\alpha}\!\!-\!\!z)}{\theta_1(\xi_{\alpha})\theta_3(\xi_{\alpha})\theta_4(\xi_{\alpha})}
-
\prod_{\alpha=1}^{2}\frac{\theta_2(\xi_{\alpha}\!\!-\!\!2z)}{\theta_2(\xi_{\alpha}\!\!+\!\!z)}
\Bigg)\nonumber\\
&&+\frac{1}{2}\sum_{\beta\neq\alpha}^{2}
\frac{\theta_1(2\xi_{\alpha}\!\!-\!\!z)}{\theta_1(2\xi_{\alpha})}
\frac{\theta_1(\xi_{\alpha}\!\!+\!\!\xi_{\beta}\!\!-\!\!z)}{\theta_1(\xi_{\alpha}\!\!+\!\!\xi_{\beta})}
\frac{\theta_1(\xi_{\beta}\!\!-\!\!\xi_{\alpha}\!\!-\!\!z)}{\theta_1(\xi_{\beta}\!\!-\!\!\xi_{\alpha})}
\frac{\theta_2(\xi_{\alpha})}{\theta_2(\xi_{\alpha}\!\!+\!\!z)}
\frac{\theta_2(\xi_{\beta}\!\!-\!\!z)}{\theta_2(\xi_{\beta})}
\nonumber\\
Z_{(01)}^{O(3),2}&=&\frac{1}{2}\frac{\theta_1(z)}{\theta_1(2z)}
\Bigg(
\prod_{\alpha=1}^{2}\frac{\theta_1(\xi_{\alpha}\!\!-\!\!z)\theta_2(\xi_{\alpha}\!\!-\!\!z)\theta_3(\xi_{\alpha}\!\!-\!\!z)}{\theta_1(\xi_{\alpha})\theta_2(\xi_{\alpha})\theta_3(\xi_{\alpha})}
-
\prod_{\alpha=1}^{2}\frac{\theta_4(\xi_{\alpha}\!\!-\!\!2z)}{\theta_4(\xi_{\alpha}\!\!+\!\!z)}
\Bigg)\nonumber\\
&&+\frac{1}{2}\sum_{\beta\neq\alpha}^{2}
\frac{\theta_1(2\xi_{\alpha}\!\!-\!\!z)}{\theta_1(2\xi_{\alpha})}
\frac{\theta_1(\xi_{\alpha}\!\!+\!\!\xi_{\beta}\!\!-\!\!z)}{\theta_1(\xi_{\alpha}\!\!+\!\!\xi_{\beta})}
\frac{\theta_1(\xi_{\beta}\!\!-\!\!\xi_{\alpha}\!\!-\!\!z)}{\theta_1(\xi_{\beta}\!\!-\!\!\xi_{\alpha})}
\frac{\theta_4(\xi_{\alpha})}{\theta_4(\xi_{\alpha}\!\!+\!\!z)}
\frac{\theta_4(\xi_{\beta}\!\!-\!\!z)}{\theta_4(\xi_{\beta})}
\nonumber\\
Z_{(11)}^{O(3),2}&=&\frac{1}{2}\frac{\theta_1(z)}{\theta_1(2z)}
\Bigg(
\prod_{\alpha=1}^{2}\frac{\theta_1(\xi_{\alpha}\!\!-\!\!z)\theta_2(\xi_{\alpha}\!\!-\!\!z)\theta_4(\xi_{\alpha}\!\!-\!\!z)}{\theta_1(\xi_{\alpha})\theta_2(\xi_{\alpha})\theta_4(\xi_{\alpha})}
-
\prod_{\alpha=1}^{2}\frac{\theta_3(\xi_{\alpha}\!\!-\!\!2z)}{\theta_3(\xi_{\alpha}\!\!+\!\!z)}
\Bigg)\nonumber\\
&&+\frac{1}{2}\sum_{\beta\neq\alpha}^{2}
\frac{\theta_1(2\xi_{\alpha}\!\!-\!\!z)}{\theta_1(2\xi_{\alpha})}
\frac{\theta_1(\xi_{\alpha}\!\!+\!\!\xi_{\beta}\!\!-\!\!z)}{\theta_1(\xi_{\alpha}\!\!+\!\!\xi_{\beta})}
\frac{\theta_1(\xi_{\beta}\!\!-\!\!\xi_{\alpha}\!\!-\!\!z)}{\theta_1(\xi_{\beta}\!\!-\!\!\xi_{\alpha})}
\frac{\theta_3(\xi_{\alpha})}{\theta_3(\xi_{\alpha}\!\!+\!\!z)}
\frac{\theta_3(\xi_{\beta}\!\!-\!\!z)}{\theta_3(\xi_{\beta})}\label{ZO32}
\end{eqnarray}
\subsubsection{$Z_{(00)}^{O(3),2}=Z^{M,2}$}
Let's start with $Z_{(00)}^{O(3),2}$. Consider $Q(z)\equiv Z_{(00)}^{O(3),2}/Z^{M,2}$. We will check that $Q(z)$ is a doubly periodic function and has no pole. It can be easily checked that $Q(z)$ is a doubly periodic function using the periodicities of the Jacobi theta functions.
\begin{eqnarray}
Q(z+1)=Q(z+\tau)=Q(z).
\end{eqnarray}
We should check that $Q$ has no poles. Since the Jacobi theta functions have no poles in $z$, there are 9 points where $Q$ has the possibility of poles  $z=2\xi_1,2\xi_2,\xi_1+\xi_2,-\xi_1,-\xi_2,\frac{a+b\tau}{2}$ ,where $a,b=0,1$. However, we only check $z=2\xi_1,\xi_1+\xi_2,-\xi_1,\frac{a+b\tau}{2}$ since $Q$ is symmetric under the change $\xi_1\leftrightarrow\xi_2$.
\begin{enumerate}
\item $z=2\xi_1$

Since $Z^{M,2} \sim \mathcal{O}((z-2\xi_1)^1)$ as $z\rightarrow 2\xi_1$, it's enough to show that $Z_{(00)}^{O(3),2}(z\!\!=\!\!2\xi_1)=0$ to check $Q$ is not divergent at $z=2\xi_1$ since  $Z_{(00)}^{O(3),2}$ admits Laurent expansion at $z=2\xi_1$.  From the eq.(\ref{ZO32}),
\begin{eqnarray}
Z_{(00)}^{O(3),2}(z\!\!=\!\!2\xi_1)&=&
\frac{1}{2}
\frac{\theta_1(2\xi_1)}{\theta_1(4\xi_1)}
\frac{\theta_1(\xi_2\!\!-\!\!4\xi_1)}{\theta_1(\xi_2\!\!+\!\!2\xi_1)}
+\frac{1}{2}
\frac{\theta_1(2\xi_1)}{\theta_1(4\xi_1)}
\frac{\theta_2(\xi_2\!\!-2\xi_1)}{\theta_2(\xi_2)}
\frac{\theta_3(\xi_2\!\!-2\xi_1)}{\theta_3(\xi_2)}
\frac{\theta_4(\xi_2\!\!-2\xi_1)}{\theta_4(\xi_2)}\nonumber\\
&&-\frac{1}{2}
\frac{\theta_1(2\xi_2\!\!-\!\!2\xi_1)}{\theta_1(2\xi_2)}
\frac{\theta_1(\xi_2)}{\theta_1(\xi_2+2\xi_1)}\nonumber\\
&=&\frac{1}{2}
\Bigg[
\frac{\theta_1(\xi_2\!\!-\!\!4\xi_1)\theta_2(\xi_2)\theta_3(\xi_2)\theta_4(\xi_2)\!\!+\!\!\theta_1(\xi_2\!\!+\!\!2\xi_1)\theta_2(\xi_2\!\!-\!\!2\xi_1)\theta_3(\xi_2\!\!-\!\!2\xi_1)\theta_4(\xi_2\!\!-\!\!2\xi_1)}{\theta_1(\xi_2\!\!+\!\!2\xi_1)\theta_2(\xi_2)\theta_3(\xi_2)\theta_4(\xi_2)}
\nonumber\\
&&\times\frac{\theta_1(2\xi_1)}{\theta_1(4\xi_1)}-\frac{\theta_1(2\xi_2\!\!-\!\!2\xi_1)\theta_2\theta_3\theta_4}{2\theta_1(\xi_2\!\!+\!\!2\xi_1)\theta_2(\xi_2)\theta_3(\xi_2)\theta_4(\xi_2)}
\Bigg]\nonumber\\
&=&
\Big[
\theta_1(\xi_2\!\!-\!\!4\xi_1)\theta_2(\xi_2)\theta_3(\xi_2)\theta_4(\xi_2)\!\!+\!\!\theta_1(\xi_2\!\!+\!\!2\xi_1)\theta_2(\xi_2\!\!-\!\!2\xi_1)\theta_3(\xi_2\!\!-\!\!2\xi_1)\theta_4(\xi_2\!\!-\!\!2\xi_1)\nonumber\\
&&-\theta_1(2\xi_2\!\!-\!\!2\xi_1)\theta_2(2\xi_1)\theta_3(2\xi_1)\theta_4(2\xi_1)
\Big]\frac{\theta_1(2\xi_1)}{\theta_1(4\xi_1)}\Big/2\theta_1(\xi_2\!\!+\!\!2\xi_1)\theta_2(\xi_2)\theta_3(\xi_2)\theta_4(\xi_2)\nonumber\\
&=&\frac{1}{2}\frac{\theta_1(2\xi_1)}{\theta_1(4\xi_1)}\frac{1}{\theta_1(\xi_2\!\!+\!\!2\xi_1)\theta_2(\xi_2)\theta_3(\xi_2)\theta_4(\xi_2)}\times\nonumber\\
&&\Bigg[
\frac{1}{2}\theta_1(\xi_2\!\!-\!\!2\xi_1)\theta_2(\xi_2\!\!-\!\!2\xi_1)\Big(\theta_3(\xi_2\!\!+\!\!2\xi_1)\theta_4(\xi_2\!\!-\!\!2\xi_1)+\theta_3(\xi_2\!\!-\!\!2\xi_1)\theta_4(\xi_2\!\!+\!\!2\xi_1)\Big)\nonumber\\
&&-\frac{1}{2}\Big(\theta_1(\xi_2\!\!+\!\!2\xi_1)\theta_2(\xi_2\!\!-\!\!2\xi_1)-\theta_1(\xi_2\!\!-\!\!2\xi_1)\theta_2(\xi_2\!\!+\!\!2\xi_1)\Big)\theta_3(\xi_2\!\!-\!\!2\xi_1)\theta_4(\xi_2\!\!-\!\!2\xi_1)\nonumber\\
&&+\theta_1(\xi_2\!\!+\!\!2\xi_1)\theta_2(\xi_2\!\!-\!\!2\xi_1)\theta_3(\xi_2\!\!-\!\!2\xi_1)\theta_4(\xi_2\!\!-\!\!2\xi_1)\nonumber\\
&&-\frac{1}{2}\Big(\theta_1(\xi_2\!\!+\!\!2\xi_1)\theta_2(\xi_2\!\!-\!\!2\xi_1)+\theta_1(\xi_2\!\!-\!\!2\xi_1)\theta_2(\xi_2\!\!+\!\!2\xi_1)\Big)\theta_3(\xi_2\!\!-\!\!2\xi_1)\theta_4(\xi_2\!\!-\!\!2\xi_1)\nonumber\\
&&-\frac{1}{2}\theta_1(\xi_2\!\!-\!\!2\xi_1)\theta_1(\xi_2\!\!-\!\!2\xi_1)\Big(\theta_3(\xi_2\!\!+\!\!2\xi_1)\theta_4(\xi_2\!\!-\!\!2\xi_1)+\theta_3(\xi_2\!\!-\!\!2\xi_1)\theta_4(\xi_2\!\!+\!\!2\xi_1)\Big)\Bigg]\nonumber\\
&=&0.
\end{eqnarray}
This shows that $Q$ has no pole at $z=2\xi_1$.
\item $z=\xi_1+\xi_2$

It's enough to show that $Z_{(00)}^{O(3),2}(z\!\!=\!\!\xi_1\!\!+\!\!\xi_2)=0$ to check $Q$ is not divergent at $z=\xi_1+\xi_2$ for the same reason as before. Just simple substitution of $z=\xi_1+\xi_2$ at eq.(\ref{ZO32}) gives $Z_{(00)}^{O(3),2}(z\!\!=\!\!\xi_1\!\!+\!\!\xi_2)=0$. So, $Q$ has no pole at $z=\xi_1+\xi_2$.

\item $z=-\xi_1$

Since $Z^{M,2}$ is non-zero at $z=-\xi_1$, we just need to check the residue of $Z_{(00)}^{O(3),2}$ at $z=-\xi_1$ is zero. Since the theta function is linear at small value $\theta_1(z) \sim z^1$, the residue evaluation is equivalent to $\lim_{z\rightarrow -\xi_1}\theta_1(\xi_1\!\!+\!\!z)Z_{(00)}^{O(3),2}$ up to some overall factor. We checked
\begin{eqnarray}
\lim_{z\rightarrow -\xi_1}\theta_1(\xi_1\!\!+\!\!z)Z_{(00)}^{O(3),2}
&=&0.
\end{eqnarray}
This shows the residue of $Z_{(00)}^{O(3),2}$ at $z=-\xi_1$ is zero so $Z_{(00)}^{O(3),2}$ is finite, which means $Q$ is finite at $z=-\xi_1$.

\item $z=\frac{a+b\tau}{2}$

Since $Z^{M,2}$ is non-zero at $z=\frac{a+b\tau}{2}$, we just need to show $\lim_{z\rightarrow \frac{a+b\tau}{2}}\theta_1(2z)Z_{(00)}^{O(3),2}=0$ as before. We checked
\begin{eqnarray}
\lim_{z\rightarrow \frac{a+b\tau}{2}}
\theta_1(2z)
Z_{(00)}^{O(3),2}=0
\end{eqnarray}
for $a,b=0,1$. This shows the residue of $Z_{(00)}^{O(3),2}$ at $z=\frac{a+b\tau}{2}$ is zero, which means $Q$ is finite at $z=\frac{a+b\tau}{2}$ for $a,b=0,1$.
\end{enumerate}

We checked that $Q$ has no pole so that $Q$ is a doubly periodic entire function of $z$ and it is independent of $z$ by the theorem. Then, we can evaluate
\begin{eqnarray}
Q(z)=\lim_{z\rightarrow 0}Q(z)=1
\end{eqnarray}
which means $Z_{(00)}^{O(3),2}=Z^{M,2}$.

\subsubsection{$Z_{(10)}^{O(3),2}=Z^{M,2}$}
We do similar works for the case $Z_{(10)}^{O(3),2}=Z^{M,2}$. Define $Q$ as $Q(z)\equiv Z_{(10)}^{O(3),2}/Z^{M,2}$. Then one can easily check that $Q(z+1)=Q(z+\tau)=Q(z)$ using the periodicity of the Jacobi theta functions which means $Q$ is a doubly periodic function of $z$. There are 9 points where $Q$ has the possibility of poles $z=2\xi_1,2\xi_2,\xi_1+\xi_2,-\xi_1+\frac{1}{2},-\xi_2+\frac{1}{2},\frac{a+b\tau}{2}$ ,where $a,b=0,1$. However, we only check $z=2\xi_1,\xi_1+\xi_2,-\xi_1+\frac{1}{2},\frac{a+b\tau}{2}$ since $Q$ is symmetric under the change $\xi_1\leftrightarrow\xi_2$.
\begin{enumerate}
\item $z=2\xi_1$

Since $Z^{M,2} \sim \mathcal{O}((z-2\xi_1)^1)$ as $z\rightarrow 2\xi_1$, it's enough to show that $Z_{(10)}^{O(3),2}(z\!\!=\!\!2\xi_1)=0$ to check $Q$ is not divergent at $z=2\xi_1$. From the eq.(\ref{ZO32}),
\begin{eqnarray}
Z_{(10)}^{O(3),2}(z\!\!=\!\!2\xi_1)&=&
-\frac{1}{2}\frac{\theta_1(2\xi_1)}{\theta_1(4\xi_1)}
\Bigg[
\frac{\theta_1(\xi_2\!\!-\!\!2\xi_1)\theta_3(\xi_2\!\!-\!\!2\xi_1)\theta_4(\xi_2\!\!-\!\!2\xi_1)}{\theta_1(\xi_2)\theta_3(\xi_2)\theta_4(\xi_2)}
\!\!+\!\!
\frac{\theta_2(\xi_2\!\!-\!\!4\xi_1)}{\theta_2(\xi_2\!\!+\!\!2\xi_1)}
\Bigg]
\nonumber\\
&&+\frac{1}{2}
\frac{\theta_1(2\xi_2\!\!-\!\!2\xi_1)}{\theta_1(2\xi_2)}
\frac{\theta_2(\xi_2)}{\theta_2(\xi_2\!\!+\!\!2\xi_1)}\nonumber\\
&=&\frac{-\theta_1(2\xi_1)}{2\theta_1(4\xi_1)\theta_1(\xi_2)\theta_3(\xi_2)\theta_4(\xi_2)\theta_2(\xi_2\!\!+\!\!2\xi_1)}
\Big[
\theta_1(\xi_2\!\!-\!\!2\xi_1)
\theta_2(\xi_2\!\!+\!\!2\xi_1)
\theta_3(\xi_2\!\!-\!\!2\xi_1)
\theta_4(\xi_2\!\!-\!\!2\xi_1)
\nonumber\\
&&+
\theta_1(\xi_2)
\theta_2(\xi_2\!\!-\!\!4\xi_1)
\theta_3(\xi_2)
\theta_4(\xi_2)
\!\!-\!\!
\theta_1(2\xi_2\!\!-\!\!2\xi_1)
\theta_2(2\xi_1)
\theta_3(2\xi_1)
\theta_4(2\xi_1)
\Big]\nonumber\\
\end{eqnarray}
The last two terms in the square bracket become
\begin{eqnarray}
&&\frac{1}{\theta_3\theta_4}
\Big\{
\theta_1(\xi_2\!\!-\!\!2\xi_1)
\theta_2(\xi_2\!\!-\!\!2\xi_1)
\theta_3(2\xi_1)
\theta_4(2\xi_1)
\!\!+\!\!
\theta_1(2\xi_1)
\theta_2(2\xi_1)
\theta_3(\xi_2\!\!-\!\!2\xi_1)
\theta_4(\xi_2\!\!-\!\!2\xi_1)
\Big\}
\theta_3(\xi_2)
\theta_4(\xi_2)
\nonumber\\
&&-\frac{1}{\theta_3\theta_4}
\Big\{
\theta_1(\xi_2)
\theta_2(\xi_2)
\theta_3(\xi_2\!\!-\!\!2\xi_1)
\theta_4(\xi_2\!\!-\!\!2\xi_1)
\!\!+\!\!
\theta_1(\xi_2\!\!-\!\!2\xi_1)
\theta_2(\xi_2\!\!-\!\!2\xi_1)
\theta_3(\xi_2)
\theta_4(\xi_2)
\Big\}
\theta_3(2\xi_1)
\theta_4(2\xi_1)
\nonumber\\
&&=-\frac{1}{\theta_3\theta_4}
\Big\{
\theta_1(\xi_2)
\theta_2(\xi_2)
\theta_3(2\xi_1)
\theta_4(2\xi_1)
\!\!-\!\!
\theta_1(2\xi_1)
\theta_2(2\xi_1)
\theta_3(\xi_2)
\theta_4(\xi_2)
\Big\}
\theta_3(\xi_2\!\!-\!\!2\xi_1)
\theta_4(\xi_2\!\!-\!\!2\xi_1)
\nonumber\\
&&=-
\theta_1(\xi_2\!\!-\!\!2\xi_1)
\theta_2(\xi_2\!\!+\!\!2\xi_1)
\theta_3(\xi_2\!\!-\!\!2\xi_1)
\theta_4(\xi_2\!\!-\!\!2\xi_1)
\end{eqnarray}
so that $Z_{(10)}^{O(3),2}(z=2\xi_1)$ becomes zero which means $Q$ has no pole at $z=2\xi_1$.

\item $z=\xi_1+\xi_2$

It's enough to show that $Z_{(10)}^{O(3),2}(z\!\!=\!\!\xi_1\!\!+\!\!\xi_2)=0$ to check $Q$ is not divergent at $z=\xi_1+\xi_2$. Just simple substitution of $z=\xi_1+\xi_2$ at eq.(\ref{ZO32}) gives $Z_{(10)}^{O(3),2}(z\!\!=\!\!\xi_1\!\!+\!\!\xi_2)=0$. So, Q has no pole at $z=\xi_1+\xi_2$.

\item $z=-\xi_1+\frac{1}{2}$

$Z^{M,2}$ is non-zero at $z=-\xi_1+\frac{1}{2}$.
So we  checked
\begin{eqnarray}
\lim_{z\rightarrow -\xi_1+\frac{1}{2}}\theta_2(\xi_1\!\!+\!\!z)Z_{(10)}^{O(3),2}
&=&
0
\end{eqnarray}
which means the residue of $Z_{(10)}^{O(3),2}$ at $z\!\!=\!\!-\xi_1\!\!+\!\!\frac{1}{2}$ is zero so that $Q$ is finite at $z\!\!=\!\!-\xi_1\!\!+\!\!\frac{1}{2}$.

\item $z=\frac{a+b\tau}{2}$

Since $Z^{M,2}$ is non-zero at $z=\frac{a+b\tau}{2}$, we just need to show $\lim_{z\rightarrow \frac{a+b\tau}{2}}\theta_1(2z)Z_{(10)}^{O(3),2}=0$. We checked
\begin{eqnarray}
\lim_{z\rightarrow \frac{a+b\tau}{2}}
\theta_1(2z)
Z_{(10)}^{O(3),2}=0
\end{eqnarray}
for $a,b=0,1$. This shows the residue of $Z_{(10)}^{O(3),2}$ at $z=\frac{a+b\tau}{2}$ is zero, which means $Q$ is finite at $z=\frac{a+b\tau}{2}$ for $a,b=0,1$.
\end{enumerate}

We checked that $Q$ has no pole so that $Q$ is a doubly periodic entire function of $z$ and it is independent of $z$ by the theorem. Then, we can evaluate
\begin{eqnarray}
Q(z)=\lim_{z\rightarrow 0}Q(z)=1
\end{eqnarray}
which means $Z_{(10)}^{O(3),2}=Z^{M,2}$.

\subsubsection{$Z_{(01)}^{O(3),2}=Z^{M,2}$}
We do similar works for the case $Z_{(01)}^{O(3),2}=Z^{M,2}$. Define $Q$ as $Q(z)\equiv Z_{(01)}^{O(3),2}/Z^{M,2}$. Then one can easily check that $Q(z+1)=Q(z+\tau)=Q(z)$ using the periodicity of the Jacobi theta functions which means $Q$ is a doubly periodic function of $z$. There are 9 points where $Q$ has the possibility of poles $z=2\xi_1,2\xi_2,\xi_1+\xi_2,-\xi_1+\frac{\tau}{2},-\xi_2+\frac{\tau}{2},\frac{a+b\tau}{2}$ ,where $a,b=0,1$. However, we only check $z=2\xi_1,\xi_1+\xi_2,-\xi_1+\frac{\tau}{2},\frac{a+b\tau}{2}$ since $Q$ is symmetric under the change $\xi_1\leftrightarrow\xi_2$.
\begin{enumerate}
\item $z=2\xi_1$

Since $Z^{M,2} \sim \mathcal{O}((z-2\xi_1)^1)$ as $z\rightarrow 2\xi_1$, it's enough to show that $Z_{(01)}^{O(3),2}(z\!\!=\!\!2\xi_1)=0$ to check $Q$ is not divergent at $z=2\xi_1$. From the eq.(\ref{ZO32}),
\begin{eqnarray}
Z_{(01)}^{O(3),2}(z\!\!=\!\!2\xi_1)&=&
-\frac{1}{2}\frac{\theta_1(2\xi_1)}{\theta_1(4\xi_1)}
\Bigg[
\frac{\theta_1(\xi_2\!\!-\!\!2\xi_1)\theta_2(\xi_2\!\!-\!\!2\xi_1)\theta_3(\xi_2\!\!-\!\!2\xi_1)}{\theta_1(\xi_2)\theta_2(\xi_2)\theta_3(\xi_2)}
\!\!+\!\!
\frac{\theta_4(\xi_2\!\!-\!\!4\xi_1)}{\theta_4(\xi_2\!\!+\!\!2\xi_1)}
\Bigg]
\nonumber\\
&&+\frac{1}{2}
\frac{\theta_1(2\xi_2\!\!-\!\!2\xi_1)}{\theta_1(2\xi_2)}
\frac{\theta_4(\xi_2)}{\theta_4(\xi_2\!\!+\!\!2\xi_1)}\nonumber\\
&=&\frac{-\theta_1(2\xi_1)}{2\theta_1(4\xi_1)\theta_1(\xi_2)\theta_2(\xi_2)\theta_3(\xi_2)\theta_4(\xi_2\!\!+\!\!2\xi_1)}
\Big[
\theta_1(\xi_2\!\!-\!\!2\xi_1)
\theta_2(\xi_2\!\!-\!\!2\xi_1)
\theta_3(\xi_2\!\!-\!\!2\xi_1)
\theta_4(\xi_2\!\!+\!\!2\xi_1)
\nonumber\\
&&+
\theta_1(\xi_2)
\theta_2(\xi_2)
\theta_3(\xi_2)
\theta_4(\xi_2\!\!-\!\!4\xi_1)
\!\!-\!\!
\theta_1(2\xi_2\!\!-\!\!2\xi_1)
\theta_2(2\xi_1)
\theta_3(2\xi_1)
\theta_4(2\xi_1)
\Big]\nonumber\\
\end{eqnarray}
The last two terms in the square bracket become
\begin{eqnarray}
&&\frac{1}{\theta_3\theta_4}
\theta_1(\xi_2)
\theta_2(\xi_2)
\Big\{
\theta_3(\xi_2\!\!-\!\!2\xi_1)
\theta_4(\xi_2\!\!-\!\!2\xi_1)
\theta_3(2\xi_1)
\theta_4(2\xi_1)
\!\!-\!\!
\theta_1(2\xi_1)
\theta_2(2\xi_1)
\theta_1(\xi_2\!\!-\!\!2\xi_1)
\theta_2(\xi_2\!\!-\!\!2\xi_1)
\Big\}
\nonumber\\
&&-\frac{1}{\theta_3\theta_4}
\Big\{
\theta_1(\xi_2)
\theta_2(\xi_2)
\theta_3(\xi_2\!\!-\!\!2\xi_1)
\theta_4(\xi_2\!\!-\!\!2\xi_1)
\!\!+\!\!
\theta_1(\xi_2\!\!-\!\!2\xi_1)
\theta_2(\xi_2\!\!-\!\!2\xi_1)
\theta_3(\xi_2)
\theta_4(\xi_2)
\Big\}
\theta_3(2\xi_1)
\theta_4(2\xi_1)
\nonumber\\
&&=-\frac{1}{\theta_3\theta_4}
\theta_1(\xi_2\!\!-\!\!2\xi_1)
\theta_2(\xi_2\!\!-\!\!2\xi_1)
\Big\{
\theta_1(\xi_2)
\theta_2(\xi_2)
\theta_1(2\xi_1)
\theta_2(2\xi_1)
\!\!+\!\!
\theta_3(2\xi_1)
\theta_4(2\xi_1)
\theta_3(\xi_2)
\theta_4(\xi_2)
\Big\}
\nonumber\\
&&=-
\theta_1(\xi_2\!\!-\!\!2\xi_1)
\theta_2(\xi_2\!\!-\!\!2\xi_1)
\theta_3(\xi_2\!\!-\!\!2\xi_1)
\theta_4(\xi_2\!\!+\!\!2\xi_1)
\end{eqnarray}
so that $Z_{(01)}^{O(3),2}(z=2\xi_1)$ becomes zero which means $Q$ has no pole at $z=2\xi_1$.

\item $z=\xi_1+\xi_2$

It's enough to show that $Z_{(01)}^{O(3),2}(z=\xi_1+\xi_2)=0$ to check $Q$ is not divergent at $z=\xi_1+\xi_2$. Just simple substitution of $z=\xi_1+\xi_2$ at eq.(\ref{ZO32}) gives $Z_{(01)}^{O(3),2}(z\!\!=\!\!\xi_1\!\!+\!\!\xi_2)=0$. So, Q has no pole at $z=\xi_1+\xi_2$.

\item $z=-\xi_1+\frac{\tau}{2}$

We checked
\begin{eqnarray}
\lim_{z\rightarrow -\xi_1+\frac{\tau}{2}}\theta_4(\xi_1\!\!+\!\!z)Z_{(01)}^{O(3),2}&=&0
\end{eqnarray}
which means the residue of $Z_{(01)}^{O(3),2}$ at $z\!\!=\!\!-\xi_1\!\!+\!\!\frac{\tau}{2}$ is zero so that $Q$ is finite at $z\!\!=\!\!-\xi_1\!\!+\!\!\frac{\tau}{2}$.

\item $z=\frac{a+b\tau}{2}$

Since $Z^{M,2}$ is non-zero at $z=\frac{a+b\tau}{2}$, we just need to show $\lim_{z\rightarrow \frac{a+b\tau}{2}}\theta_1(2z)Z_{(01)}^{O(3),2}=0$. We checked
\begin{eqnarray}
\lim_{z\rightarrow \frac{a+b\tau}{2}}
\theta_1(2z)
Z_{(01)}^{O(3),2}=0
\end{eqnarray}
for $a,b=0,1$. This shows the residue of $Z_{(01)}^{O(3),2}$ at $z=\frac{a+b\tau}{2}$ is zero, which means $Q$ is finite at $z=\frac{a+b\tau}{2}$ for $a,b=0,1$.
\end{enumerate}

We checked that $Q$ has no pole so that $Q$ is a doubly periodic entire function of $z$ and it is independent of $z$ by the theorem. Then, we can evaluate
\begin{eqnarray}
Q(z)=\lim_{z\rightarrow 0}Q(z)=1
\end{eqnarray}
which means $Z_{(01)}^{O(3),2}=Z^{M,2}$.

\subsubsection{$Z_{(11)}^{O(3),2}=Z^{M,2}$}
We do similar works for the case $Z_{(11)}^{O(3),2}=Z^{M,2}$. Define $Q$ as $Q(z)\equiv Z_{(11)}^{O(3),2}/Z^{M,2}$. Then one can easily check that $Q(z+1)=Q(z+\tau)=Q(z)$ using the periodicity of the Jacobi theta functions which means $Q$ is a doubly periodic function of $z$. There are 9 points where $Q$ has the possibility of poles $z=2\xi_1,2\xi_2,\xi_1+\xi_2,-\xi_1+\frac{1+\tau}{2},-\xi_2+\frac{1+\tau}{2},\frac{a+b\tau}{2}$ ,where $a,b=0,1$. However, we only check $z=2\xi_1,\xi_1+\xi_2,-\xi_1+\frac{1+\tau}{2},\frac{a+b\tau}{2}$ since $Q$ is symmetric under the change $\xi_1\leftrightarrow\xi_2$.
\begin{enumerate}
\item $z=2\xi_1$

Since $Z^{M,2} \sim \mathcal{O}((z-2\xi_1)^1)$ as $z\rightarrow 2\xi_1$, it's enough to show that $Z_{(11)}^{O(3),2}(z\!\!=\!\!2\xi_1)=0$ to check $Q$ is not divergent at $z=2\xi_1$. From the eq.(\ref{ZO32}),
\begin{eqnarray}
Z_{(11)}^{O(3),2}(z\!\!=\!\!2\xi_1)&=&
-\frac{1}{2}\frac{\theta_1(2\xi_1)}{\theta_1(4\xi_1)}
\Bigg[
\frac{\theta_1(\xi_2\!\!-\!\!2\xi_1)\theta_2(\xi_2\!\!-\!\!2\xi_1)\theta_4(\xi_2\!\!-\!\!2\xi_1)}{\theta_1(\xi_2)\theta_2(\xi_2)\theta_4(\xi_2)}
\!\!+\!\!
\frac{\theta_3(\xi_2\!\!-\!\!4\xi_1)}{\theta_3(\xi_2\!\!+\!\!2\xi_1)}
\Bigg]
\nonumber\\
&&+\frac{1}{2}
\frac{\theta_1(2\xi_2\!\!-\!\!2\xi_1)}{\theta_1(2\xi_2)}
\frac{\theta_3(\xi_2)}{\theta_3(\xi_2\!\!+\!\!2\xi_1)}\nonumber\\
&=&\frac{-\theta_1(2\xi_1)}{2\theta_1(4\xi_1)\theta_1(\xi_2)\theta_2(\xi_2)\theta_4(\xi_2)\theta_3(\xi_2\!\!+\!\!2\xi_1)}
\Big[
\theta_1(\xi_2\!\!-\!\!2\xi_1)
\theta_2(\xi_2\!\!-\!\!2\xi_1)
\theta_3(\xi_2\!\!+\!\!2\xi_1)
\theta_4(\xi_2\!\!-\!\!2\xi_1)
\nonumber\\
&&+
\theta_1(\xi_2)
\theta_2(\xi_2)
\theta_3(\xi_2\!\!-\!\!4\xi_1)
\theta_4(\xi_2)
\!\!-\!\!
\theta_1(2\xi_2\!\!-\!\!2\xi_1)
\theta_2(2\xi_1)
\theta_3(2\xi_1)
\theta_4(2\xi_1)
\Big]\nonumber\\
\end{eqnarray}
The last two terms in the square bracket become
\begin{eqnarray}
&&\frac{1}{\theta_3\theta_4}
\theta_1(\xi_2)
\theta_2(\xi_2)
\Big\{
\theta_3(\xi_2\!\!-\!\!2\xi_1)
\theta_4(\xi_2\!\!-\!\!2\xi_1)
\theta_3(2\xi_1)
\theta_4(2\xi_1)
\!\!+\!\!
\theta_1(2\xi_1)
\theta_2(2\xi_1)
\theta_1(\xi_2\!\!-\!\!2\xi_1)
\theta_2(\xi_2\!\!-\!\!2\xi_1)
\Big\}
\nonumber\\
&&-\frac{1}{\theta_3\theta_4}
\Big\{
\theta_1(\xi_2)
\theta_2(\xi_2)
\theta_3(\xi_2\!\!-\!\!2\xi_1)
\theta_4(\xi_2\!\!-\!\!2\xi_1)
\!\!+\!\!
\theta_1(\xi_2\!\!-\!\!2\xi_1)
\theta_2(\xi_2\!\!-\!\!2\xi_1)
\theta_3(\xi_2)
\theta_4(\xi_2)
\Big\}
\theta_3(2\xi_1)
\theta_4(2\xi_1)
\nonumber\\
&&=-\frac{1}{\theta_3\theta_4}
\theta_1(\xi_2\!\!-\!\!2\xi_1)
\theta_2(\xi_2\!\!-\!\!2\xi_1)
\Big\{
\theta_3(2\xi_1)
\theta_4(2\xi_1)
\theta_3(\xi_2)
\theta_4(\xi_2)
\!\!-\!\!
\theta_1(\xi_2)
\theta_2(\xi_2)
\theta_1(2\xi_1)
\theta_2(2\xi_1)
\Big\}
\nonumber\\
&&=-
\theta_1(\xi_2\!\!-\!\!2\xi_1)
\theta_2(\xi_2\!\!-\!\!2\xi_1)
\theta_3(\xi_2\!\!+\!\!2\xi_1)
\theta_4(\xi_2\!\!-\!\!2\xi_1)
\end{eqnarray}
so that $Z_{(11)}^{O(3),2}(z=2\xi_1)$ becomes zero which means $Q$ has no pole at $z=2\xi_1$.

\item $z=\xi_1+\xi_2$

It's enough to show that $Z_{(11)}^{O(3),2}(z\!\!=\!\!\xi_1\!\!+\!\!\xi_2)=0$ to check $Q$ is not divergent at $z=\xi_1+\xi_2$. Just simple substitution of $z=\xi_1+\xi_2$ at eq.(\ref{ZO32}) gives $Z_{(11)}^{O(3),2}(z\!\!=\!\!\xi_1\!\!+\!\!\xi_2)=0$. So, Q has no pole at $z=\xi_1+\xi_2$.

\item $z=-\xi_1+\frac{1+\tau}{2}$

$Z^{M,2}$ is non-zero at $z=-\xi_1+\frac{1+\tau}{2}$. We checked
\begin{eqnarray}
\lim_{z\rightarrow -\xi_1+\frac{1+\tau}{2}}\theta_3(\xi_1\!\!+\!\!z)Z_{(11)}^{O(3),2}&=&0,
\end{eqnarray}
which means the residue of $Z_{(11)}^{O(3),2}$ at $z\!\!=\!\!-\xi_1\!\!+\!\!\frac{1+\tau}{2}$ is zero so that $Q$ is finite at $z\!\!=\!\!-\xi_1\!\!+\!\!\frac{1+\tau}{2}$.

\item $z=\frac{a+b\tau}{2}$

Since $Z^{M,2}$ is non-zero at $z=\frac{a+b\tau}{2}$, we just need to show $\lim_{z\rightarrow \frac{a+b\tau}{2}}\theta_1(2z)Z_{(11)}^{O(3),2}=0$. We checked
\begin{eqnarray}
\lim_{z\rightarrow \frac{a+b\tau}{2}}
\theta_1(2z)
Z_{(11)}^{O(3),2}=0
\end{eqnarray}
for $a,b=0,1$. This shows the residue of $Z_{(11)}^{O(3),2}$ at $z=\frac{a+b\tau}{2}$ is zero, which means $Q$ is finite at $z=\frac{a+b\tau}{2}$ for $a,b=0,1$.
\end{enumerate}

We checked that $Q$ has no pole so that $Q$ is a doubly periodic entire function of $z$ and it is independent of $z$ by the theorem. Then, we can evaluate
\begin{eqnarray}
Q(z)=\lim_{z\rightarrow 0}Q(z)=1
\end{eqnarray}
which means $Z_{(11)}^{O(3),2}=Z^{M,2}$.

To summarize we checked analytically that
\begin{eqnarray}
Z_{(00)}^{O(3),2}=Z_{(10)}^{O(3),2}=Z_{(01)}^{O(3),2}=Z_{(11)}^{O(3),2}=Z^{M,2}
\end{eqnarray}
which gives
\begin{eqnarray}
Z^{A,SO(3),2}&=&Z_{(00)}^{O(3),2}=Z^{M,2}\\
Z^{A,O_{+}(3),2}&=&\frac{1}{2}\Big(Z_{(00)}^{O(3),2}+Z_{(10)}^{O(3),2}+Z_{(01)}^{O(3),2}+Z_{(11)}^{O(3),2}\Big)=2Z^{M,2}\\
Z^{A,O_{-}(3),2}&=&\frac{1}{2}\Big(-Z_{(00)}^{O(3),2}+Z_{(10)}^{O(3),2}+Z_{(01)}^{O(3),2}+Z_{(11)}^{O(3),2}\Big)=Z^{M,2}.
\end{eqnarray}

\subsection{$SO(4), O_{\pm}(4)$ theories with 3 fundamentals}

Using the Jacobi theta function's relations we can get
\begin{eqnarray}
Z_{(00)}^{O(4),N}&=&
\frac{1}{4}
\Bigg[
\frac{1}{2}
\Bigg(\!\!
\frac{\theta_1(z)}{\theta_1(2z)}
\!\!\Bigg)^2
\!\!\Bigg\{\!\!
\prod_{\alpha=1}^{N}\!\!
\frac{\theta_1(\xi_{\alpha}\!\!-\!\!2z)}{\theta_1(\xi_{\alpha}\!\!+\!\!z)}
\frac{\theta_1(\xi_{\alpha}\!\!-\!\!z)}{\theta_1(\xi_{\alpha})}
\!\!+\!\!
\prod_{\alpha=1}^{N}\!\!
\frac{\theta_2(\xi_{\alpha}\!\!-\!\!2z)}{\theta_2(\xi_{\alpha}\!\!+\!\!z)}
\frac{\theta_2(\xi_{\alpha}\!\!-\!\!z)}{\theta_2(\xi_{\alpha})}
\!\!+\!\!
\prod_{\alpha=1}^{N}\!\!
\frac{\theta_3(\xi_{\alpha}\!\!-\!\!2z)}{\theta_3(\xi_{\alpha}\!\!+\!\!z)}
\frac{\theta_3(\xi_{\alpha}\!\!-\!\!z)}{\theta_3(\xi_{\alpha})}\nonumber\\
&&
\!\!+\!\!
\prod_{\alpha=1}^{N}
\frac{\theta_4(\xi_{\alpha}\!\!-\!\!2z)}{\theta_4(\xi_{\alpha}\!\!+\!\!z)}
\frac{\theta_4(\xi_{\alpha}\!\!-\!\!z)}{\theta_4(\xi_{\alpha})}
\Bigg\}
-
\sum_{\alpha=1}^{N}
\frac{\theta_1(2\xi_{\alpha}\!\!-\!\!z)}{\theta_1(2\xi_{\alpha})}
\frac{\theta_1(2\xi_{\alpha}\!\!+\!\!z)}{\theta_1(2\xi_{\alpha}\!\!+\!\!2z)}
\prod_{\beta\neq\alpha}
\frac{\theta_1(\xi_{\beta}\!\!-\!\!\xi_{\alpha}\!\!-\!\!2z)}{\theta_1(\xi_{\beta}\!\!-\!\!\xi_{\alpha})}
\frac{\theta_1(\xi_{\alpha}\!\!+\!\!\xi_{\beta}\!\!-\!\!z)}{\theta_1(\xi_{\alpha}\!\!+\!\!\xi_{\beta}\!\!+\!\!z)}\nonumber\\
&&
+
\sum_{\alpha=1}^{N}\sum_{\beta\neq\alpha}
\frac{\theta_1(2\xi_{\alpha}\!\!-\!\!z)}{\theta_1(2\xi_{\alpha})}
\frac{\theta_1(2\xi_{\beta}\!\!-\!\!z)}{\theta_1(2\xi_{\beta})}
\frac{\theta_1(\xi_{\alpha}\!\!+\!\!\xi_{\beta}\!\!-\!\!z)}{\theta_1(\xi_{\alpha}\!\!+\!\!\xi_{\beta}\!\!+\!\!z)}
\prod_{\gamma\neq\alpha,\beta}
\prod_{\lambda=\alpha,\beta}
\frac{\theta_1(\xi_{\gamma}\!\!-\!\!\xi_{\lambda}\!\!-\!\!z)}{\theta_1(\xi_{\gamma}\!\!-\!\!\xi_{\lambda})}
\frac{\theta_1(\xi_{\gamma}\!\!+\!\!\xi_{\lambda}\!\!-\!\!z)}{\theta_1(\xi_{\gamma}\!\!+\!\!\xi_{\lambda})}
\Bigg]\nonumber\\
&&
+(-1)^{N+1}
\frac{1}{2}
\Bigg(
\frac{\theta_1(z)}{\theta_1(2z)}
\Bigg)^2
\prod_{\alpha=1}^{N}
\frac{\theta_1(2\xi_{\alpha}\!\!-\!\!2z)}{\theta_1(2\xi_{\alpha})}
\end{eqnarray}
\begin{eqnarray}
Z_{(10)}^{O(4),N}&=&
\frac{1}{4}
\sum_{\alpha=1}^{N}
\frac{\theta_1(2\xi_{\alpha}\!\!-\!\!z)}{\theta_1(2\xi_{\alpha})}
\frac{\theta_2}{\theta_2(z)}
\prod_{\beta\neq\alpha}
\frac{\theta_1(\xi_{\beta}\!\!-\!\!\xi_{\alpha}\!\!-\!\!z)}{\theta_1(\xi_{\beta}\!\!-\!\!\xi_{\alpha})}
\frac{\theta_1(\xi_{\alpha}\!\!+\!\!\xi_{\beta}\!\!-\!\!z)}{\theta_1(\xi_{\alpha}\!\!+\!\!\xi_{\beta})}\times\nonumber\\
&&\Bigg(\!\!
\frac{\theta_1(\xi_{\alpha})}{\theta_1(\xi_{\alpha}\!\!+\!\!z)}
\frac{\theta_2(\xi_{\alpha})}{\theta_2(\xi_{\alpha}\!\!+\!\!z)}
\!\prod_{\beta\neq\alpha}\!
\frac{\theta_1(\xi_{\beta}\!\!-\!\!z)}{\theta_1(\xi_{\beta})}
\frac{\theta_2(\xi_{\beta}\!\!-\!\!z)}{\theta_2(\xi_{\beta})}
\!\!+\!\!(-1)^{N+1}\!
\frac{\theta_3(\xi_{\alpha})}{\theta_3(\xi_{\alpha}\!\!+\!\!z)}
\frac{\theta_4(\xi_{\alpha})}{\theta_4(\xi_{\alpha}\!\!+\!\!z)}
\!\prod_{\beta\neq\alpha}\!
\frac{\theta_3(\xi_{\beta}\!\!-\!\!z)}{\theta_3(\xi_{\beta})}
\frac{\theta_4(\xi_{\beta}\!\!-\!\!z)}{\theta_4(\xi_{\beta})}
\!\!\Bigg)\nonumber\\
&&-\frac{1}{4}
\frac{\theta_1(z)}{\theta_1(2z)}
\frac{\theta_2(z)}{\theta_2(2z)}\times
\Bigg(\!
\prod_{\alpha=1}^{N}
\frac{\theta_1(\xi_{\alpha}\!\!-\!\!2z)}{\theta_1(\xi_{\alpha}\!\!+\!\!z)}
\frac{\theta_2(\xi_{\alpha}\!\!-\!\!z)}{\theta_2(\xi_{\alpha})}
\!\!+(-1)^{N+1}\!\!
\prod_{\alpha=1}^{N}
\frac{\theta_3(\xi_{\alpha}\!\!-\!\!2z)}{\theta_3(\xi_{\alpha}\!\!+\!\!z)}
\frac{\theta_4(\xi_{\alpha}\!\!-\!\!z)}{\theta_4(\xi_{\alpha})}\nonumber\\
&&\qquad\qquad\qquad\qquad\qquad
+\!\!
\prod_{\alpha=1}^{N}
\frac{\theta_2(\xi_{\alpha}\!\!-\!\!2z)}{\theta_2(\xi_{\alpha}\!\!+\!\!z)}
\frac{\theta_1(\xi_{\alpha}\!\!-\!\!z)}{\theta_1(\xi_{\alpha})}
\!\!+(-1)^{N+1}\!\!
\prod_{\alpha=1}^{N}
\frac{\theta_4(\xi_{\alpha}\!\!-\!\!2z)}{\theta_4(\xi_{\alpha}\!\!+\!\!z)}
\frac{\theta_3(\xi_{\alpha}\!\!-\!\!z)}{\theta_3(\xi_{\alpha})}
\!\!\Bigg)\nonumber\\
\end{eqnarray}
\begin{eqnarray}
Z_{(01)}^{O(4),N}&=&
\frac{1}{4}
\sum_{\alpha=1}^{N}
\frac{\theta_1(2\xi_{\alpha}\!\!-\!\!z)}{\theta_1(2\xi_{\alpha})}
\frac{\theta_4}{\theta_4(z)}
\prod_{\beta\neq\alpha}
\frac{\theta_1(\xi_{\beta}\!\!-\!\!\xi_{\alpha}\!\!-\!\!z)}{\theta_1(\xi_{\beta}\!\!-\!\!\xi_{\alpha})}
\frac{\theta_1(\xi_{\alpha}\!\!+\!\!\xi_{\beta}\!\!-\!\!z)}{\theta_1(\xi_{\alpha}\!\!+\!\!\xi_{\beta})}\times\nonumber\\
&&\Bigg(\!\!
\frac{\theta_1(\xi_{\alpha})}{\theta_1(\xi_{\alpha}\!\!+\!\!z)}
\frac{\theta_4(\xi_{\alpha})}{\theta_4(\xi_{\alpha}\!\!+\!\!z)}
\!\prod_{\beta\neq\alpha}\!
\frac{\theta_1(\xi_{\beta}\!\!-\!\!z)}{\theta_1(\xi_{\beta})}
\frac{\theta_4(\xi_{\beta}\!\!-\!\!z)}{\theta_4(\xi_{\beta})}
\!\!+\!\!(-1)^{N+1}\!
\frac{\theta_2(\xi_{\alpha})}{\theta_2(\xi_{\alpha}\!\!+\!\!z)}
\frac{\theta_3(\xi_{\alpha})}{\theta_3(\xi_{\alpha}\!\!+\!\!z)}
\!\prod_{\beta\neq\alpha}\!
\frac{\theta_2(\xi_{\beta}\!\!-\!\!z)}{\theta_2(\xi_{\beta})}
\frac{\theta_3(\xi_{\beta}\!\!-\!\!z)}{\theta_3(\xi_{\beta})}
\!\!\Bigg)\nonumber\\
&&-\frac{1}{4}
\frac{\theta_1(z)}{\theta_1(2z)}
\frac{\theta_4(z)}{\theta_4(2z)}\times
\Bigg(\!
\prod_{\alpha=1}^{N}
\frac{\theta_1(\xi_{\alpha}\!\!-\!\!2z)}{\theta_1(\xi_{\alpha}\!\!+\!\!z)}
\frac{\theta_4(\xi_{\alpha}\!\!-\!\!z)}{\theta_4(\xi_{\alpha})}
\!\!+(-1)^{N+1}\!\!
\prod_{\alpha=1}^{N}
\frac{\theta_2(\xi_{\alpha}\!\!-\!\!2z)}{\theta_2(\xi_{\alpha}\!\!+\!\!z)}
\frac{\theta_3(\xi_{\alpha}\!\!-\!\!z)}{\theta_3(\xi_{\alpha})}\nonumber\\
&&\qquad\qquad\qquad\qquad\qquad
+\!\!
\prod_{\alpha=1}^{N}
\frac{\theta_4(\xi_{\alpha}\!\!-\!\!2z)}{\theta_4(\xi_{\alpha}\!\!+\!\!z)}
\frac{\theta_1(\xi_{\alpha}\!\!-\!\!z)}{\theta_1(\xi_{\alpha})}
\!\!+(-1)^{N+1}\!\!
\prod_{\alpha=1}^{N}
\frac{\theta_3(\xi_{\alpha}\!\!-\!\!2z)}{\theta_3(\xi_{\alpha}\!\!+\!\!z)}
\frac{\theta_2(\xi_{\alpha}\!\!-\!\!z)}{\theta_2(\xi_{\alpha})}
\!\!\Bigg)\nonumber\\
\end{eqnarray}
\begin{eqnarray}
Z_{(11)}^{O(4),N}&=&
\frac{1}{4}
\sum_{\alpha=1}^{N}
\frac{\theta_1(2\xi_{\alpha}\!\!-\!\!z)}{\theta_1(2\xi_{\alpha})}
\frac{\theta_3}{\theta_3(z)}
\prod_{\beta\neq\alpha}
\frac{\theta_1(\xi_{\beta}\!\!-\!\!\xi_{\alpha}\!\!-\!\!z)}{\theta_1(\xi_{\beta}\!\!-\!\!\xi_{\alpha})}
\frac{\theta_1(\xi_{\alpha}\!\!+\!\!\xi_{\beta}\!\!-\!\!z)}{\theta_1(\xi_{\alpha}\!\!+\!\!\xi_{\beta})}\times\nonumber\\
&&\Bigg(\!\!
\frac{\theta_1(\xi_{\alpha})}{\theta_1(\xi_{\alpha}\!\!+\!\!z)}
\frac{\theta_3(\xi_{\alpha})}{\theta_3(\xi_{\alpha}\!\!+\!\!z)}
\!\prod_{\beta\neq\alpha}\!
\frac{\theta_1(\xi_{\beta}\!\!-\!\!z)}{\theta_1(\xi_{\beta})}
\frac{\theta_3(\xi_{\beta}\!\!-\!\!z)}{\theta_3(\xi_{\beta})}
\!\!+\!\!(-1)^{N+1}\!
\frac{\theta_2(\xi_{\alpha})}{\theta_2(\xi_{\alpha}\!\!+\!\!z)}
\frac{\theta_4(\xi_{\alpha})}{\theta_4(\xi_{\alpha}\!\!+\!\!z)}
\!\prod_{\beta\neq\alpha}\!
\frac{\theta_2(\xi_{\beta}\!\!-\!\!z)}{\theta_2(\xi_{\beta})}
\frac{\theta_4(\xi_{\beta}\!\!-\!\!z)}{\theta_4(\xi_{\beta})}
\!\!\Bigg)\nonumber\\
&&-\frac{1}{4}
\frac{\theta_1(z)}{\theta_1(2z)}
\frac{\theta_3(z)}{\theta_3(2z)}\times
\Bigg(\!
\prod_{\alpha=1}^{N}
\frac{\theta_1(\xi_{\alpha}\!\!-\!\!2z)}{\theta_1(\xi_{\alpha}\!\!+\!\!z)}
\frac{\theta_3(\xi_{\alpha}\!\!-\!\!z)}{\theta_3(\xi_{\alpha})}
\!\!+(-1)^{N+1}\!\!
\prod_{\alpha=1}^{N}
\frac{\theta_2(\xi_{\alpha}\!\!-\!\!2z)}{\theta_2(\xi_{\alpha}\!\!+\!\!z)}
\frac{\theta_4(\xi_{\alpha}\!\!-\!\!z)}{\theta_4(\xi_{\alpha})}\nonumber\\
&&\qquad\qquad\qquad\qquad\qquad
+\!\!
\prod_{\alpha=1}^{N}
\frac{\theta_3(\xi_{\alpha}\!\!-\!\!2z)}{\theta_3(\xi_{\alpha}\!\!+\!\!z)}
\frac{\theta_1(\xi_{\alpha}\!\!-\!\!z)}{\theta_1(\xi_{\alpha})}
\!\!+(-1)^{N+1}\!\!
\prod_{\alpha=1}^{N}
\frac{\theta_4(\xi_{\alpha}\!\!-\!\!2z)}{\theta_4(\xi_{\alpha}\!\!+\!\!z)}
\frac{\theta_2(\xi_{\alpha}\!\!-\!\!z)}{\theta_2(\xi_{\alpha})}
\!\!\Bigg).\nonumber\\
\end{eqnarray}

\subsubsection{$Z_{(00)}^{O(4),3}=Z^{M,3}$}

Consider $Q(z) \equiv Z_{(00)}^{O(4),3}/Z^{M,3}$. It can be easily checked that $Q(z)$ is a doubly periodic function under $z\rightarrow z+1, z\rightarrow z+\tau$. Again, we check that $Q(z)$ has no pole. There could be potential poles at $z=-\xi_{\alpha}+\frac{a+b\tau}{2},\frac{a+b\tau}{2},-\xi_{\alpha}-\xi_{\beta},\xi_{\alpha}+\xi_{\beta},2\xi_{\alpha}$ for $\alpha\neq\beta, a,b = 0,1$. Since $Q(z)$ is symmetric under $\xi_{\alpha}\leftrightarrow\xi_{\beta}$, we only need to check for $z=-\xi_1+\frac{a+b\tau}{2},\frac{a+b\tau}{2},-\xi_1-\xi_2,\xi_1+\xi_2,2\xi_1$.

\begin{enumerate}
\item $z=-\xi_1+\frac{a+b\tau}{2}$

The potential divergence at this point comes from $Z_{(00)}^{O(4),3}$ since $Z^{M,3}$ is nonzero at this point. So, it's sufficient to check that the residue of $Z_{(00)}^{O(4),3}$ at this point is zero to show $Q(z)$ has no pole. We checked
\begin{eqnarray}
\lim_{z\rightarrow-\xi_1+\frac{a+b\tau}{2}}\theta_i(\xi_1+z)Z_{(00)}^{O(4),3}=0
\end{eqnarray}
for $(a,b,i)=(0,0,1),(1,0,2),(0,1,4),(1,1,3)$ which means the residue of $Z_{(00)}^{O(4),3}$ at $z=-\xi_1+\frac{a+b\tau}{2}$ is zero for $a,b=0,1$ so that $Q(z)$ has no pole at $z=-\xi_1+\frac{a+b\tau}{2}$.

\item $z=\frac{a+b\tau}{2}$

By the same reason, it's sufficient to check that the residue of $Z_{(00)}^{O(4),3}$ at this point is zero to show $Q(z)$ has no pole.
\begin{eqnarray}
\lim_{z\rightarrow 0}Z_{(00)}^{O(4),3} = 1
\end{eqnarray}
\begin{eqnarray}
\lim_{\epsilon\rightarrow 0}\lim_{z\rightarrow\frac{1}{2}+\epsilon}\theta_1(2z)Z_{(00)}^{O(4),3}
&=&
-\lim_{\epsilon\rightarrow 0}
\frac{\theta_2(\epsilon)^2}{\theta_1(2\epsilon)}
\Bigg[
\frac{1}{8}
\Bigg\{
\prod_{\alpha=1}^{3}
\frac{\theta_2(\xi_{\alpha}\!\!-\!\!\epsilon)}{\theta_2(\xi_{\alpha}\!\!+\!\!\epsilon)}
\frac{\theta_1(\xi_{\alpha}\!\!-\!\!2\epsilon)}{\theta_1(\xi_{\alpha})}
+
\prod_{\alpha=1}^{3}
\frac{\theta_1(\xi_{\alpha}\!\!-\!\!\epsilon)}{\theta_1(\xi_{\alpha}\!\!+\!\!\epsilon)}
\frac{\theta_2(\xi_{\alpha}\!\!-\!\!2\epsilon)}{\theta_2(\xi_{\alpha})}
\nonumber\\
&&+\!\!
\prod_{\alpha=1}^{3}\!\!
\frac{\theta_3(\xi_{\alpha}\!\!-\!\!\epsilon)}{\theta_3(\xi_{\alpha}\!\!+\!\!\epsilon)}
\frac{\theta_4(\xi_{\alpha}\!\!-\!\!2\epsilon)}{\theta_4(\xi_{\alpha})}
\!\!+\!\!
\prod_{\alpha=1}^{3}\!\!
\frac{\theta_4(\xi_{\alpha}\!\!-\!\!\epsilon)}{\theta_4(\xi_{\alpha}\!\!+\!\!\epsilon)}
\frac{\theta_3(\xi_{\alpha}\!\!-\!\!2\epsilon)}{\theta_3(\xi_{\alpha})}
\Bigg\}
\!\!-\!\!\frac{1}{2}
\!\!\prod_{\alpha=1}^{3}\!\!
\frac{\theta_1(2\xi_{\alpha}\!\!-\!\!2\epsilon)}{\theta_1(2\xi_{\alpha})}
\Bigg]
\nonumber\\
\label{from2}
&=&0
\end{eqnarray}
where we used $\theta_i(\xi_{\alpha}+\epsilon)\sim\theta_i(\xi_{\alpha})+\epsilon\theta_i^{\prime}(\xi_{\alpha})$ for $i=1,2,3,4$ to see the square bracket is order of $\mathcal{O}(\epsilon^2)$.
\begin{eqnarray}
[\cdots]&=&\frac{1}{8}
\Bigg\{\
4-4\epsilon\sum_{\alpha=1}^{3}\sum_{i=1}^{4}
\Bigg(
\frac{\theta_i^{\prime}(\xi_{\alpha})}{\theta_i(\xi_{\alpha})}
\Bigg)
\Bigg\}
-\frac{1}{2}
\Bigg\{
1-\epsilon\sum_{\alpha=1}^{3}\sum_{i=1}^{4}
\Bigg(
\frac{\theta_i^{\prime}(\xi_{\alpha})}{\theta_i(\xi_{\alpha})}
\Bigg)
\Bigg\}
+
\mathcal{O}(\epsilon^2)\nonumber\\
&\sim&\mathcal{O}(\epsilon^2).
\end{eqnarray}
Similarly,
\begin{eqnarray}
\lim_{z\rightarrow \frac{\tau}{2}}\theta_1(2z)Z_{(00)}^{O(4),3}
&=&
\lim_{\epsilon\rightarrow 0}
\frac{\theta_4(\epsilon)^2}{\theta_1(2\epsilon)}
\frac{c}{q^{\frac{5}{4}}}
\Bigg[
\frac{1}{8}
\Bigg\{
\prod_{\alpha=1}^{3}
\frac{\theta_4(\xi_{\alpha}\!\!-\!\!\epsilon)}{\theta_4(\xi_{\alpha}\!\!+\!\!\epsilon)}
\frac{\theta_1(\xi_{\alpha}\!\!-\!\!2\epsilon)}{\theta_1(\xi_{\alpha})}
+
\prod_{\alpha=1}^{3}
\frac{\theta_3(\xi_{\alpha}\!\!-\!\!\epsilon)}{\theta_3(\xi_{\alpha}\!\!+\!\!\epsilon)}
\frac{\theta_2(\xi_{\alpha}\!\!-\!\!2\epsilon)}{\theta_2(\xi_{\alpha})}
\nonumber\\
&&+
\!\!\prod_{\alpha=1}^{3}\!\!
\frac{\theta_2(\xi_{\alpha}\!\!-\!\!\epsilon)}{\theta_2(\xi_{\alpha}\!\!+\!\!\epsilon)}
\frac{\theta_3(\xi_{\alpha}\!\!-\!\!2\epsilon)}{\theta_3(\xi_{\alpha})}
\!\!+\!\!
\prod_{\alpha=1}^{3}\!\!
\frac{\theta_1(\xi_{\alpha}\!\!-\!\!\epsilon)}{\theta_1(\xi_{\alpha}\!\!+\!\!\epsilon)}
\frac{\theta_4(\xi_{\alpha}\!\!-\!\!2\epsilon)}{\theta_4(\xi_{\alpha})}
\Bigg\}
\!\!-\!\!\frac{1}{2}
\prod_{\alpha=1}^{3}
\frac{\theta_1(2\xi_{\alpha}\!\!-\!\!2\epsilon)}{\theta_1(2\xi_{\alpha})}
\Bigg]
\nonumber\\
&=&0
\end{eqnarray}
\begin{eqnarray}
\lim_{z\rightarrow \frac{1+\tau}{2}}\theta_1(2z)Z_{(00)}^{O(4),3}
&=&
\lim_{\epsilon\rightarrow 0}
\frac{\theta_3(\epsilon)^2}{\theta_1(2\epsilon)}
\frac{c}{-q^{\frac{5}{4}}}
\Bigg[
\frac{1}{8}
\Bigg\{
\prod_{\alpha=1}^{3}
\frac{\theta_3(\xi_{\alpha}\!\!-\!\!\epsilon)}{\theta_3(\xi_{\alpha}\!\!+\!\!\epsilon)}
\frac{\theta_1(\xi_{\alpha}\!\!-\!\!2\epsilon)}{\theta_1(\xi_{\alpha})}
+
\prod_{\alpha=1}^{3}
\frac{\theta_4(\xi_{\alpha}\!\!-\!\!\epsilon)}{\theta_4(\xi_{\alpha}\!\!+\!\!\epsilon)}
\frac{\theta_2(\xi_{\alpha}\!\!-\!\!2\epsilon)}{\theta_2(\xi_{\alpha})}
\nonumber\\
&&+
\!\!\prod_{\alpha=1}^{3}\!\!
\frac{\theta_1(\xi_{\alpha}\!\!-\!\!\epsilon)}{\theta_1(\xi_{\alpha}\!\!+\!\!\epsilon)}
\frac{\theta_3(\xi_{\alpha}\!\!-\!\!2\epsilon)}{\theta_3(\xi_{\alpha})}
\!\!+\!\!
\prod_{\alpha=1}^{3}\!\!
\frac{\theta_2(\xi_{\alpha}\!\!-\!\!\epsilon)}{\theta_2(\xi_{\alpha}\!\!+\!\!\epsilon)}
\frac{\theta_4(\xi_{\alpha}\!\!-\!\!2\epsilon)}{\theta_4(\xi_{\alpha})}
\Bigg\}
\!\!-\!\!\frac{1}{2}
\prod_{\alpha=1}^{3}
\frac{\theta_1(2\xi_{\alpha}\!\!-\!\!2\epsilon)}{\theta_1(2\xi_{\alpha})}
\Bigg]
\nonumber\\
&=&0
\label{to2}
\end{eqnarray}
where $c=e^{4\pi i (\xi_1\!+\!\xi_2\!+\!\xi_3\!-\!3\epsilon)}$. So, the residue of $Z_{(00)}^{O(4),3}$ at $z=\frac{a+b\tau}{2}$ is zero for $a,b=0,1$ which means $Q(z)$ has no pole at $z=\frac{a+b\tau}{2}$.

\item $z=-\xi_1-\xi_2$

By the same reason, it's sufficient to check that the residue of $Z_{(00)}^{O(4),3}$ at this point is zero to show $Q(z)$ has no pole. We checked
\begin{eqnarray}
\lim_{z\rightarrow-\xi_1-\xi_2}\theta_1(\xi_1+\xi_2+z)Z_{(00)}^{O(4),3}
&=&0.
\end{eqnarray}
So, the residue of $Z_{(00)}^{O(4),3}$ at $z=-\xi_1-\xi_2$ is zero which means $Q(z)$ has no pole at $z=-\xi_1-\xi_2$.

\item $z=\xi_1+\xi_2$

The divergence at this point comes from $\frac{1}{Z^{M,3}}$ and the order of divergence of it is $\frac{1}{z-\xi_1-\xi_2}$ so that we only need to check that $Z_{(00)}^{O(4),3}|_{z=\xi_1+\xi_2}=0$ to show $Q(z)$ has no pole at here.
\begin{eqnarray}
Z_{(00)}^{O(4),3}|_{z=\xi_1+\xi_2}
&=&
\frac{1}{8}
\Bigg(
\frac{\theta_1(\xi_1\!\!+\!\!\xi_2)}{\theta_1(2\xi_1\!\!+\!\!2\xi_2)}
\Bigg)^2
\Bigg\{
\frac{\theta_1(\xi_1\!\!+\!\!\xi_2\!\!-\!\!\xi_3)}{\theta_1(\xi_1\!\!+\!\!\xi_2\!\!+\!\!\xi_3)}
\frac{\theta_1(2\xi_1\!\!+\!\!2\xi_2\!\!-\!\!\xi_3)}{\theta_1(\xi_3)}
+
\sum_{i=2}^{4}(\theta_1\leftrightarrow\theta_i)
\Bigg\}\nonumber\\
&&+\frac{1}{4}
\frac{\theta_1(\xi_1\!\!+\!\!\xi_2\!\!-\!\!2\xi_3)}{\theta_1(2\xi_3)}
\frac{\theta_1(\xi_1\!\!+\!\!\xi_2\!\!+\!\!2\xi_3)}{\theta_1(2\xi_1\!\!+\!\!2\xi_2\!\!+\!\!2\xi_3)}
\!\!-\!\!
\frac{1}{2}
\Bigg(
\frac{\theta_1(\xi_1\!\!+\!\!\xi_2)}{\theta_1(2\xi_1\!\!+\!\!2\xi_2)}
\Bigg)^2
\frac{\theta_1(2\xi_1\!\!+\!\!2\xi_2\!\!-\!\!2\xi_3)}{\theta_1(2\xi_3)}\nonumber\\
&=&
\frac{1}{2}
\Bigg(
\frac{\theta_1(\xi_1\!\!+\!\!\xi_2)}{\theta_1(2\xi_1\!\!+\!\!2\xi_2)}
\Bigg)^2
\frac{1}{(\theta_2\theta_3\theta_4)^2}
\frac{1}{\theta_1(2\xi_3)\theta_1(2\xi_1\!\!+\!\!2\xi_2\!\!+\!\!2\xi_3)}\times
\nonumber\\
&&
\Big[
\theta_1(\eta)\theta_2(\xi)\theta_3(\xi)\theta_4(\xi)\cdot
\theta_1(\xi)\theta_2(\eta)\theta_3(\eta)\theta_4(\eta)
\nonumber\\
&&
+\theta_1(\xi)\theta_2(\eta)\theta_3(\xi)\theta_4(\xi)\cdot
\theta_1(\eta)\theta_2(\xi)\theta_3(\eta)\theta_4(\eta)
\nonumber\\
&&+\theta_1(\xi)\theta_2(\xi)\theta_3(\eta)\theta_4(\xi)\cdot
\theta_1(\eta)\theta_2(\eta)\theta_3(\xi)\theta_4(\eta)
\nonumber\\
&&
+\theta_1(\xi)\theta_2(\xi)\theta_3(\xi)\theta_4(\eta)\cdot
\theta_1(\eta)\theta_2(\eta)\theta_3(\eta)\theta_4(\xi)
\Big]\nonumber\\
&&
-
\frac{1}{2}
\Bigg(
\frac{\theta_1(\xi_1\!\!+\!\!\xi_2)}{\theta_1(2\xi_1\!\!+\!\!2\xi_2)}
\Bigg)^2
\frac{\theta_1(2\xi_1\!\!+\!\!2\xi_2\!\!-\!\!2\xi_3)}{\theta_1(2\xi_3)}\nonumber\\
&=&
\frac{1}{2}\Bigg(
\frac{\theta_1(\xi_1\!\!+\!\!\xi_2)}{\theta_1(2\xi_1\!\!+\!\!2\xi_2)}
\Bigg)^2
\Bigg[
\frac{\theta_1(2\xi)\theta_1(2\eta)}{\theta_1(2\xi_3)\theta_1(2\xi_1\!\!+\!\!2\xi_2\!\!+\!\!2\xi_3)}
-
\frac{\theta_1(2\xi_1\!\!+\!\!2\xi_2\!\!-\!\!2\xi_3)}{\theta_1(2\xi_3)}
\Bigg]
\nonumber\\
&=&0.
\label{xi1xi2}
\end{eqnarray}
where we set $\eta\equiv\xi_1+\xi_2-\xi_3$ and $\xi\equiv\xi_1+\xi_2+\xi_3$. This result shows that $Q(z)$ has no pole at $z=\xi_1+\xi_2$.

\item $z=2\xi_1$

The divergence at this point comes from $\frac{1}{Z^{M,3}}$ and the order of divergence of it is $\frac{1}{z-2\xi_1}$ so that we only need to check that $Z_{(00)}^{O(4),3}|_{z=2\xi_1}=0$ to show $Q(z)$ has no pole.
\begin{eqnarray}
Z_{(00)}^{O(4),3}|_{z=2\xi_1}
&=&
\frac{1}{8}
\Bigg(
\frac{\theta_1(2\xi_1)}{\theta_1(4\xi_1)}
\Bigg)^2
\Bigg[
\Bigg\{
\frac{\theta_1(2\xi_1\!\!-\!\!\xi_2)\theta_1(4\xi_1\!\!-\!\!\xi_2)}{\theta_1(2\xi_1\!\!+\!\!\xi_2)\theta_1(\xi_2)}
\frac{\theta_1(2\xi_1\!\!-\!\!\xi_3)\theta_1(4\xi_1\!\!-\!\!\xi_3)}{\theta_1(2\xi_1\!\!+\!\!\xi_3)\theta_1(\xi_3)}
\!+\!
\sum_{i=2}^{4}(\theta_1\leftrightarrow\theta_i)
\Bigg\}
\nonumber\\
&&\!-\!4
\frac{\theta_1(4\xi_1\!\!-\!\!2\xi_2)}{\theta_1(2\xi_2)}
\frac{\theta_1(4\xi_1\!\!-\!\!2\xi_3)}{\theta_1(2\xi_3)}
\Bigg]
-
\frac{1}{4}
\frac{\theta_1(2\xi_1\!\!-\!\!\xi_2\!\!-\!\!\xi_3)}{\theta_1(2\xi_1\!\!+\!\!\xi_2\!\!+\!\!\xi_3)}
\Bigg\{
2
\frac{\theta_1(2\xi_1\!\!-\!\!2\xi_2)}{\theta_1(2\xi_2)}
\frac{\theta_1(2\xi_1\!\!-\!\!2\xi_3)}{\theta_1(2\xi_3)}
+
\nonumber\\
&&
\frac{\theta_1(2\xi_1\!\!-\!\!2\xi_2)}{\theta_1(2\xi_2)}
\frac{\theta_1(2\xi_1\!\!+\!\!2\xi_2)}{\theta_1(4\xi_1\!\!+\!\!2\xi_2)}
\frac{\theta_1(4\xi_1\!\!+\!\!\xi_2\!\!-\!\!\xi_3)}{\theta_1(\xi_2\!\!-\!\!\xi_3)}
\!\!-\!\!
\frac{\theta_1(2\xi_1\!\!-\!\!2\xi_3)}{\theta_1(2\xi_3)}
\frac{\theta_1(2\xi_1\!\!+\!\!2\xi_3)}{\theta_1(4\xi_1\!\!+\!\!2\xi_3)}
\frac{\theta_1(4\xi_1\!\!-\!\!\xi_2\!\!+\!\!\xi_3)}{\theta_1(\xi_2\!\!-\!\!\xi_3)}
\Bigg\}.
\nonumber\\
\end{eqnarray}
We define a function of $2\xi_1$ that is doubly periodic by dividing $Z_{(00)}^{O(4),3}|_{z=2\xi_1}$ with the first term in the second curly brackets.
\begin{eqnarray}
P_{(00)}(2\xi_1)
\equiv
Z_{(00)}^{O(4),3}|_{z=2\xi_1}\times
\frac{\theta_1(2\xi_1\!\!+\!\!\xi_2\!\!+\!\!\xi_3)}{\theta_1(2\xi_1\!\!-\!\!\xi_2\!\!-\!\!\xi_3)}
\frac{\theta_1(2\xi_2)}{\theta_1(2\xi_1\!\!-\!\!2\xi_2)}
\frac{\theta_1(2\xi_3)}{\theta_1(2\xi_1\!\!-\!\!2\xi_3)}.
\end{eqnarray}
We will show $P_{(00)}=0$ by checking that it has no pole. The potential pole points are $2\xi_1=2\xi_{\alpha},\xi_2+\xi_3,-\xi_{\alpha}+\frac{a+b\tau}{2},\frac{a+b\tau}{2}$ for $\alpha=2,3$ and $a,b=0,1$ and only $\alpha=2$ will be checked since $P_{(00)}$ is symmetric under $\xi_2\leftrightarrow\xi_3$. We checked
\begin{eqnarray}
\lim_{2\xi_1\rightarrow2\xi_2}
\theta_1(2\xi_1-2\xi_2)P_{(00)}
&=&0,
\label{00f}
\end{eqnarray}
where we used the eq.(\ref{xi1xi2}) with $\xi_1\rightarrow\xi_2$,
\begin{eqnarray}
\lim_{2\xi_1\rightarrow\xi_2+\xi_3}
\theta_1(2\xi_1\!\!-\!\!\xi_2\!\!-\!\!\xi_3)P_{(00)}
&=&
0
\\
\lim_{2\xi_1\rightarrow-\xi_2+\frac{a+b\tau}{2}}
\theta_i(2\xi_1\!\!+\!\!\xi_2)P_{(00)}
&=&0
\\
\lim_{2\xi_1\rightarrow\frac{a+b\tau}{2}}\theta_1(4\xi_1)P_{(00)}=0
\label{00t}
\end{eqnarray}
for $(a,b,i)=(0,0,1),(1,0,2),(0,1,4),(1,1,3)$ and $a,b=0,1$. We used eq.(\ref{from2})-eq.(\ref{to2}) with changing $z\rightarrow 2\xi_1$.

 Eq.(\ref{00f})-(\ref{00t}) show the residues of $P_{(00)}(2\xi_1)$ at the all potential poles are zero so that $P_{(00)}(2\xi_1)$ is a doubly periodic entire function of $2\xi_1$. So, we have
\begin{eqnarray}
P_{(00)}(2\xi_1)=\lim_{2\xi_1\rightarrow 0}P_{(00)}(2\xi_1)=0.
\label{00re0}
\end{eqnarray}
Multiplying $\frac{\theta_1(2\xi_1+\xi_2+\xi_3)}{\theta_1(2\xi_1-\xi_2-\xi_3)}
\frac{\theta_1(2\xi_2)}{\theta_1(2\xi_1-2\xi_2)}
\frac{\theta_1(2\xi_3)}{\theta_1(2\xi_1-2\xi_3)}$, we have $Z_{(00)}^{O(4),3}|_{z=2\xi_1}=0$ except for a point $2\xi_1=-\xi_2-\xi_3$. Since the residue of $Z_{(00)}^{O(4),3}|_{z=2\xi_1}$ at $2\xi_1=-\xi_2-\xi_3$ is zero(from the eq.(\ref{00re0})), $Z_{(00)}^{O(4),3}|_{z=2\xi_1}$ is continuous at this point. So, we finally have
\begin{eqnarray}
Z_{(00)}^{O(4),3}|_{z=2\xi_1}=0
\end{eqnarray}
for all $2\xi_1$. This result shows that $Q(z)$ has no pole at $z=2\xi_1$.
\end{enumerate}

We checked that $Q$ has no pole. So, $Q$ is a doubly periodic entire function of $z$ and it is independent of $z$ by the theorem. Then, we can easily evaluate
\begin{eqnarray}
Q(z)=\lim_{z\rightarrow 0}Q(z)=1
\end{eqnarray}
which means $Z_{(00)}^{O(4),3}=Z^{M,3}$.

\subsubsection{$Z_{(10)}^{O(4),3}=Z^{M,3}$}
Consider $Q(z) \equiv Z_{(10)}^{O(4),3}/Z^{M,3}$ which is a doubly periodic function under $z\rightarrow z+1, z\rightarrow z+\tau$. Again, we check that $Q(z)$ has no pole. There are many potential poles: $z=-\xi_{\alpha}+\frac{a+b\tau}{2},\frac{a+b\tau}{2},\frac{1}{4}+\frac{a+b\tau}{2},\xi_{\alpha}+\xi_{\beta},2\xi_{\alpha}$ for $\alpha\neq\beta, a,b = 0,1$. Since $Q(z)$ is symmetric under $\xi_{\alpha}\leftrightarrow\xi_{\beta}$, we only need to check for $z=-\xi_1+\frac{a+b\tau}{2},\frac{a+b\tau}{2},\frac{1}{4}+\frac{a+b\tau}{2},\xi_1+\xi_2,2\xi_1$.
\begin{enumerate}
\item $z=-\xi_1+\frac{a+b\tau}{2}$

The potential divergence at this point comes from $Z_{(10)}^{O(4),3}$ and $Z^{M,3}$ is nonzero at this point. So, it's sufficient to check that the residue of $Z_{(10)}^{O(4),3}$ at this point iz zero to show $Q(z)$ has no pole. We checked
\begin{eqnarray}
\lim_{z\rightarrow-\xi_1+\frac{a+b\tau}{2}}\theta_i(\xi_1+z)Z_{(10)}^{O(4),3}
&=&0
\label{from10}
\end{eqnarray}
where $(a,b,i)=(0,0,1),(1,0,2),(0,1,4),(1,1,3)$. So, the residue of $Z_{(10)}^{O(4),3}$ at $z=-\xi_1+\frac{a+b\tau}{2}$ is zero for $a,b=0,1$ which means $Q(z)$ has no pole at $z=-\xi_1+\frac{a+b\tau}{2}$.

\item $z=\frac{a+b\tau}{2}$

By the same reason, it's sufficient to check that the residue of $Z_{(10)}^{M,3}$ at this point is zero to show $Q(z)$ has no pole. We checked
\begin{eqnarray}
\lim_{z\rightarrow\frac{a+b\tau}{2}}\theta_1(2z)Z_{(10)}^{O(4),3}
&=&0
\end{eqnarray}
for $a,b,=0,1$. So, the residue of $Z_{(10)}^{O(4),3}$ at $z=\frac{a+b\tau}{2}$ is zero which means $Q(z)$ has no pole at $z=\frac{a+b\tau}{2}$.

\item $z=\frac{1}{4}+\frac{a+b\tau}{2}$

By the same reason, it's sufficient to check that the residue of $Z_{(10)}^{O(4),3}$ at this point is zero to show $Q(z)$ has no pole. We checked
\begin{eqnarray}
\lim_{z\rightarrow\frac{1}{4}+\frac{a+b\tau}{2}}
\theta_2(2z)Z_{(10)}^{O(4),3}
&=&0.\label{to10}
\end{eqnarray}
So, the residue of $Z_{(10)}^{O(4),3}$ at $z=\frac{1}{4}+\frac{a+b\tau}{2}$ is zero for $a,b=0,1$ which means $Q(z)$ has no pole at $z=\frac{1}{4}+\frac{a+b\tau}{2}$.

\item $z=\xi_1+\xi_2$

The divergence at this point comes from $\frac{1}{Z^{M,3}}$ and the order of divergence of it is $\frac{1}{z-\xi_1-\xi_2}$ so that we only need to check that $Z_{(10)}^{O(4),3}|_{z=\xi_1+\xi_2}=0$ to show $Q(z)$ has no pole.
\begin{eqnarray}
Z_{(10)}^{O(4),3}|_{z=\xi_1+\xi_2}
&=&
\frac{\theta_1(2\xi_3\!\!-\!\!z)\theta_1(2\xi_3\!\!+\!\!z)}{2\theta_1(2\xi_3)\theta_1(2\xi_3\!\!+\!\!2z)}
-\frac{\theta_1(z)\theta_2(z)}{4\theta_1(2z)\theta_2(2z)}\Bigg(
\frac{\theta_2(\xi_3\!\!-\!\!2z)\theta_1(\xi_3\!\!-\!\!z)}{\theta_2(\xi_3\!\!+\!\!z)\theta_1(\xi_3)}
\nonumber\\
&&+
\frac{\theta_1(\xi_3\!\!-\!\!2z)\theta_2(\xi_3\!\!-\!\!z)}{\theta_1(\xi_3\!\!+\!\!z)\theta_2(\xi_3)}
\!\!+\!\!
\frac{\theta_3(\xi_3\!\!-\!\!2z)\theta_4(\xi_3\!\!-\!\!z)}{\theta_3(\xi_3\!\!+\!\!z)\theta_4(\xi_3)}
\!\!+\!\!
\frac{\theta_4(\xi_3\!\!-\!\!2z)\theta_3(\xi_3\!\!-\!\!z)}{\theta_4(\xi_3\!\!+\!\!z)\theta_3(\xi_3)}
\Bigg)
\nonumber\\
&=&
\frac{\theta_1(2\xi_3\!\!-\!\!z)\theta_1(2\xi_3\!\!+\!\!z)}{2\theta_1(2\xi_3)\theta_1(2\xi_3\!\!+\!\!2z)}
-\frac{\theta_2(\theta_3\theta_4)^2}{8\theta_2(2z)}\times
\nonumber\\
&&
\frac{\theta_4(2\xi_3\!\!+\!\!z)\theta_4(2\xi_3\!\!-\!\!z)\theta_3(2z)\theta_3-\theta_3(2\xi_3\!\!+\!\!z)\theta_3(2\xi_3\!\!-\!\!z)\theta_4(2z)\theta_4}{\theta_1(\xi_3\!\!+\!\!z)\theta_2(\xi_3\!\!+\!\!z)\theta_3(\xi_3\!\!+\!\!z)\theta_4(\xi_3\!\!+\!\!z)\theta_1(\xi_3)\theta_2(\xi_3)\theta_3(\xi_3)\theta_4(\xi_3)}
\nonumber\\
&=&
\frac{\theta_1(2\xi_3\!\!-\!\!z)\theta_1(2\xi_3\!\!+\!\!z)}{2\theta_1(2\xi_3)\theta_1(2\xi_3\!\!+\!\!2z)}
\nonumber\\
&&-\frac{\theta_4(2\xi_3\!\!+\!\!z)\theta_4(2\xi_3\!\!-\!\!z)\theta_3(2z)\theta_3\!\!-\!\!\theta_3(2\xi_3\!\!+\!\!z)\theta_3(2\xi_3\!\!-\!\!z)\theta_4(2z)\theta_4}{2\theta_1(2\xi_3)\theta_1(2\xi_3\!\!+\!\!2z)\theta_2\theta_2(2z)}
\nonumber\\
&=&
\frac{\theta_1(2\xi_3\!\!-\!\!z)\theta_1(2\xi_3\!\!+\!\!z)}{2\theta_1(2\xi_3)\theta_1(2\xi_3\!\!+\!\!2z)}
-\frac{\theta_1(2\xi_3\!\!+\!\!z)\theta_2\theta_1(2\xi_3\!\!-\!\!z)\theta_2(2z)}{2\theta_1(2\xi_3)\theta_1(2\xi_3\!\!+\!\!2z)\theta_2\theta_2(2z)}
\nonumber\\
&=&0.
\label{10xi1xi2}
\end{eqnarray}
This result shows that $Q(z)$ has no pole at $z=\xi_1+\xi_2$.

\item $z=2\xi_1$

The divergence at this point comes from $\frac{1}{Z^{M,3}}$ and the order of divergence of it is $\frac{1}{z-2\xi_1}$ so that we only need to check that $Z_{(10)}^{O(4),3}|_{z=2\xi_1}=0$ to show $Q(z)$ has no pole.

\begin{eqnarray}
Z_{(10)}^{O(4),3}|_{z=2\xi_1}
&=&
\frac{1}{4}
\frac{\theta_1(2\xi_2\!\!-\!\!2\xi_1)}{\theta_1(2\xi_2)}
\frac{\theta_2}{\theta_2(2\xi_1)}
\frac{\theta_1(\xi_3\!\!-\!\!\xi_2\!\!-\!\!2\xi_1)\theta_1(\xi_2\!\!+\!\!\xi_3\!\!-\!\!2\xi_1)}{\theta_1(\xi_3\!\!-\!\!\xi_2)\theta_1(\xi_2\!\!+\!\!\xi_3)}\times\nonumber\\
&&
\Bigg(
\!\!-\!\!
\frac{\theta_1(\xi_2)\theta_2(\xi_2)\theta_1(\xi_3\!\!-\!\!2\xi_1)\theta_2(\xi_3\!\!-\!\!2\xi_1)}
{\theta_1(\xi_2\!\!+\!\!2\xi_1)\theta_2(\xi_2\!\!+\!\!2\xi_1)\theta_1(\xi_3)\theta_2(\xi_3)}
\!\!+\!\!
\frac{\theta_3(\xi_2)\theta_4(\xi_2)\theta_3(\xi_3\!\!-\!\!2\xi_1)\theta_4(\xi_3\!\!-\!\!2\xi_1)}
{\theta_3(\xi_2\!\!+\!\!2\xi_1)\theta_4(\xi_2\!\!+\!\!2\xi_1)\theta_3(\xi_3)\theta_4(\xi_3)}
\Bigg)
\nonumber\\
&&+(\xi_2\leftrightarrow\xi_3)
\nonumber\\
&&-\frac{1}{4}
\frac{\theta_1(2\xi_1)\theta_2(2\xi_1)}{\theta_1(4\xi_1)\theta_2(4\xi_1)}
\Bigg(
\!\!-\!\!
\prod_{\alpha=2}^{3}
\frac{\theta_1(\xi_{\alpha}\!\!-\!\!4\xi_1)\theta_2(\xi_{\alpha}\!\!-\!\!2\xi_1)}{\theta_1(\xi_{\alpha}\!\!+\!\!2\xi_1)\theta_2(\xi_{\alpha})}
\!\!+\!\!
\prod_{\alpha=2}^{3}
\frac{\theta_3(\xi_{\alpha}\!\!-\!\!4\xi_1)\theta_4(\xi_{\alpha}\!\!-\!\!2\xi_1)}{\theta_3(\xi_{\alpha}\!\!+\!\!2\xi_1)\theta_4(\xi_{\alpha})}
\nonumber\\
&&
\!\!-\!\!
\prod_{\alpha=2}^{3}
\frac{\theta_2(\xi_{\alpha}\!\!-\!\!4\xi_1)\theta_1(\xi_{\alpha}\!\!-\!\!2\xi_1)}{\theta_2(\xi_{\alpha}\!\!+\!\!2\xi_1)\theta_1(\xi_{\alpha})}
\!\!+\!\!
\prod_{\alpha=2}^{3}
\frac{\theta_4(\xi_{\alpha}\!\!-\!\!4\xi_1)\theta_3(\xi_{\alpha}\!\!-\!\!2\xi_1)}{\theta_4(\xi_{\alpha}\!\!+\!\!2\xi_1)\theta_3(\xi_{\alpha})}
\Bigg).
\end{eqnarray}
We define a function of $2\xi_1$ that is doubly periodic by multiplying $Z_{(10)}^{O(4),3}|_{z=2\xi_1}$ by appropriate factor as
\begin{eqnarray}
P_{(10)}(2\xi_1)
\equiv
Z_{(10)}^{O(4),3}|_{z=2\xi_1}
\times
\frac{\theta_2(\xi_2\!+\!\xi_3\!+\!2\xi_1)}{\theta_1(2\xi_1\!-\!\xi_2\!-\!\xi_3)\theta_1(2\xi_1\!-\!2\xi_2)\theta_1(2\xi_1\!-\!2\xi_3)}.
\end{eqnarray}
We will show $P_{(10)}=0$ by checking that it has no pole. The potential pole points are $2\xi_1=\xi_2+\xi_3,2\xi_{\alpha},-\xi_{\alpha}+\frac{a+b\tau}{2},\frac{a+b\tau}{2},\frac{1}{4}+\frac{a+b\tau}{2}$ for $\alpha=2,3$ and $a,b=0,1$ and only $\alpha=2$ will be checked since $P_{(10)}$ is symmetric under $\xi_2\leftrightarrow\xi_3$. We checked
\begin{eqnarray}
\lim_{2\xi_1\rightarrow\xi_2+\xi_3}\theta_1(2\xi_1\!\!-\!\!\xi_2\!\!-\!\!\xi_3)P_{(10)}
=0
\\
\lim_{2\xi_1\rightarrow 2\xi_2}\theta_1(2\xi_1\!\!-\!\!2\xi_2)P_{(10)}
=0,
\end{eqnarray}
where we used the result of eq.(\ref{10xi1xi2}),
\begin{eqnarray}
\lim_{2\xi_1\rightarrow-\xi_2+\frac{a+b\tau}{2}}
\theta_i(2\xi_1\!\!+\!\!\xi_2)P_{(10)}=0
\\
\lim_{2\xi_1\rightarrow \frac{a+b\tau}{2}}\theta_1(4\xi_1)P_{(10)}=0
\\
\lim_{2\xi_1\rightarrow \frac{1}{4}+\frac{a+b\tau}{2}}\theta_2(4\xi_1)P_{(10)}=0
\end{eqnarray}
for $(a,b,i)=(0,0,1),(1,0,2),(0,1,4),(1,1,3)$ and $a,b=0,1$. Those are basically the same calculations that we did from eq.(\ref{from10}) to eq.(\ref{to10}). Thus we proved $P_{(10)}$ is a doubly periodic entire function of $2\xi_1$.
\begin{eqnarray}
P_{(10)}(2\xi_1)=\lim_{2\xi_1\rightarrow 0}P_{(10)}(2\xi_1)=0.
\end{eqnarray}
Multiplying $\frac{\theta_1(2\xi_1\!-\!\xi_2\!-\!\xi_3)\theta_1(2\xi_1\!-\!2\xi_2)\theta_1(2\xi_1\!-\!2\xi_3)}{\theta_2(2\xi_1\!+\!\xi_2\!+\!\xi_3)}$, we have $Z_{(10)}^{O(4),3}|_{z=2\xi_1}=0$ except for a point $2\xi_1=\frac{1}{2}-\xi_2-\xi_3$. Since the residue of $Z_{(10)}^{O(4),3}|_{z=2\xi_1}$ at $2\xi_1=\frac{1}{2}-\xi_2-\xi_3$ is zero, $Z_{(10)}^{O(4),3}|_{z=2\xi_1}$ is continuous at this point. So, we finally have
\begin{eqnarray}
Z_{(10)}^{O(4),3}|_{z=2\xi_1}=0
\end{eqnarray}
for all $2\xi_1$. This result shows that $Q(z)$ has no pole at $z=2\xi_1$.
\end{enumerate}

We checked that $Q$ has no pole. Since $Q$ is a doubly periodic entire function of $z$ and it is independent of $z$ by the theorem. Then, we can easily evaluate
\begin{eqnarray}
Q(z)=\lim_{z\rightarrow 0}Q(z)=1
\end{eqnarray}
which means $Z_{(10)}^{O(4),3}=Z^{M,3}$.

\subsubsection{$Z_{(01)}^{O(4),3}=Z^{M,3}$}
Consider $Q(z) \equiv Z_{(01)}^{O(4),3}/Z^{M,3}$ which is a doubly periodic function under $z\rightarrow z+1, z\rightarrow z+\tau$. Again, we check that $Q(z)$ has no pole. There are many probable poles: $z=-\xi_{\alpha}+\frac{a+b\tau}{2},\frac{a+b\tau}{2},\frac{\tau}{4}+\frac{a+b\tau}{2},\xi_{\alpha}+\xi_{\beta},2\xi_{\alpha}$ for $\alpha\neq\beta, a,b = 0,1$. Since $Q(z)$ is symmetric under $\xi_{\alpha}\leftrightarrow\xi_{\beta}$, we only need to check for $z=-\xi_1+\frac{a+b\tau}{2},\frac{a+b\tau}{2},\frac{\tau}{4}+\frac{a+b\tau}{2},\xi_1+\xi_2,2\xi_1$.
\begin{enumerate}
\item $z=-\xi_1+\frac{a+b\tau}{2}$

The divergence at this point comes from $Z_{(01)}^{O(4),3}$ and $Z^{M,3}$ is nonzero at this point. So, it's sufficient to check that the residue of $Z_{(01)}^{O(4),3}$ at this point iz zero to show $Q(z)$ has no pole. We checked
\begin{eqnarray}
\lim_{z\rightarrow-\xi_1+\frac{a+b\tau}{2}}\theta_i(\xi_1+z)Z_{(01)}^{O(4),3}
&=&
0\label{from01}
\end{eqnarray}
for $(a,b,i)=(0,0,1),(1,0,2),(0,1,4),(1,1,3)$. So the residue of $Z_{(01)}^{O(4),3}$ at $z=-\xi_1+\frac{a+b\tau}{2}$ is zero for $a,b=0,1$ which means $Q(z)$ has no pole at $z=-\xi_1+\frac{a+b\tau}{2}$.

\item $z=\frac{a+b\tau}{2}$

By the same reason, it's sufficient to check that the residue of $Z_{(01)}^{M,3}$ at this point is zero to show $Q(z)$ has no pole. We checked
\begin{eqnarray}
\lim_{z\rightarrow\frac{a+b\tau}{2}}\theta_1(2z)Z_{(01)}^{O(4),3}
&=&0
\end{eqnarray}
for $a,b,=0,1$. So, the residue of $Z_{(01)}^{O(4),3}$ at $z=\frac{a+b\tau}{2}$ is zero which means $Q(z)$ has no pole at $z=\frac{a+b\tau}{2}$.

\item $z=\frac{\tau}{4}+\frac{a+b\tau}{2}$

By the same reason, it's sufficient to check that the residue of $Z_{(01)}^{O(4),3}$ at this point is zero to show $Q(z)$ has no pole. We checked
\begin{eqnarray}
\lim_{z\rightarrow\frac{\tau}{4}+\frac{a+b\tau}{2}}
\theta_4(2z)Z_{(01)}^{O(4),3}
&=&0.\label{to01}
\end{eqnarray}
So the residue of $Z_{(01)}^{O(4),3}$ at $z=\frac{\tau}{4}+\frac{a+b\tau}{2}$ is zero for $a,b=0,1$ which means $Q(z)$ has no pole at $z=\frac{\tau}{4}+\frac{a+b\tau}{2}$.

\item $z=\xi_1+\xi_2$

The divergence at this point comes from $\frac{1}{Z^{M,3}}$ and the order of divergence of it is $\frac{1}{z-\xi_1-\xi_2}$ so that we only need to check that $Z_{(01)}^{O(4),3}|_{z=\xi_1+\xi_2}=0$ to show $Q(z)$ has no pole.

\begin{eqnarray}
Z_{(01)}^{O(4),3}|_{z=\xi_1+\xi_2}
&=&
\frac{\theta_1(2\xi_3\!\!-\!\!z)\theta_1(2\xi_3\!\!+\!\!z)}{2\theta_1(2\xi_3)\theta_1(2\xi_3\!\!+\!\!2z)}
-\frac{\theta_1(z)\theta_4(z)}{4\theta_1(2z)\theta_4(2z)}\Bigg(
\frac{\theta_1(\xi_3\!\!-\!\!2z)\theta_4(\xi_3\!\!-\!\!z)}{\theta_1(\xi_3\!\!+\!\!z)\theta_4(\xi_3)}
\nonumber\\
&&+
\frac{\theta_2(\xi_3\!\!-\!\!2z)\theta_3(\xi_3\!\!-\!\!z)}{\theta_2(\xi_3\!\!+\!\!z)\theta_3(\xi_3)}
\!\!+\!\!
\frac{\theta_4(\xi_3\!\!-\!\!2z)\theta_1(\xi_3\!\!-\!\!z)}{\theta_4(\xi_3\!\!+\!\!z)\theta_1(\xi_3)}
\!\!+\!\!
\frac{\theta_3(\xi_3\!\!-\!\!2z)\theta_2(\xi_3\!\!-\!\!z)}{\theta_3(\xi_3\!\!+\!\!z)\theta_2(\xi_3)}
\Bigg)
\nonumber\\
&=&
\frac{\theta_1(2\xi_3\!\!-\!\!z)\theta_1(2\xi_3\!\!+\!\!z)}{2\theta_1(2\xi_3)\theta_1(2\xi_3\!\!+\!\!2z)}
-\frac{\theta_1(z)\theta_4(z)}{4\theta_1(2z)\theta_4(2z)}\times
\nonumber\\
&&
\frac{2\theta_1(\xi_3\!\!-\!\!\frac{z}{2})\theta_2(\xi_3\!\!-\!\!\frac{z}{2})\theta_3(\xi_3\!\!-\!\!\frac{z}{2})\theta_4(\xi_3\!\!-\!\!\frac{z}{2})\theta_1(2\xi_3\!\!+\!\!z)\theta_2(z)\theta_3(z)\theta_4(2z)}
{\theta_1(\xi_3\!\!+\!\!z)\theta_2(\xi_3\!\!+\!\!z)\theta_3(\xi_3\!\!+\!\!z)\theta_4(\xi_3\!\!+\!\!z)\theta_1(\xi_3)\theta_2(\xi_3)\theta_3(\xi_3)\theta_4(\xi_3)}
\nonumber\\
&=&
\frac{\theta_1(2\xi_3\!\!-\!\!z)\theta_1(2\xi_3\!\!+\!\!z)}{2\theta_1(2\xi_3)\theta_1(2\xi_3\!\!+\!\!2z)}
-\frac{\theta_1(z)\cdots\theta_4(z)\theta_1(2\xi_3\!\!+\!\!z)\theta_1(\xi_3\!\!-\!\!\frac{z}{2})\cdots\theta_4(\xi_3\!\!-\!\!\frac{z}{2})}{2\theta_1(2z)\theta_1(\xi_3\!\!+\!\!z)\cdots\theta_4(\xi_3\!\!+\!\!z)\theta_1(\xi_3)\cdots\theta_4(\xi_3)}
\nonumber\\
&=&
\frac{\theta_1(2\xi_3\!\!-\!\!z)\theta_1(2\xi_3\!\!+\!\!z)}{2\theta_1(2\xi_3)\theta_1(2\xi_3\!\!+\!\!2z)}
-\frac{\theta_1(2\xi_3\!\!+\!\!z)\theta_1(2\xi_3\!\!-\!\!z)}{2\theta_1(2\xi_3\!\!+\!\!2z)\theta_1(2\xi_3)}
\nonumber\\
&=&0.
\label{01xi1xi2}
\end{eqnarray}
This result shows that $Q(z)$ has no pole at $z=\xi_1+\xi_2$.

\item $z=2\xi_1$

The divergence at this point comes from $\frac{1}{Z^{M,3}}$ and the order of divergence of it is $\frac{1}{z-2\xi_1}$ so that we only need to check that $Z_{(01)}^{O(4),3}|_{z=2\xi_1}=0$ to show $Q(z)$ has no pole.

\begin{eqnarray}
Z_{(01)}^{O(4),3}|_{z=2\xi_1}
&=&
\frac{1}{4}
\frac{\theta_1(2\xi_2\!\!-\!\!2\xi_1)}{\theta_1(2\xi_2)}
\frac{\theta_4}{\theta_4(2\xi_1)}
\frac{\theta_1(\xi_3\!\!-\!\!\xi_2\!\!-\!\!2\xi_1)\theta_1(\xi_2\!\!+\!\!\xi_3\!\!-\!\!2\xi_1)}{\theta_1(\xi_3\!\!-\!\!\xi_2)\theta_1(\xi_2\!\!+\!\!\xi_3)}\times\nonumber\\
&&
\Bigg(
\!\!-\!\!
\frac{\theta_1(\xi_2)\theta_4(\xi_2)\theta_1(\xi_3\!\!-\!\!2\xi_1)\theta_4(\xi_3\!\!-\!\!2\xi_1)}
{\theta_1(\xi_2\!\!+\!\!2\xi_1)\theta_4(\xi_2\!\!+\!\!2\xi_1)\theta_1(\xi_3)\theta_4(\xi_3)}
\!\!+\!\!
\frac{\theta_2(\xi_2)\theta_3(\xi_2)\theta_2(\xi_3\!\!-\!\!2\xi_1)\theta_3(\xi_3\!\!-\!\!2\xi_1)}
{\theta_2(\xi_2\!\!+\!\!2\xi_1)\theta_3(\xi_2\!\!+\!\!2\xi_1)\theta_2(\xi_3)\theta_3(\xi_3)}
\Bigg)
\nonumber\\
&&+(\xi_2\leftrightarrow\xi_3)
\nonumber\\
&&-\frac{1}{4}
\frac{\theta_1(2\xi_1)\theta_4(2\xi_1)}{\theta_1(4\xi_1)\theta_4(4\xi_1)}
\Bigg(
\!\!-\!\!
\prod_{\alpha=2}^{3}
\frac{\theta_1(\xi_{\alpha}\!\!-\!\!4\xi_1)\theta_4(\xi_{\alpha}\!\!-\!\!2\xi_1)}{\theta_1(\xi_{\alpha}\!\!+\!\!2\xi_1)\theta_4(\xi_{\alpha})}
\!\!+\!\!
\prod_{\alpha=2}^{3}
\frac{\theta_2(\xi_{\alpha}\!\!-\!\!4\xi_1)\theta_3(\xi_{\alpha}\!\!-\!\!2\xi_1)}{\theta_2(\xi_{\alpha}\!\!+\!\!2\xi_1)\theta_3(\xi_{\alpha})}
\nonumber\\
&&
\!\!-\!\!
\prod_{\alpha=2}^{3}
\frac{\theta_4(\xi_{\alpha}\!\!-\!\!4\xi_1)\theta_1(\xi_{\alpha}\!\!-\!\!2\xi_1)}{\theta_4(\xi_{\alpha}\!\!+\!\!2\xi_1)\theta_1(\xi_{\alpha})}
\!\!+\!\!
\prod_{\alpha=2}^{3}
\frac{\theta_3(\xi_{\alpha}\!\!-\!\!4\xi_1)\theta_2(\xi_{\alpha}\!\!-\!\!2\xi_1)}{\theta_3(\xi_{\alpha}\!\!+\!\!2\xi_1)\theta_2(\xi_{\alpha})}
\Bigg).
\end{eqnarray}
We define a function of $2\xi_1$ that is doubly periodic by multiplying $Z_{(01)}^{O(4),3}|_{z=2\xi_1}$ by appropriate factor as
\begin{eqnarray}
P_{(01)}(2\xi_1)
\equiv
Z_{(01)}^{O(4),3}|_{z=2\xi_1}
\times
\frac{\theta_4(\xi_2\!+\!\xi_3\!+\!2\xi_1)}{\theta_1(2\xi_1\!-\!\xi_2\!-\!\xi_3)\theta_1(2\xi_1\!-\!2\xi_2)\theta_1(2\xi_1\!-\!2\xi_3)}.
\end{eqnarray}
We will show $P_{(01)}=0$ by checking that it has no pole. The potential pole points are $2\xi_1=\xi_2+\xi_3,2\xi_{\alpha},-\xi_{\alpha}+\frac{a+b\tau}{2},\frac{a+b\tau}{2},\frac{\tau}{4}+\frac{a+b\tau}{2}$ for $\alpha=2,3$ and $a,b=0,1$ and only $\alpha=2$ will be checked since $P_{(01)}$ is symmetric under $\xi_2\leftrightarrow\xi_3$. We checked
\begin{eqnarray}
\lim_{2\xi_1\rightarrow\xi_2+\xi_3}\theta_1(2\xi_1\!\!-\!\!\xi_2\!\!-\!\!\xi_3)P_{(01)}
=0
\\
\lim_{2\xi_1\rightarrow 2\xi_2}\theta_1(2\xi_1\!\!-\!\!2\xi_2)P_{(01)}
=0,
\end{eqnarray}
where we used the result of eq.(\ref{01xi1xi2}),
\begin{eqnarray}
\lim_{2\xi_1\rightarrow-\xi_2+\frac{a+b\tau}{2}}
\theta_i(2\xi_1\!\!+\!\!\xi_2)P_{(01)}&=&0
\\
\lim_{2\xi_1\rightarrow \frac{a+b\tau}{2}}\theta_1(4\xi_1)P_{(01)}=0
\\
\lim_{2\xi_1\rightarrow \frac{\tau}{4}+\frac{a+b\tau}{2}}\theta_4(4\xi_1)P_{(01)}=0
\end{eqnarray}
for $(a,b,i)=(0,0,1),(1,0,2),(0,1,4),(1,1,3)$ and $a,b=0,1$. Those are basically the same calculations that we did from eq.(\ref{from01}) to eq.(\ref{to01}). So we proved $P_{(01)}$ is a doubly periodic entire function of $2\xi_1$.
\begin{eqnarray}
P_{(01)}(2\xi_1)=\lim_{2\xi_1\rightarrow 0}P_{(01)}(2\xi_1)=0.
\end{eqnarray}
Multiplying $\frac{\theta_1(2\xi_1\!-\!\xi_2\!-\!\xi_3)\theta_1(2\xi_1\!-\!2\xi_2)\theta_1(2\xi_1\!-\!2\xi_3)}{\theta_4(2\xi_1\!+\!\xi_2\!+\!\xi_3)}$, we have $Z_{(10)}^{O(4),3}|_{z=2\xi_1}=0$ except for a point $2\xi_1=\frac{\tau}{2}-\xi_2-\xi_3$. Since the residue of $Z_{(01)}^{O(4),3}|_{z=2\xi_1}$ at $2\xi_1=\frac{\tau}{2}-\xi_2-\xi_3$ is zero, $Z_{(01)}^{O(4),3}|_{z=2\xi_1}$ is continuous at this point. So, we finally have
\begin{eqnarray}
Z_{(01)}^{O(4),3}|_{z=2\xi_1}=0
\end{eqnarray}
for all $2\xi_1$. This result shows that $Q(z)$ has no pole at $z=2\xi_1$.
\end{enumerate}

We checked that $Q$ has no pole. Hence $Q$ is a doubly periodic entire function of $z$ and it is independent of $z$ by the theorem. Then we can easily evaluate
\begin{eqnarray}
Q(z)=\lim_{z\rightarrow 0}Q(z)=1
\end{eqnarray}
which means $Z_{(01)}^{O(4),3}=Z^{M,3}$.

\subsubsection{$Z_{(11)}^{O(4),3}=Z^{M,3}$}
Consider $Q(z) \equiv Z_{(11)}^{O(4),3}/Z^{M,3}$ which is a doubly periodic function under $z\rightarrow z+1, z\rightarrow z+\tau$. Again, we check that $Q(z)$ has no pole. There are potential poles at $z=-\xi_{\alpha}+\frac{a+b\tau}{2},\frac{a+b\tau}{2},\frac{1+\tau}{4}+\frac{a+b\tau}{2},\xi_{\alpha}+\xi_{\beta},2\xi_{\alpha}$ for $\alpha\neq\beta, a,b = 0,1$. Since $Q(z)$ is symmetric under $\xi_{\alpha}\leftrightarrow\xi_{\beta}$, we only need to check for $z=-\xi_1+\frac{a+b\tau}{2},\frac{a+b\tau}{2},\frac{1+\tau}{4}+\frac{a+b\tau}{2},\xi_1+\xi_2,2\xi_1$.

\begin{enumerate}
\item $z=-\xi_1+\frac{a+b\tau}{2}$

The divergence at this point comes from $Z_{(11)}^{O(4),3}$ and $Z^{M,3}$ is nonzero at this point. So, it's sufficient to check that the residue of $Z_{(11)}^{O(4),3}$ at this point iz zero to show $Q(z)$ has no pole. We checked
\begin{eqnarray}
\lim_{z\rightarrow-\xi_1+\frac{a+b\tau}{2}}\theta_i(\xi_1+z)Z_{(11)}^{O(4),3}
&=&
0\label{from11}
\end{eqnarray}
for $(a,b,i)=(0,0,1),(1,0,2),(0,1,4),(1,1,3)$. So the residue of $Z_{(11)}^{O(4),3}$ at $z=-\xi_1+\frac{a+b\tau}{2}$ is zero for $a,b=0,1$ which means $Q(z)$ has no pole at $z=-\xi_1+\frac{a+b\tau}{2}$.

\item $z=\frac{a+b\tau}{2}$

By the same reason, it's sufficient to check that the residue of $Z_{(11)}^{M,3}$ at this point is zero to show $Q(z)$ has no pole. We checked
\begin{eqnarray}
\lim_{z\rightarrow \frac{a+b\tau}{2}}\theta_1(2z)Z_{(11)}^{O(4),3}
&=&0
\end{eqnarray}
for $a,b,=0,1$. So the residue of $Z_{(11)}^{O(4),3}$ at $z=\frac{a+b\tau}{2}$ is zero which means $Q(z)$ has no pole at $z=\frac{a+b\tau}{2}$.

\item $z=\frac{1+\tau}{4}+\frac{a+b\tau}{2}$

By the same reason, it's sufficient to check that the residue of $Z_{(11)}^{O(4),3}$ at this point is zero to show $Q(z)$ has no pole. We checked
\begin{eqnarray}
\lim_{z\rightarrow\frac{1+\tau}{4}+\frac{a+b\tau}{2}}
\theta_3(2z)Z_{(11)}^{O(4),3}
&=&0.\label{to11}
\end{eqnarray}
Thus the residue of $Z_{(11)}^{O(4),3}$ at $z=\frac{1+\tau}{4}+\frac{a+b\tau}{2}$ is zero for $a,b=0,1$ which means $Q(z)$ has no pole at $z=\frac{1+\tau}{4}+\frac{a+b\tau}{2}$.

\item $z=\xi_1+\xi_2$

The divergence at this point comes from $\frac{1}{Z^{M,3}}$ and the order of divergence of it is $\frac{1}{z-\xi_1-\xi_2}$ so that we only need to check that $Z_{(11)}^{O(4),3}|_{z=\xi_1+\xi_2}=0$ to show $Q(z)$ has no pole.
\begin{eqnarray}
Z_{(11)}^{O(4),3}|_{z=\xi_1+\xi_2}
&=&
\frac{\theta_1(2\xi_3\!\!-\!\!z)\theta_1(2\xi_3\!\!+\!\!z)}{2\theta_1(2\xi_3)\theta_1(2\xi_3\!\!+\!\!2z)}
-\frac{\theta_1(z)\theta_3(z)}{4\theta_1(2z)\theta_3(2z)}\Bigg(
\frac{\theta_1(\xi_3\!\!-\!\!2z)\theta_3(\xi_3\!\!-\!\!z)}{\theta_1(\xi_3\!\!+\!\!z)\theta_3(\xi_3)}
\nonumber\\
&&+
\frac{\theta_2(\xi_3\!\!-\!\!2z)\theta_4(\xi_3\!\!-\!\!z)}{\theta_2(\xi_3\!\!+\!\!z)\theta_4(\xi_3)}
\!\!+\!\!
\frac{\theta_3(\xi_3\!\!-\!\!2z)\theta_1(\xi_3\!\!-\!\!z)}{\theta_3(\xi_3\!\!+\!\!z)\theta_1(\xi_3)}
\!\!+\!\!
\frac{\theta_4(\xi_3\!\!-\!\!2z)\theta_2(\xi_3\!\!-\!\!z)}{\theta_4(\xi_3\!\!+\!\!z)\theta_2(\xi_3)}
\Bigg)
\nonumber\\
&=&
\frac{\theta_1(2\xi_3\!\!-\!\!z)\theta_1(2\xi_3\!\!+\!\!z)}{2\theta_1(2\xi_3)\theta_1(2\xi_3\!\!+\!\!2z)}
-\frac{\theta_1(z)\theta_3(z)}{4\theta_1(2z)\theta_3(2z)}\times
\nonumber\\
&&
\frac{2\theta_1(\xi_3\!\!-\!\!\frac{z}{2})\theta_2(\xi_3\!\!-\!\!\frac{z}{2})\theta_3(\xi_3\!\!-\!\!\frac{z}{2})\theta_4(\xi_3\!\!-\!\!\frac{z}{2})\theta_1(2\xi_3\!\!+\!\!z)\theta_2(z)\theta_4(z)\theta_3(2z)}
{\theta_1(\xi_3\!\!+\!\!z)\theta_2(\xi_3\!\!+\!\!z)\theta_3(\xi_3\!\!+\!\!z)\theta_4(\xi_3\!\!+\!\!z)\theta_1(\xi_3)\theta_2(\xi_3)\theta_3(\xi_3)\theta_4(\xi_3)}
\nonumber\\
&=&
\frac{\theta_1(2\xi_3\!\!-\!\!z)\theta_1(2\xi_3\!\!+\!\!z)}{2\theta_1(2\xi_3)\theta_1(2\xi_3\!\!+\!\!2z)}
-\frac{\theta_1(z)\cdots\theta_4(z)\theta_1(2\xi_3\!\!+\!\!z)\theta_1(\xi_3\!\!-\!\!\frac{z}{2})\cdots\theta_4(\xi_3\!\!-\!\!\frac{z}{2})}{2\theta_1(2z)\theta_1(\xi_3\!\!+\!\!z)\cdots\theta_4(\xi_3\!\!+\!\!z)\theta_1(\xi_3)\cdots\theta_4(\xi_3)}
\nonumber\\
&=&
\frac{\theta_1(2\xi_3\!\!-\!\!z)\theta_1(2\xi_3\!\!+\!\!z)}{2\theta_1(2\xi_3)\theta_1(2\xi_3\!\!+\!\!2z)}
-\frac{\theta_1(2\xi_3\!\!+\!\!z)\theta_1(2\xi_3\!\!-\!\!z)}{2\theta_1(2\xi_3\!\!+\!\!2z)\theta_1(2\xi_3)}
\nonumber\\
&=&0.
\label{11xi1xi2}
\end{eqnarray}
This result shows that $Q(z)$ has no pole at $z=\xi_1+\xi_2$.

\item $z=2\xi_1$

The divergence at this point comes from $\frac{1}{Z^{M,3}}$ and the order of divergence of it is $\frac{1}{z-2\xi_1}$ so that we only need to check that $Z_{(11)}^{O(4),3}|_{z=2\xi_1}=0$ to show $Q(z)$ has no pole.

\begin{eqnarray}
Z_{(11)}^{O(4),3}|_{z=2\xi_1}
&=&
\frac{1}{4}
\frac{\theta_1(2\xi_2\!\!-\!\!2\xi_1)}{\theta_1(2\xi_2)}
\frac{\theta_3}{\theta_3(2\xi_1)}
\frac{\theta_1(\xi_3\!\!-\!\!\xi_2\!\!-\!\!2\xi_1)\theta_1(\xi_2\!\!+\!\!\xi_3\!\!-\!\!2\xi_1)}{\theta_1(\xi_3\!\!-\!\!\xi_2)\theta_1(\xi_2\!\!+\!\!\xi_3)}\times\nonumber\\
&&
\Bigg(
\!\!-\!\!
\frac{\theta_1(\xi_2)\theta_3(\xi_2)\theta_1(\xi_3\!\!-\!\!2\xi_1)\theta_3(\xi_3\!\!-\!\!2\xi_1)}
{\theta_1(\xi_2\!\!+\!\!2\xi_1)\theta_3(\xi_2\!\!+\!\!2\xi_1)\theta_1(\xi_3)\theta_3(\xi_3)}
\!\!+\!\!
\frac{\theta_2(\xi_2)\theta_4(\xi_2)\theta_2(\xi_3\!\!-\!\!2\xi_1)\theta_4(\xi_3\!\!-\!\!2\xi_1)}
{\theta_2(\xi_2\!\!+\!\!2\xi_1)\theta_4(\xi_2\!\!+\!\!2\xi_1)\theta_2(\xi_3)\theta_4(\xi_3)}
\Bigg)
\nonumber\\
&&+(\xi_2\leftrightarrow\xi_3)
\nonumber\\
&&-\frac{1}{4}
\frac{\theta_1(2\xi_1)\theta_3(2\xi_1)}{\theta_1(4\xi_1)\theta_3(4\xi_1)}
\Bigg(
\!\!-\!\!
\prod_{\alpha=2}^{3}
\frac{\theta_1(\xi_{\alpha}\!\!-\!\!4\xi_1)\theta_3(\xi_{\alpha}\!\!-\!\!2\xi_1)}{\theta_1(\xi_{\alpha}\!\!+\!\!2\xi_1)\theta_3(\xi_{\alpha})}
\!\!+\!\!
\prod_{\alpha=2}^{3}
\frac{\theta_2(\xi_{\alpha}\!\!-\!\!4\xi_1)\theta_4(\xi_{\alpha}\!\!-\!\!2\xi_1)}{\theta_2(\xi_{\alpha}\!\!+\!\!2\xi_1)\theta_4(\xi_{\alpha})}
\nonumber\\
&&
\!\!-\!\!
\prod_{\alpha=2}^{3}
\frac{\theta_3(\xi_{\alpha}\!\!-\!\!4\xi_1)\theta_1(\xi_{\alpha}\!\!-\!\!2\xi_1)}{\theta_1(\xi_{\alpha}\!\!+\!\!2\xi_1)\theta_1(\xi_{\alpha})}
\!\!+\!\!
\prod_{\alpha=2}^{3}
\frac{\theta_4(\xi_{\alpha}\!\!-\!\!4\xi_1)\theta_2(\xi_{\alpha}\!\!-\!\!2\xi_1)}{\theta_4(\xi_{\alpha}\!\!+\!\!2\xi_1)\theta_2(\xi_{\alpha})}
\Bigg).
\end{eqnarray}
We define a function of $2\xi_1$ that is doubly periodic by multiplying $Z_{(11)}^{O(4),3}|_{z=2\xi_1}$ by appropriate factor as
\begin{eqnarray}
P_{(11)}(2\xi_1)
\equiv
Z_{(11)}^{O(4),3}|_{z=2\xi_1}
\times
\frac{\theta_3(\xi_2\!+\!\xi_3\!+\!2\xi_1)}{\theta_1(2\xi_1\!-\!\xi_2\!-\!\xi_3)\theta_1(2\xi_1\!-\!2\xi_2)\theta_1(2\xi_1\!-\!2\xi_3)}.
\end{eqnarray}
We will show $P_{(11)}=0$ by checking that it has no pole. The potential pole points are $2\xi_1=\xi_2+\xi_3,2\xi_{\alpha},-\xi_{\alpha}+\frac{a+b\tau}{2},\frac{a+b\tau}{2},\frac{1+\tau}{4}+\frac{a+b\tau}{2}$ for $\alpha=2,3$ and $a,b=0,1$ and only $\alpha=2$ will be checked since $P_{(11)}$ is symmetric under $\xi_2\leftrightarrow\xi_3$. We checked
\begin{eqnarray}
\lim_{2\xi_1\rightarrow\xi_2+\xi_3}\theta_1(2\xi_1\!\!-\!\!\xi_2\!\!-\!\!\xi_3)P_{(11)}
=0
\\
\lim_{2\xi_1\rightarrow 2\xi_2}\theta_1(2\xi_1\!\!-\!\!2\xi_2)P_{(11)}
=0,
\end{eqnarray}
where we used the result of eq.(\ref{11xi1xi2}),
\begin{eqnarray}
\lim_{2\xi_1\rightarrow-\xi_2+\frac{a+b\tau}{2}}
\theta_i(2\xi_1\!\!+\!\!\xi_2)P_{(11)}=0
\\
\lim_{2\xi_1\rightarrow \frac{a+b\tau}{2}}\theta_1(4\xi_1)P_{(11)}=0
\\
\lim_{2\xi_1\rightarrow \frac{1+\tau}{4}+\frac{a+b\tau}{2}}\theta_3(4\xi_1)P_{(11)}=0
\end{eqnarray}
for $(a,b,i)=(0,0,1),(1,0,2),(0,1,4),(1,1,3)$ and $a,b=0,1$. Those are basically the same calculations that we did from eq.(\ref{from11}) to eq.(\ref{to11}). So, we proved $P_{(11)}$ is a doubly periodic entire function of $2\xi_1$.
\begin{eqnarray}
P_{(11)}(2\xi_1)=\lim_{2\xi_1\rightarrow 0}P_{(11)}(2\xi_1)=0.
\end{eqnarray}
Multiplying $\frac{\theta_1(2\xi_1\!-\!\xi_2\!-\!\xi_3)\theta_1(2\xi_1\!-\!2\xi_2)\theta_1(2\xi_1\!-\!2\xi_3)}{\theta_3(2\xi_1\!+\!\xi_2\!+\!\xi_3)}$, we have $Z_{(11)}^{O(4),3}|_{z=2\xi_1}=0$ except for a point $2\xi_1=\frac{1+\tau}{2}-\xi_2-\xi_3$. Since the residue of $Z_{(11)}^{O(4),3}|_{z=2\xi_1}$ at $2\xi_1=\frac{1+\tau}{2}-\xi_2-\xi_3$ is zero, $Z_{(11)}^{O(4),3}|_{z=2\xi_1}$ is continuous at this point. So, we finally have
\begin{eqnarray}
Z_{(11)}^{O(4),3}|_{z=2\xi_1}=0
\end{eqnarray}
for all $2\xi_1$. This result shows that $Q(z)$ has no pole at $z=2\xi_1$.
\end{enumerate}

We checked that $Q$ has no pole. Hence $Q$ is a doubly periodic entire function of $z$ and it is independent of $z$ by the theorem. Then we can easily evaluate
\begin{eqnarray}
Q(z)=\lim_{z\rightarrow 0}Q(z)=1
\end{eqnarray}
which means $Z_{(11)}^{O(4),3}=Z^{M,3}$.

Thus we proved analytically
\begin{eqnarray}
Z_{(00)}^{O(4),3}=Z_{(10)}^{O(4),3}=Z_{(01)}^{O(4),3}=Z_{(11)}^{O(4),3}=Z^{M,3}
\end{eqnarray}
so that we have
\begin{eqnarray}
Z^{A,SO(4),3}&=&Z_{(00)}^{O(4),3}=Z^{M,3}
\\
Z^{A,O_{+}(4),3}&=&\frac{1}{2}\Big(Z_{(00)}^{O(4),3}+Z_{(10)}^{O(4),3}+Z_{(01)}^{O(4),3}+Z_{(11)}^{O(4),3}\Big)=2Z^{M,3}
\\
Z^{A,O_{-}(4),3}&=&\frac{1}{2}\Big(Z_{(00)}^{O(4),3}-Z_{(10)}^{O(4),3}-Z_{(01)}^{O(4),3}-Z_{(11)}^{O(4),3}\Big)=-Z^{M,3}.
\end{eqnarray}

\section*{Acknowledgements}

We thank for Hee-Cheol Kim for helpful discussions.
JP is supported in part
by the NRF Grant 2018R1A2B6007159.
The work of HK was supported by Basic Science Research Program through the National Research Foundation of Korea(NRF) funded by the Ministry of Education(2017R1A6A3A03009422) and by the Center for Mathematical Sciences and Applications at Harvard University.

\end{document}